\documentclass[twocolumn, times,tighten]{aastex62} % manuscript
\usepackage{multirow}

\def\lsol       {\ensuremath{\mathrm{L}_{\odot}}}
\def\msol       {\ensuremath{\mathrm{M}_{\odot}}}
\def\msolyr     {\ensuremath{\mathrm{M}_{\odot}\,\mathrm{yr}^{-1}}}

\def\mujybeam {\ensuremath{\mu\mathrm{Jy}\,\mathrm{beam}^{-1}}}
\def\mujy       {\ensuremath{\mu\mathrm{Jy}}}

\def\jykms {\ensuremath{\text{Jy\,km\,s}^{-1}}}
\def\kms {\ensuremath{\text{km\,s}^{-1}}}

\def\kkmspc {\ensuremath{\text{K}\,\text{km}\,\text{s}^{-1}\,\text{pc}^{2}}}

\shortauthors{Novak et al.}

\graphicspath{{./}{figures/}}

\def\banados {{Ba{\~n}ados}}

\def\cii      {\ensuremath{\text{[C\textsc{ii}]}_{158\,\mu\text{m}}}}
\def\nii   {\ensuremath{\text{\rm [N\textsc{ii}]}_{122\,\mu\mathrm{m}}}}
\def\oiii     {\ensuremath{\text{\rm [O\textsc{iii}]}_{88\,\mu\mathrm{m}}}}
\def\oi	{\ensuremath{\text{\rm [O\textsc{i}]}_{146\,\mu\mathrm{m}}}}
\def\niilow {\ensuremath{\text{\rm [N\textsc{ii}]}_{205\,\mu\mathrm{m}}}}
\def\ci {\ensuremath{\text{\rm [C\textsc{i}]}_{369\,\mu\mathrm{m}}}}

\def\pisco {J1342+0928}

\begin{document}

\title{\Large \bf An ALMA multi-line survey of the interstellar medium of the redshift 7.5 quasar host galaxy \pisco}

\correspondingauthor{Mladen Novak}
\email{novak@mpia.de}

\author[0000-0001-8695-825X]{Mladen Novak}
\affiliation{Max-Planck-Institut f\"{u}r Astronomie, K\"{o}nigstuhl 17, 69117 Heidelberg, Germany}

\author[0000-0002-2931-7824]{Eduardo Ba\~nados}
\affiliation{Max-Planck-Institut f\"{u}r Astronomie, K\"{o}nigstuhl 17, 69117 Heidelberg, Germany}
\affiliation{The Observatories of the Carnegie Institution for Science, 813 Santa Barbara Street, Pasadena, CA 91101, USA}

\author[0000-0002-2662-8803]{Roberto Decarli}
\affiliation{INAF - Osservatorio di Astrofisica e Scienza dello Spazio, via Gobetti 93/3, I-40129, Bologna, Italy}

\author[0000-0003-4793-7880]{Fabian Walter}
\affiliation{Max-Planck-Institut f\"{u}r Astronomie, K\"{o}nigstuhl 17, 69117 Heidelberg, Germany}
\affiliation{National Radio Astronomy Observatory, Pete V. Domenici Array Science Center, P.O. Box O, Socorro, NM 87801, USA}

\author[0000-0001-9024-8322]{Bram Venemans}
\affiliation{Max-Planck-Institut f\"{u}r Astronomie, K\"{o}nigstuhl 17, 69117 Heidelberg, Germany}

\author[0000-0002-9838-8191]{Marcel Neeleman}
\affiliation{Max-Planck-Institut f\"{u}r Astronomie, K\"{o}nigstuhl 17, 69117 Heidelberg, Germany}

\author[0000-0002-6822-2254]{Emanuele Paolo Farina}
\affiliation{Max-Planck-Institut f\"{u}r Astronomie, K\"{o}nigstuhl 17, 69117 Heidelberg, Germany}
%\affil{University of California Santa Barbara, Santa Barbara, CA, USA}

\author[0000-0002-5941-5214]{Chiara Mazzucchelli}
\affil{European Southern Observatory, Alonso de C\'ordova 3107, Vitacura, Regi\'on Metropolitana, Chile}

\author{Chris Carilli}
\affil{National Radio Astronomy Observatory, Pete V. Domenici Array Science Center, P.O. Box O, Socorro, NM 87801, USA}

\author{Xiaohui Fan}
\affil{Steward Observatory, The University of Arizona, 933 N.\ Cherry Ave., Tucson, AZ 85721, USA}

\author{Hans--Walter Rix}
\affiliation{Max-Planck-Institut f\"{u}r Astronomie, K\"{o}nigstuhl 17, 69117 Heidelberg, Germany}

\author{Feige Wang}
\affil{Department of Physics, University of California, Santa Barbara, CA 93106-9530, USA}

\begin{abstract}

We use ALMA observations of the host galaxy of the quasar ULAS~\pisco\ at $z=7.54$ to study the dust continuum and far infrared lines emitted from its interstellar medium. The Rayleigh-Jeans tail of the dust continuum is well sampled with eight different spectral setups, and from a modified black body fit we obtain an emissivity coefficient of $\beta=1.85\pm0.3$. 
Assuming a  standard dust temperature of 47\,K we derive a dust mass of $M_{\mathrm{dust}}=0.35\times10^8$\,M$_\sun$ and a star formation rate of $150\pm30\,\msolyr$.
We have $>4\sigma$ detections of the \cii, \oiii\ and \nii\ atomic fine structure lines and limits on the \ci, \oi\ and \niilow\ emission. We also report multiple limits of CO rotational lines with $J_{\rm up}\ge7$, as well as a tentative $3.3\sigma$  detection of the stack of four CO lines ($J_{\rm up}=11, 10, 8$ and 7). 
%These measurements allows us to constrain various physical properties of the host galaxy.
We find line deficits that are in agreement with local ultra luminous infrared galaxies.
Comparison of the \nii\ and \cii\ lines indicates that the  \cii\ emission arises predominantly from the neutral medium, and we estimate that the photo-disassociation regions in \pisco\ have densities $\lesssim 5\times10^4$\,cm$^{-3}$. The data suggest that $\sim16\%$ of hydrogen is in ionized form and that the H\textsc{ii} regions have high electron densities of $n_e>180\,\mathrm{cm}^{-3}$.
Our observations favor a low gas-to-dust ratio of $<100$, and a metallicity of the interstellar medium comparable to the Solar value. All the measurements presented here suggest that the host galaxy of \pisco\ is highly enriched in metal and dust, despite being observed just 680\,Myr after the Big Bang.

\end{abstract}


\section{Introduction} \label{sec:intro}

Observations of the early universe present a critical piece of information for the overall picture of galaxy evolution. Technical advancements continually push the observational boundaries and fainter, more distant, objects become detectable with a reasonable investment of telescope time. Quasars, being the most luminous  non-transient shining light sources, present natural targets for early universe investigations.
The large energy output of the quasar arises from  rapid accretion of material ($\gtrsim 10\,\msolyr$) onto a supermassive black hole (SMBH; mass greater than $\gtrsim10^8\,\msol$) centered in the galaxy host \citep[e.g.,][]{derosa14}.

Redshifts higher than $z\gtrsim6$ correspond to the first Gyr of the universe, which is of particular interest as it overlaps with the last  phase change of the universe: the reionization epoch \citep[see e.g.,][]{becker15}. Several hundreds quasars were found at these cosmic times, owing to large survey programs \citep[e.g.,][and references therein]{fan06,banados16, jiang16, matsuoka18}. These observations of high-redshift quasars constrain the black hole seed masses and their growth in the early universe \citep[e.g.,][]{volonteri12}, but also challenge the models of galaxy mass buildup \citep[see also][]{sijacki15, vlugt19}.

A study of atomic fine structure emission lines can provide a plethora of information on the physical properties of the interstellar medium (ISM, see \citealt{carilliwalter13} for a review). For high-redshift galaxies, several far infrared (FIR) emission lines of the most abundant atom/ion species, carbon (C), oxygen (O) and nitrogen (N),  are conveniently shifted into the millimeter  and sub-millimeter atmospheric windows accessible to facilities such as the NOrthern Extended Millimeter Array (NOEMA) and  the Atacama Large Millimeter Array (ALMA).
The singly ionized carbon line, \cii, the brightest FIR emission line emitted in both neutral and ionized medium, is an important coolant of the ISM and a good tracer of  gas kinematics. These properties led to systematic targeting of the \cii\ resulting in detections for dozens of $z>6$ quasars \cite[e.g.,][]{wang13, willott15, venemans17b, decarli18}.
Although \cii\ remains the main diagnostic line of the ISM for high-redshift quasars, other fine-structure lines can be observed to provide additional constraints on the physical properties of the ISM in these sources.

Recently, several studies were aimed at observing doubly ionized oxygen high-redshift galaxies and quasars, specifically the \oiii\ line \citep[e.g.,][]{carniani17,walter18, hashimoto18b, marrone18}. Given the high ionization energy required to produce O$^{++}$ emission, it originates exclusively in an ionized medium around early type stars or in the presence of an active galactic nuclei (AGN). This line shows a promising future for high-redshift galaxy observations \citep[see also][]{hashimoto18a}.
Singly ionized nitrogen gives rise to two emission lines \nii\ and \niilow, which can provide additional ionization diagnostics \citep[see e.g.,][]{herrera-camus16}, especially when used in conjunction with other lines. Observations of multiple fine structure lines at high redshifts are still rare \cite[e.g.,][]{tadaki19}.

%Several observations of high redshift galaxies exist targeting these transitions \citep[e.g.,][]{pavesi16, lu18}.

Beside the above mentioned fine structure lines, useful ISM diagnostic lines
can also originate in molecules \citep[see e.g.][]{carilliwalter13}. The most abundant molecule, H$_2$, is not practically observable and we rely on tracers such as carbon monoxide ($^{12}$CO) to measure the gas content of galaxies \citep[for a review see][]{bolatto13}. When multiple CO rotational line detections are available, it is possible to put constraints on the kinetic temperature of the gas \citep[see e.g.,][]{daddi15}.
Several studies of CO excitation ladders for high-redshift quasars were performed showing its usefulness in analyzing molecular gas excitation \citep[e.g.,][]{weiss07,riechers09,carniani19}.
In denser regions, where it is more difficult to observe common gas tracers due to optical depth effects, the water (H$_2$O) and the hydroxyl (OH, OH$^+$) molecules can provide valuable insight into the state of the ISM. Their emission probes heated regions of star formation or in the presence of an AGN \citep[e.g.,][]{liu17}, but it can also be tied to shocked regions and molecular outflows 
\cite[e.g.,][]{fischer10,gonzales13}. Due to the complex energy level diagram of water, multiple lines are necessary for a valid interpretation \cite[see e.g.,][]{vanderwerf11,riechers13}.

In this work, we present a multi-line search survey targeting the quasar host galaxy ULAS~\pisco\ (hereafter \pisco), which to date holds the record as the most distant quasar observed.
It was discovered by \cite{banados18}, where they report
% a redshift of $z=7.527$ based on the $\text{Mg{\sc ii}}_{2.4\,\mu\mathrm{m}}$ emission line,
 an absolute  AB magnitude at 1\,450\,\AA\  of M$_{1450}= -26.8$, bolometric luminosity of $L_{\mathrm{bol}}=10^{13}\lsol$, and a SMBH mass of $8\times10^8\,\msol$. It was followed-up with NOEMA by \cite{venemans17c} in order to constrain the dust continuum and the ionized carbon emission.
These observations resulted in the detection of bright \cii\ emission, and upper limits of several CO
emission lines. For the \cii\ line, a redshift of $z=7.5413$, and a line-width of 380\,\kms\ were reported, which we adopt throughout. 
We here present ALMA observations targeting various emission lines of carbon, oxygen and nitrogen species, as well as molecular lines of carbon monoxide, hydroxyl and water in order to explore, in detail, the conditions of the ISM in the most distant quasar host galaxy known.

Throughout the paper we assume the concordance lambda cold dark matter ($\Lambda$CDM) cosmology with the Hubble constant of  $H_0=70$\,km\,s$^{-1}$\,Mpc$^{-1}$, dark energy density of $\Omega_\Lambda=0.7$, and matter density of $\Omega_{\mathrm{m}}=0.3$. 
At the redshift of the source ($z=7.5413$) the age of the universe is 0.68\,Gyr, and an angular size of 1$\arcsec$ corresponds to 5.0\,kpc.

% lum dist DL= 75179.8 Mpc

\section{Data} \label{sec:data}

\subsection{Observations and data reduction}

We have observed the quasar \pisco\ at $z=7.5413$  with ALMA using between 41 and 46 antennas with 12\,m diameter, eight different frequency setups reaching an effective bandwidth of 60\,GHz, located between 93.5\,GHz (band~3) and 412\,GHz (band~8), to cover 24 molecular and fine-structure emission lines (program ID: 2017.1.00396.S, PI: \banados). Table~\ref{tab:obs} summarizes all observational setups along with the total (science) time spent on source.

\begin{table*}
	\caption{Frequency setups of our ALMA observations of \pisco\ and properties of the derived continuum maps, which exclude emission line channels.
		Flux densities were measured inside a $2.6\arcsec$ diameter aperture and were corrected for the residual, as explained in Sect.~\ref{sec:fluxes}.
		Previously available radio data is listed at the end for completeness \citep[see][]{venemans17c}. }
	\label{tab:obs}
	\centering
	%	\scriptsize 
	\begin{tabular}{cccccccc}
		\hline
		Band & $\nu_{\rm obs}$\footnote{Center of the entire frequency setup.} & Beam & rms  & Aperture $S_\nu$ & Covered emission lines & Observation & Time on \\
		&		GHz & arcsec$^2$ & \mujybeam & \mujy			&		& date(s) 		& source \\
		\hline
		3  		& 101.3 & $0.92\times 0.72$   & 7.5  & ﻿$ <22.4\footnote{Upper limits correspond to $3\sigma$ of the local rms noise.} $  &	{\rm CO} (7-6),  \ci, {\rm CO} (8-7), {\rm OH}$^+$	& 2018	Jan 16					&	67\,min\\
		4  		& 141.6 & $1.07\times 0.92$   & 9.2  &  $ 63.8 \pm  21 $									 &	{\rm CO} (10-9),  {\rm CO} (11-10), {\rm H$_2$O}				&	  2018	 Mar 13/14	&	82\,min        \\
		5 (a) 	& 176.0 & $1.69\times 1.07$  & 19   &  $  137 \pm  31 $ 									 &	\niilow							&	  2018		Aug 23/24								&	71\,min                      \\
		5 (b) 	& 195.3 & $1.90\times 1.33$  & 26   &  $  214 \pm  36 $ 									 &	{\rm CO} (14-13),  {\rm CO} (15-14), {\rm H$_2$O}				&	 2018	Jul 4		&	46\,min           \\
		6 (a) 	& 221.8 & $1.11\times 0.88$  & 19   &  $  257 \pm  42 $ 									 &	{\rm CO} (16-15), {\rm CO} (17-16), {\rm OH}, {\rm OH}$^+$			&	2018 Apr 12		&	39\,min           \\
		6 (b)   & 231.5 & $0.27\times 0.19$  & 7.7  &  $  394 \pm  74 $ 									 &	\cii,  \oi, {\rm H$_2$O}											&	2017 Dec 26			&	114\,min            \\
		7 		& 293.7 & $1.23\times 0.94$   & 24   &  $  583 \pm  49 $									 &	\nii							&  2018 May	8/15 											&	56\,min                       \\
		8 		& 403.9 & $0.98\times 0.59$   & 110  &  $  993 \pm 320 $									 &	\oiii								&	2018	May 12									&	22\,min                          \\
		\hline
				FIRST & 1.4 & $5.4\times 5.4$ & 144&  $<432$\\
		VLA&41&$2.2\times 2.0$& 5.7 &$<17.1$ \\
		\hline
	\end{tabular}
\end{table*}

The default calibration pipeline was used to process the raw data with the Common Astronomy Software Applications (CASA) package \citep{mcmullin07}. 
Visibility data of various execution blocks were merged together and imaged in CASA using {\sc tclean}. We used natural weighting to maximize sensitivity in the resulting maps. 
Several imaging products were created, as follows. For each setup we  produced a cube with a channel width corresponding to 50\,km\,s$^{-1}$ at the central frequency of the setup. We  then subtracted the continuum estimated from all channels excluding the ones where emission lines might be detected using the {\sc uvcontsub} task (covered emission lines are listed in Table~\ref{tab:obs}).

Only the \cii\ observations have enough signal-to-noise to model the shape of the line, specifically its width and precise redshift. These measurements were already performed with the low resolution NOEMA observations by \cite{venemans17c}.
For other lines we assume that the full width at half maximum (FWHM) remains the same (further discussed in Sect.~\ref{sec:lines}). 
To maximize the measured signal of the other lines, we imaged integrated emission line maps (also known as moment zero maps) over a velocity width of 455\,km\,s$^{-1}$, corresponding to 1.2 times the FWHM of the \cii\ line, centered at the targeted line frequency redshifted to $z=7.5413$. 
A Gaussian fit to our \cii\ spectrum measured inside an aperture (as explained in the following section) across 50\,\kms\ wide channels would result in a value of  $z=7.5400 \pm 0.0005$, with line flux and width that are also in agreement with already published values. For consistency reasons, we adopt the redshift and the line width values reported in \cite{venemans17c} and use them throughout.
%line flux of $1.22\pm 0.15$\,Jy\,\kms, FWHM of $318 \pm 29$\,\kms, and a redshift of $7.5400 \pm 0.0003$.

Additional  455\,km\,s$^{-1}$ wide channels were imaged on both sides of the moment zero line map in order to visually check that the continuum subtraction performed as expected.
We also image the continuum part of each spectral setup (i.e., excluding line emission).
In every map cleaning was performed inside a circle of 4$\arcsec$ diameter located at the phase center, and another circle positioned on the foreground source in the field 10$\arcsec$ away (in the NE direction), down to $2\sigma$ of the rms noise inside the cube. The foreground source was cleaned to prevent its sidelobes from biasing our flux measurements, as it is detected in every continuum band that we have observed. We also detect several emission lines in this source, and their observed frequencies imply that the source is not related to \pisco. Further analysis of the foreground source is outside the scope of this paper.

\subsection{Measuring flux densities}
\label{sec:fluxes}

To ensure that we are probing the same spatial scales independent of the resolution of the maps, we extract flux densities for both the continuum and line emissions  inside an aperture with a diameter of $2.6\arcsec$, corresponding to 13\,kpc at the redshift of the source. This approach yields measurements that describe average properties of the galaxy, which enables comparison in a consistent way. This  aperture size was chosen in such a way to encompass all recoverable \cii\ extended emission, which was detected at the highest signal-to-noise (S/N) ratio and also at the highest resolution (see Table~\ref{tab:obs} and Appendix~\ref{sec:residual_scaling}).  No significant emission was recovered for the continuum or other emission lines beyond the chosen aperture.
We employ the residual scaling method when computing aperture flux densities in order to mitigate the issue of ill-defined units in interferometric maps and effects of sidelobes. We refer the reader to Appendix~\ref{sec:residual_scaling} for further technical details. 
Error estimate of the measured flux density is calculated as $\sigma\sqrt{N}$, where $\sigma$ is the local rms (in units of Jy\,beam$^{-1}$) and $N$ is the number of independent clean beams that fill the aperture.

%pz 22
%px,py 514 506
%deltav 52.01747781641135
%Gauss fit
%line peak [Jy]:  0.003602102345749136 0.0005063385736493974
%nu [GHz]:  222.5453916148006 0.016348034755921965
%fwhm [km/s]:  320.31406647688806 52.04346394289653
%z:  7.540010584850066 0.000553884105200903
%integral [Jy.km/s] 1.2280925077271387 0.2638478772156026

\section{Results and discussion} \label{sec:res}

\subsection{The dust continuum emission} \label{sec:disc_cont}

\begin{figure}
	\centering
	\includegraphics[width=\linewidth]{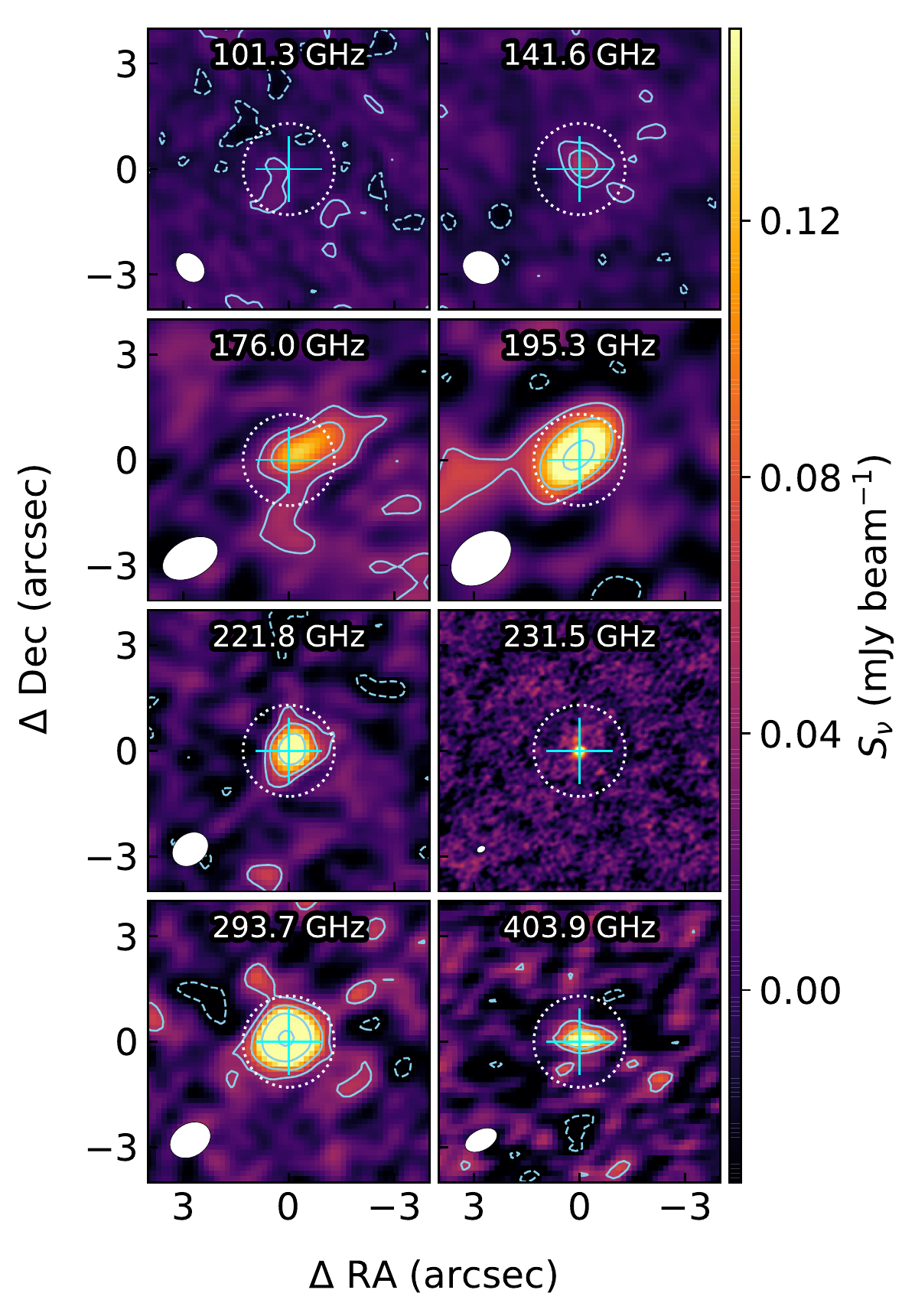}
	
	\caption{Continuum maps of eight different spectral setups of \pisco\ (frequencies are given at  the top of each panel). The respective synthesized beam sizes are shown in the lower left corner of each panel. The colorbar is multiplied by a factor of 0.5 (4) inside the 231.5 (403.9)\,GHz panel to facilitate better contrast. The white dotted circle represents the $2.6\arcsec$ diameter aperture used for flux extraction. Full (dashed) contours outline +(-) $2\sigma$, $4\sigma$, $8\sigma$ and $16\sigma$, where $\sigma$ is the rms noise (see Table~\ref{tab:obs}). Cross marks the dust continuum center obtained from the high resolution data ($13^{\text{h}}42^{\text{m}}8.098^{\text{s}}$ $+9^\circ28\arcmin38.35\arcsec$, ICRS).}
	\label{fig:cont}
\end{figure}

\begin{figure}
	\centering
	\includegraphics[width=\linewidth]{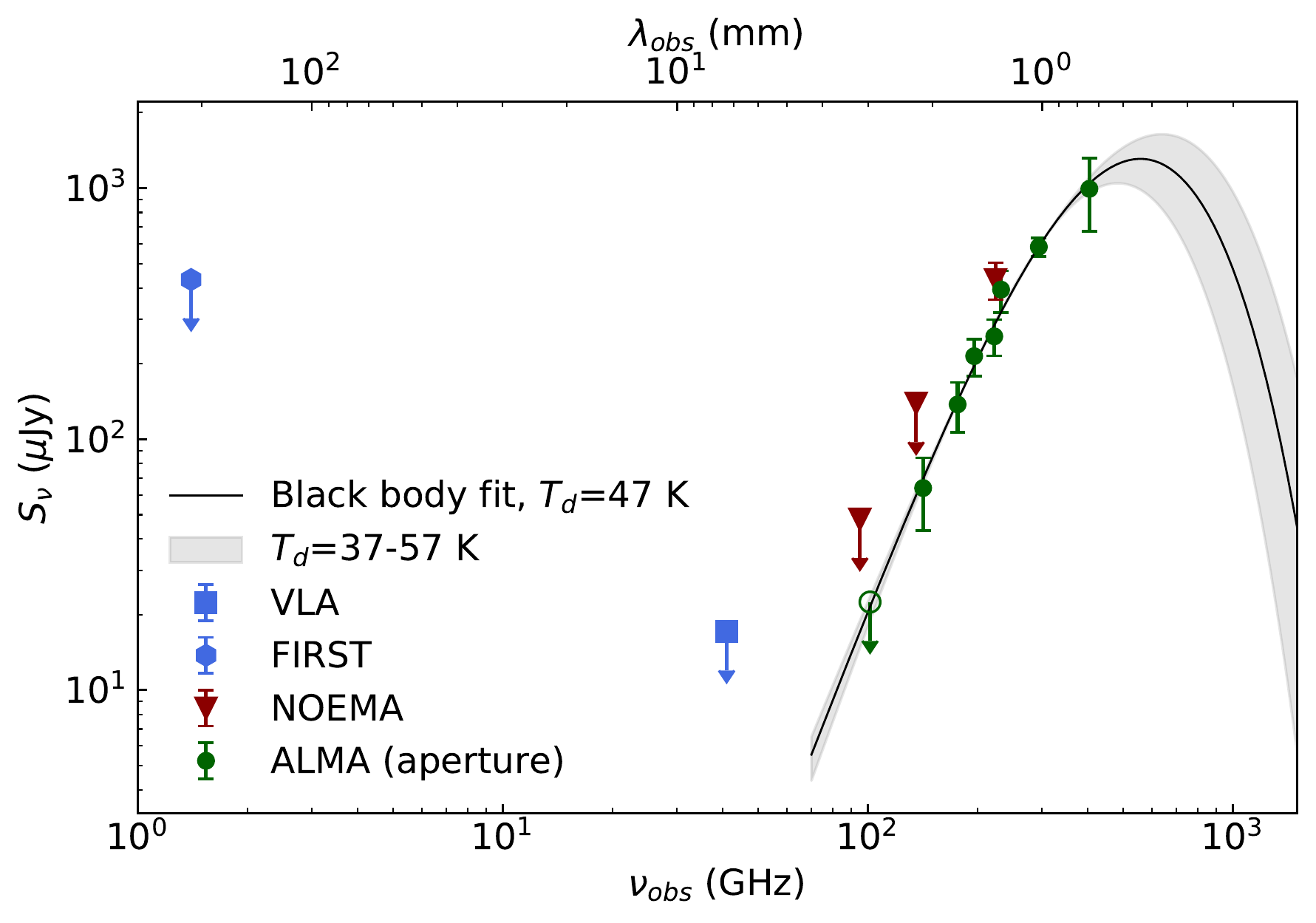}
	\caption{Spectral energy distribution of the dust continuum emission of \pisco. Our  ALMA measurements are shown with green points. Only $3\sigma$ upper limits are available for the radio continuum \citep{venemans17c}. A modified black body fit with intrinsic dust temperature of 47\,K is shown with black line, while the grey shaded area shows the fit for a wider range of temperatures (see text for details).}
	\label{fig:cont_sed}
\end{figure}

The continuum emission of \pisco\ was detected in seven out of eight observed spectral setups
with a peak surface brightness $\mathrm{S/N}>4$, while only an upper limit is available in the lowest frequency band. Continuum maps are  shown in Fig.~\ref{fig:cont}, while flux  density measurements and beam sizes are summarized in Table~\ref{tab:obs}. 
Continuum emission is spatially coincident in all observed bands, and we define its center at the peak of the high resolution 
231.5\, GHz observations: $13^{\text{h}}42^{\text{m}}8.098^{\text{s}}$ $+9^\circ28\arcmin38.35\arcsec$ (ICRS), also shown with a cross in Fig.~\ref{fig:cont}. 
The peak of the continuum coincides with the center measured in the {\it J}-band image obtained with the Magellan Baade telescope  and corrected to \emph{Gaia} DR2 astrometry, within the uncertainty of 50 mas.
Two setups in band 5 (176 and 195.3\,GHz) were observed at a lower resolution and appear slightly extended at $2\sigma$ significance. We attribute this behavior to noise fluctuations, and note that our nominal aperture integral is still consistent within the error bars with a larger aperture measurement that would encompass more of this extended flux.
We show the measured \pisco\ spectral energy distribution (SED) in Fig.~\ref{fig:cont_sed} along with $3\sigma$ upper limits on the radio continuum from the Faint Images of the Radio Sky at Twenty-cm (FIRST) survey and a targeted  Karl G. Jansky Very Large Array (VLA) observation \citep[see][]{venemans17c}.

To constrain the FIR properties of \pisco, we fit a modified black body to our observations under the assumption that the dust optical depth is small at FIR wavelengths (optically thin regime). In this case the observed flux density originating from the heated dust can be expressed as
\begin{equation}
S_{\nu_{\mathrm{obs}}}=f_{\mathrm{cmb}}(1+z) D_L^{-2} \kappa_{\nu_{\mathrm{rest}}} M_{\mathrm{dust}} B_{\nu_{\mathrm{rest}}}(T_{\mathrm{dust},z}),
\end{equation}
where $f_{\mathrm{cmb}}$ is a measure of contrast against the cosmic microwave background (CMB) radiation, $D_L$ is the luminosity distance, $\kappa_{\nu_{\mathrm{rest}}}$ is the dust mass opacity coefficient, 
$M_{\mathrm{dust}}$ is the dust mass, $B_{\nu_{\mathrm{rest}}}$ is the black body radiation spectrum, $T_{\mathrm{dust},z}$ is the temperature of the dust, which includes heating by the CMB at redshift $z$, and rest and observed frequencies are related via $\nu_{\mathrm{rest}}=(1+z)\nu_{\mathrm{obs}}$, with all values given in SI units.
The opacity coefficient is defined as $\kappa_0$ (units of m$^2$\,kg$^{-1}$) at a reference frequency $\nu_0$ with a power-law dependence on frequency
\begin{equation}
\kappa_{\nu_{\mathrm{rest}}}=  \kappa_{\nu_0} (\nu_{\mathrm{rest}}/\nu_0)^\beta,
\label{eq:beta}
\end{equation}
where $\beta$ is the dust spectral emissivity index.  Given the high redshift of \pisco, we correct for both the contrast and the CMB heating as prescribed by \cite{dacunha13}, namely 
\begin{equation}
f_{\mathrm{cmb}}=1 - B_{\nu_{\mathrm{rest}}}(T_{\mathrm{cmb},z}) / B_{\nu_{\mathrm{rest}}}(T_{\mathrm{dust},z})
\end{equation}
\begin{equation}
T_{\mathrm{dust},z}= \left(T_{\mathrm{dust}}^{\beta+4} + T_{\mathrm{cmb},z=0}^{\beta+4}\left[(1+z)^{\beta+4}-1\right]\right)^{\frac{1}{4+\beta}},
\end{equation}
where $T_{\mathrm{dust}}$ is the intrinsic dust temperature the source would have at redshift zero, $\beta$ is the same as in Eq.~\ref{eq:beta}, and $T_{\mathrm{cmb},z}=2.73 (1+z)\,\mathrm{K}=23.3\,$K is the CMB temperature at the redshift of \pisco.

We adopt the opacity coefficient of $\kappa_{\nu_0}=2.64$\,m$^2$\,kg$^{-1}$ at $\nu_0=c/(125\,\mu\mathrm{m})$ \citep{dunne03}. This parameter directly scales the dust mass and provides the dominant uncertainty on the dust mass estimate, which is at least a factor of two. 
All of our continuum measurements lie on the Rayleigh-Jeans tail and do not cover the peak of the dust SED, therefore the fitting parameters are degenerate and the dust SED shape and its temperature cannot be fully constrained.
Fixing the dust temperature  to 47\,K, as is often assumed for quasar hosts in the literature \citep[see e.g.,][]{beelen06},  results in $\beta=1.85 \pm 0.3$ and $M_{\mathrm{dust}}=(0.35\pm0.02)\times10^8$\,M$_\sun$. The integral of the SED yields the far infrared (FIR, $42.5-122.5\,\mu$m) and total infrared (TIR, $8-1000\,\mu$m) luminosities of $L_{\mathrm{FIR}}=(1.0\pm0.2)\times10^{12}$\,L$_\sun$ and $L_{\mathrm{TIR}}=(1.5\pm0.3)\times10^{12}$\,L$_\sun$. With this luminosity \pisco\ can be classified as an ultra luminous infrared galaxy (ULIRG, $L_{\mathrm{TIR}}>10^{12}$\,L$_\sun$).
The TIR luminosity can be converted to the star formation rate (SFR) using the \cite{kennicutt98} relation scaled to \cite{chabrier03} initial mass function (IMF;  yielding 1.7 times smaller SFR value than the Salpeter IMF) as
\begin{equation}
	\mathrm{SFR}[\mathrm{M}_\sun\,\mathrm{yr}^{-1}]=10^{-10} {\mathrm{L}_{\mathrm{TIR}}} [\mathrm{L}_\sun],
\end{equation}
resulting in SFR of $150\pm30\,\msolyr$ for \pisco. The calibration from \cite{kennicutt12} would give $14\%$ smaller values.
A range of temperatures\footnote{Larger temperature yields smaller $\beta$ and $M_{\mathrm{dust}}$, but larger IR integrated luminosities.} between 37 and 57\,K yields $\beta=1.6-2.2$, $M_{\mathrm{dust}}=(0.2-0.8)\times10^8$\,M$_\sun$, $L_{\mathrm{FIR}}=(0.8-1.3)\times10^{12}$\,L$_\sun$, $L_{\mathrm{TIR}}=(1.0-2.2)\times10^{12}$\,L$_\sun$ and $\mathrm{SFR}=(100-230)\,\mathrm{M}_\sun\,\mathrm{yr}^{-1}$, consistent with previous estimates from NOEMA observations \citep{venemans17c}.

\subsection{Emission lines}
\label{sec:lines}

\begin{figure}
	
	\includegraphics[width=\linewidth]{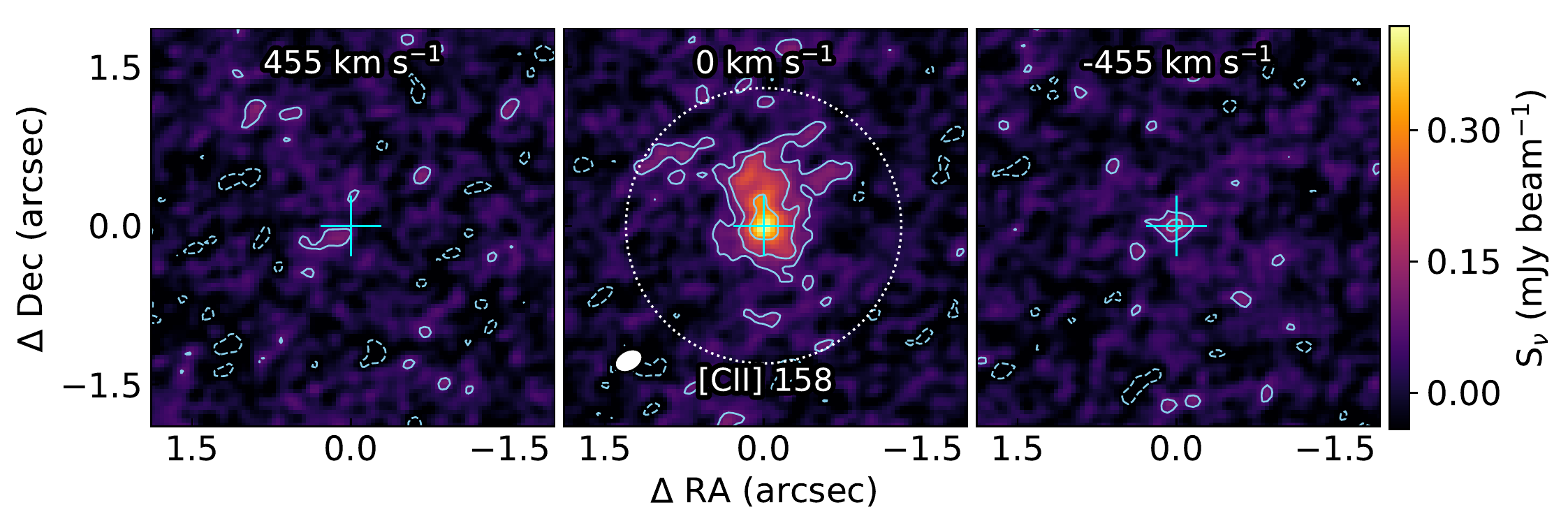}
	\includegraphics[width=0.73\linewidth]{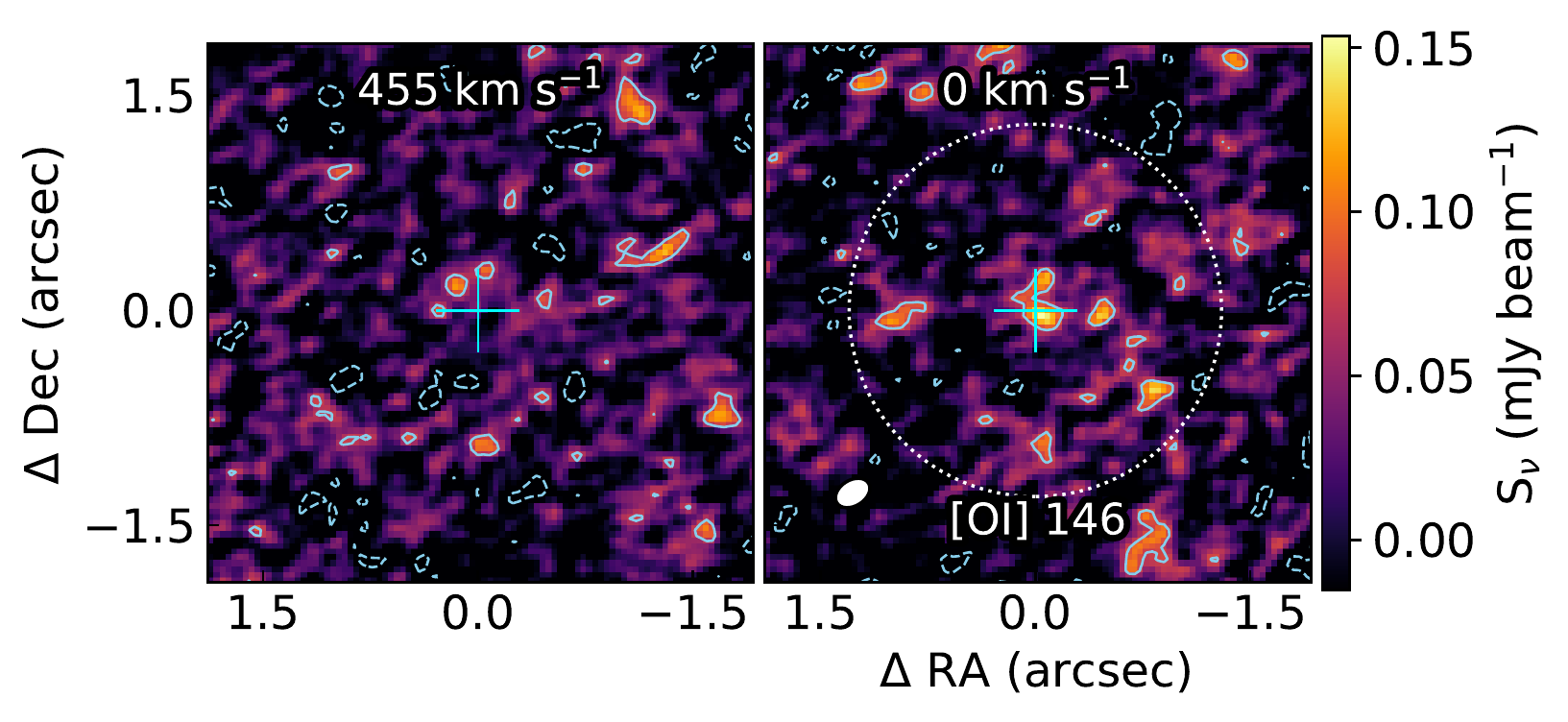}
	\includegraphics[width=\linewidth]{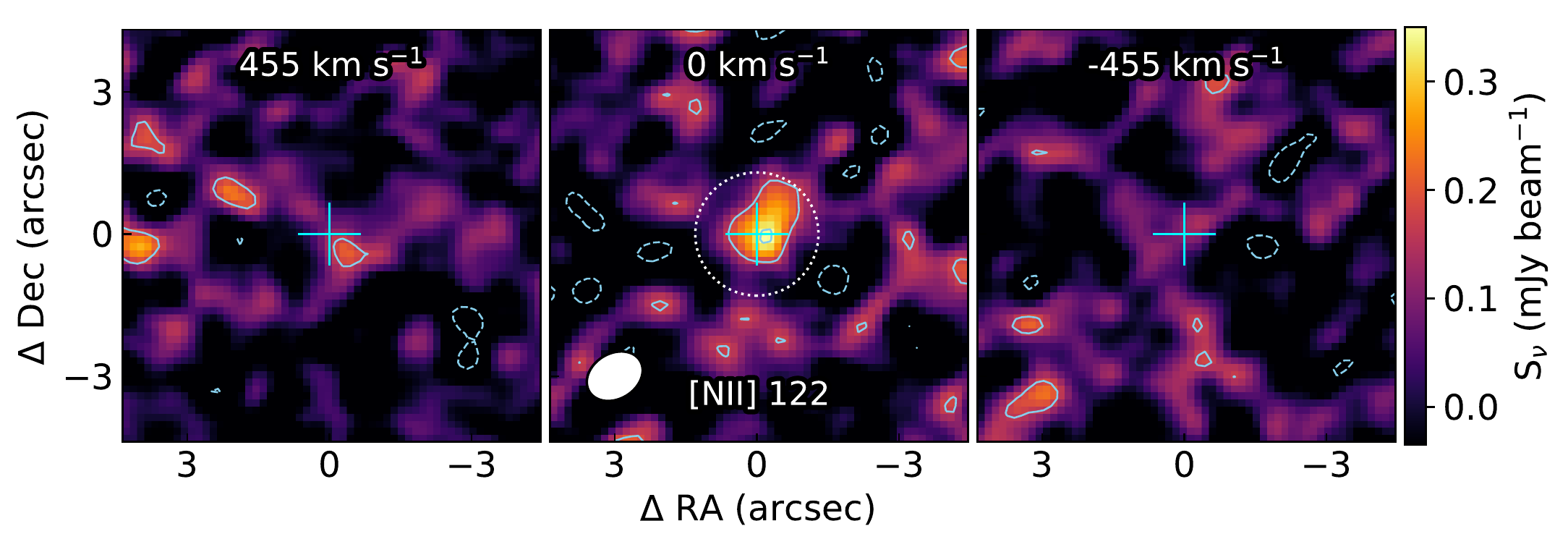}
	\includegraphics[width=\linewidth]{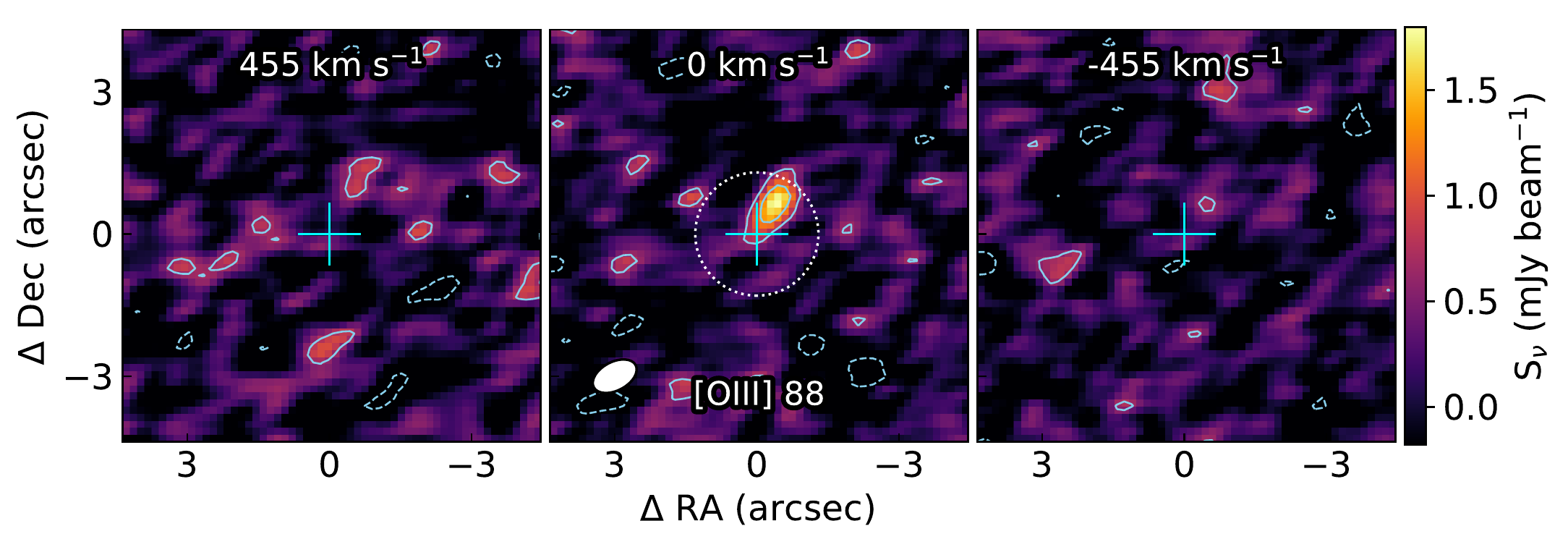}
	\includegraphics[width=\linewidth]{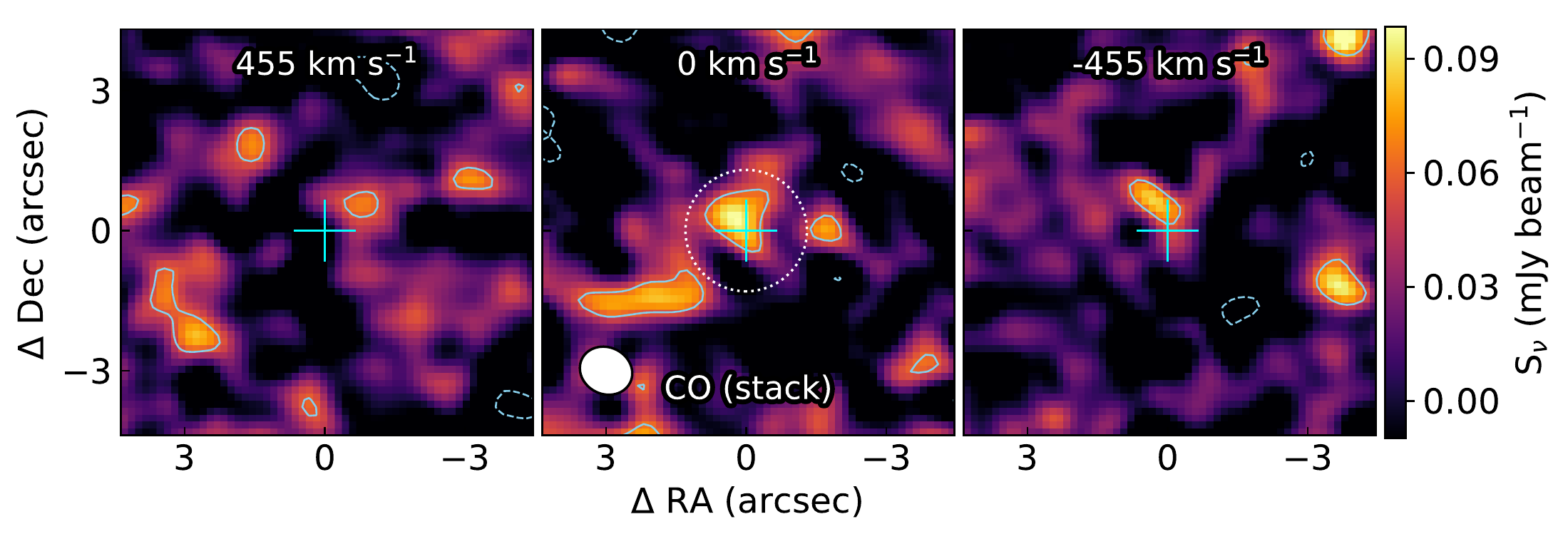}
	\caption{Cutouts of \pisco\ showing emission line in the central panel. Left and right panels demonstrate the absence of the continuum emission due to its successful subtraction.
		Each panel contains flux integrated over a width of 455\,km\,s$^{-1}$ ($1.2\times$FWHM of the \cii\ line), with the exception of  \oi\ where the central panel is 336\,km\,s$^{-1}$, due to its proximity to the sideband edge. Aperture used for flux integration is shown as the white dotted circle (diameter of 2.6$\arcsec$). The synthesized beam is shown in the lower left corner of the middle panel.
		Full (dashed) contours represent the +(-) $2\sigma$, $4\sigma$ and $8\sigma$ emission significance. The rms values in the middle panel are, from top to bottom: 35, 39, 85, 318 and 30\,\mujybeam.
		The cross marks the dust continuum emission center in the high resolution data (same as in Fig.~\ref{fig:cont}).
		The bottom three panels show a stack of four CO lines in bands 3 and 4 ($J_{\rm up}=11, 10, 8$ and 7).}
	\label{fig:line_det}
\end{figure}

Our observational spectral setups cover six atomic fine structure lines (from C, N and O), eight CO rotational lines, and multiple water and hydroxyl lines  (see Table~\ref{tab:lines}). Because of the low S/N of all lines except the \cii, we only analyze the integrated spectrum moment zero maps throughout. We show spectra of all detections and non-detections in Appendix~\ref{sec:non_det}.
Different atomic and molecular transitions trace different phases of the ISM and their line widths do not have to be equal. Nevertheless, we still adopt the same integration window for all potential lines as this provides the least biased measurement given the low S/N.
Line luminosities were derived using equations given in \cite{carilliwalter13} \citep[see also][]{solomon97}, namely
\begin{equation}
\frac{L_{\mathrm{line}}^\prime}{\mathrm{K}\,\mathrm{km}\,\mathrm{s}^{-1}\,\mathrm{pc}^{2}}=\frac{3.25\times10^7}{(1+z)^3}\frac{S_{\mathrm{line}}\Delta v}{\mathrm{Jy}\,\mathrm{km}\,\mathrm{s}^{-1}} \left(\frac{D_L}{\mathrm{Mpc}}\right)^2\left(\frac{\nu_{\mathrm{obs}}}{\mathrm{GHz}}\right)^{-2},
\end{equation}
commonly used for molecular (specifically CO) transitions, and
\begin{equation}
\frac{L_{\mathrm{line}}}{L_\sun}=1.04\times10^{-3}\frac{S_{\mathrm{line}}\Delta v}{\mathrm{Jy}\,\mathrm{km}\,\mathrm{s}^{-1}}\left(\frac{D_L}{\mathrm{Mpc}}\right)^2\frac{\nu_{\mathrm{obs}}}{\mathrm{GHz}},
\label{eq:lprime}
\end{equation}
often used when comparing lines and the underlying continuum.

\subsubsection{Fine structure lines}

We observe \cii, \nii,  \oiii, and \oi\ fine structure lines at peak surface brightness S/N of 12, 4.1, 5.7, and 3.9, respectively. These measurements were taken from the moment zero maps integrated over 455\,\kms\  (336\,\kms\ in the case of \oi), with subtracted continuum.  We show these maps in Fig.~\ref{fig:line_det} and list measured aperture flux densities in Table~\ref{tab:lines}.

The \cii\ observations were taken at a high resolution ($\approx0.2\arcsec$), where the emission is resolved into a morphologically complex structure, possibly a merger, which is discussed in the accompanying paper (\banados\ et al., in prep.). In this work we use only the total integrated emission to study relations between different lines and the underlying continuum averaged across the entire galaxy.
The \oi\ emission line is also observed at high resolution. However, the surface brightness sensitivity allows measurement of only the peak emission arising from the central region. 
This line lies on the edge of the sideband and is not fully recovered, and a smaller frequency range had to be used for integration across the line. 
Since the peak surface brightness of \oi\ is $3.9\sigma$, we consider this line a tentative detection,
and report a lower limit on the total emission measured from this single resolution element. Additionally, we provide  a $3\sigma$ upper limit on the integrated flux inside the aperture, where $\sigma$ is the uncertainty of the aperture integration.
The \nii\ emission line is observed at lower resolution ($\sim1.25\arcsec$) and is spatially coincident with the continuum emission. 
We detect the \oiii\  line at a resolution of $\sim1\arcsec$ with a peak that is slightly offset ($0.6\arcsec$) from the central region and the underlying continuum, however the line emission is still covered in its entirety by the aperture. Finally, no significant emission was observed for the \ci\ and \niilow\ lines and only upper limits can be obtained, which we set to $3\sigma$ of the noise level.

\subsubsection{Molecular lines}

We do not detect any individual molecular (CO, H$_2$O, OH) transitions, so we report their $3\sigma$ upper limits ($\sigma$ is the rms noise in the line map) in Table~\ref{tab:lines}. Given the number of available lines that are covered with our observations, we have attempted to recover an average measurement through stacking. When the four lowest available CO transitions covered in bands 3 and 4 are stacked together (namely transitions 7-6, 8-7, 10-9 and 11-10)  we recover a $3.3\sigma$ peak at the expected spatial and spectral coordinates, with a line flux of $0.045\pm0.013$\,\jykms\ extracted at the peak of the stacked emission\footnote{We note that this value is consistent with the aperture integration within the errors, but has a higher S/N.}.  
Data in the higher frequency bands have higher noise or poorer resolution, and the expected CO brightness is lower for higher J transitions, therefore we do not include them in our CO stack (we stacked them separately, but no significant detection was recovered).
For the sake of convenience, we will treat the four line stack as a tentative CO (9-8) detection, because this transition frequency corresponds to the mean frequency of the stacked lines. 
However, the effective frequency of the stacked emission is strongly dependent on the actual shape of the CO spectral energy distribution, which is presently unknown.  

We stacked two water lines in band 4 (260 and 258\,$\mu$m), resulting only in a $3\sigma$ upper limit on line flux density of 0.045\,\jykms. The remaining two available water lines were observed at significantly different resolution, and convolving all water lines to the same beam would result in a degraded signal quality.

\subsection{Properties of the ISM}

In the following sections we focus on specific lines and their ratios to derive  various properties of the ISM in \pisco.

\begin{table}
	\centering
	\caption{Fine structure and molecular lines measurements in \pisco.
Flux densities were measured inside a $2.6\arcsec$ diameter aperture, unless stated otherwise, and corrected for the residual. Upper limits correspond to $3\sigma$ of the local rms noise.}
	
	\label{tab:lines}

\begin{tabular}{cccc}
		\hline
	Line & $S_{\text{line}}\Delta v $\footnote{In all cases $\Delta v=455$\,\kms, except for \oi\ where $\Delta v=336$\,\kms\ due to availability of fewer channels.} & $L_{\text{line}}$ & $L^\prime_{\text{line}} $ \\
	& \jykms& $10^8\,L_\sun$ &$10^8$\,K\,km\,s$^{-1}$\,pc$^{2}$\\
	\hline
	
	{\rm [CI]}$_{369\,\mu\mathrm{m}}$ & $<0.074$ & $<0.41$ & $<24.0$ \\
		{\rm [NII]}$_{205\,\mu\mathrm{m}}$ & $<0.079$ & $<0.8$ & $<8.0$ \\
		{\rm [CII]}$_{158\,\mu\mathrm{m}}$ & $1.07\pm0.15$ & $14.0\pm1.9$ & $63.8\pm8.8$ \\
	{\rm [OI]}$_{146\,\mu\mathrm{m}}$\footnote{Given insufficient surface brightness sensitivity of the higher resolution map we report a $3\sigma$ upper limit on the aperture integrated emission, and a lower limit from the $3.9\sigma$ significant single beam measurement.} & $0.052 -0.40 $ & $0.73 - 5.7$ & $2.62 - 20$ \\
	{\rm [NII]}$_{122\,\mu\mathrm{m}}$ & $0.21\pm0.078$ & $3.55\pm1.3$ & $7.46\pm2.8$ \\
	{\rm [OIII]}$_{88\,\mu\mathrm{m}}$ & $1.14\pm0.41$ & $26.5\pm9.7$ & $21.2\pm7.7$ \\
	
	\hline
		CO stack\footnote{Stack of four CO lines in bands 3 and 4: CO (7-6), (8-7), (10-9), (11-10).} & $0.045\pm0.013$\footnote{Peak surface brightness and rms is reported (it is consistent with aperture integration within the errors, but has a  higher S/N).} & $0.32\pm0.1$ & $9.0\pm2.7$ \\

		CO (7-6) & $<0.078$ & $<0.43$ & $<26.0$ \\
	CO (8-7) & $<0.08$ & $<0.51$ & $<20.0$ \\
	CO (10-9) & $<0.069$ & $<0.55$ & $<11.0$ \\
	CO (11-10) & $<0.073$ & $<0.64$ & $<9.8$ \\
	CO (14-13) & $<0.24$ & $<2.7$ & $<20.0$ \\
	CO (15-14) & $<0.12$ & $<1.4$ & $<8.3$ \\
	CO (16-15) & $<0.11$ & $<1.4$ & $<6.8$ \\
	CO (17-16) & $<0.097$ & $<1.3$ & $<5.5$ \\

	\hline
		OH$^+_{330\,\mu\mathrm{m}}$ & $<0.073$ & $<0.46$ & $<19.0$ \\	
	OH$^+_{153.5\,\mu\mathrm{m}}$ & $<0.095$ & $<1.3$ & $<5.4$ \\
	OH$^+_{153.0\,\mu\mathrm{m}}$ & $<0.097$ & $<1.3$ & $<5.4$ \\
		OH$^+_{152.4\,\mu\mathrm{m}}$ & $<0.092$ & $<1.2$ & $<5.1$ \\

		\hline
			OH$_{163.4\,\mu\mathrm{m}}$ & $<0.084$ & $<1.1$ & $<5.4$ \\
	OH$_{163.1\,\mu\mathrm{m}}$ & $<0.095$ & $<1.2$ & $<6.0$ \\

\hline
H$_2$O$^{3_{12}-2_{21}}_{260\,\mu\mathrm{m}}$ & $<0.066$ & $<0.53$ & $<11.0$ \\
	H$_2$O$^{3_{21}-3_{12}}_{258\,\mu\mathrm{m}}$ & $<0.059$ & $<0.47$ & $<9.4$ \\
		H$_2$O$^{3_{03}-2_{12}}_{175\,\mu\mathrm{m}}$ & $<0.11$ & $<1.2$ & $<7.7$ \\
	H$_2$O$^{3_{22}-3_{13}}_{156\,\mu\mathrm{m}}$ & $<0.038$ & $<0.5$ & $<2.2$ \\

	\hline
\end{tabular}

\end{table}

\subsubsection{The FIR line deficit}

\begin{figure}
	\includegraphics[width=\linewidth]{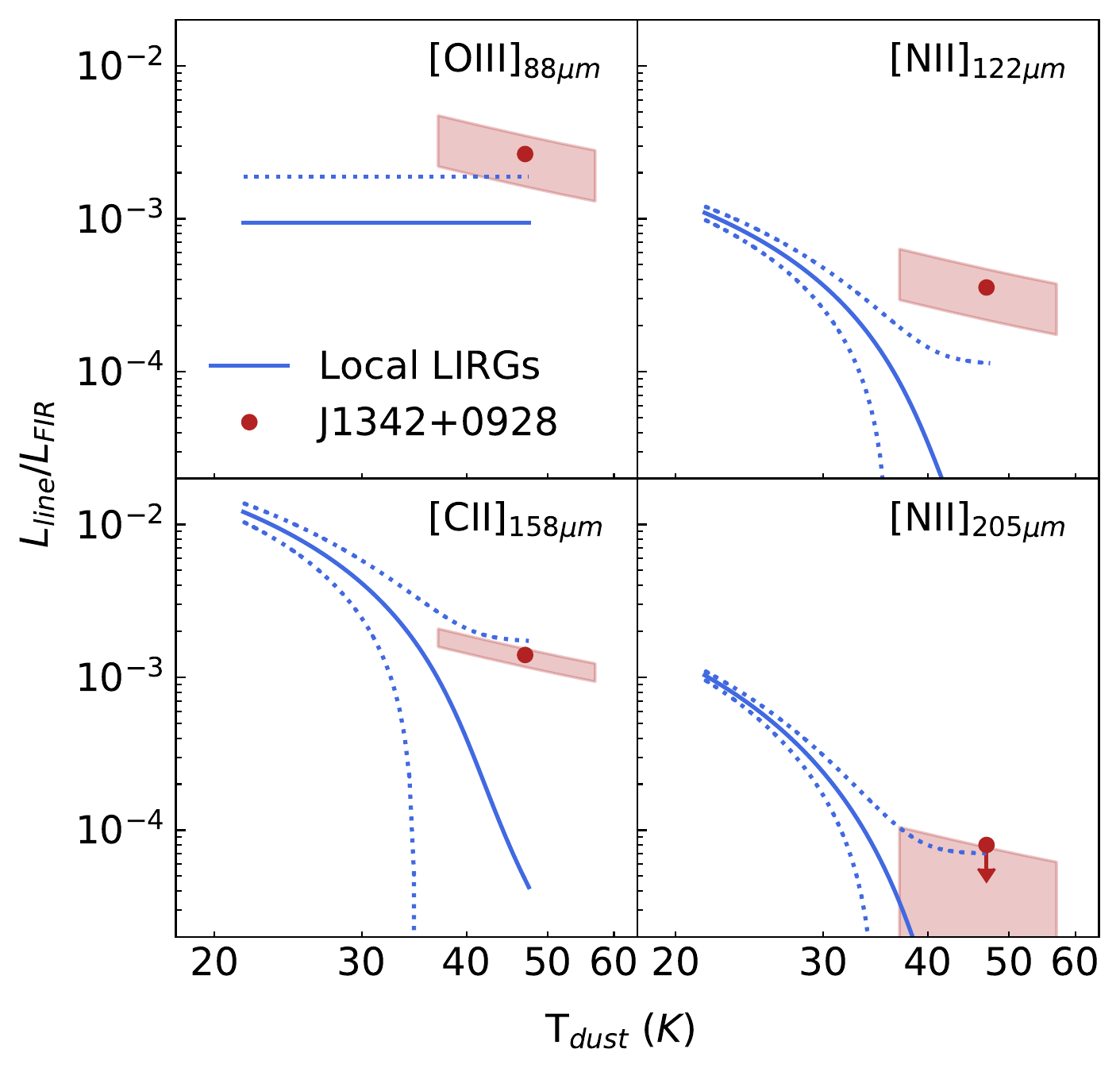}
	\caption{Ratio between various fine structure lines and the FIR ($42.5-122.5\,\mu$m) luminosities as a function of estimated dust temperature demonstrating the line deficit. Our \pisco\ measurements are shown with a red point. The accompanying red shaded area encompasses the considered dust temperature range in the horizontal direction and the flux density measurement error in the vertical direction. Blue line shows the fit to the local trend in LIRGs, while the dotted lines outline its $1\sigma$ dispersion (only the upper bound is constrained for the \oiii\ case),  as reported in  \cite{diaz-santos17}.}
	\label{fig:deficit}
\end{figure}

An empirical observation showing a decrease in the \cii\ line-to-FIR continuum ratio with increasing dust temperature and IR luminosities is referred to as the line deficit \citep[e.g.,][]{malhotra97,diaz-santos13}.
The same trend was observed in other fine structure lines, thanks to systematic studies with \emph{Herschel} \cite[e.g.,][]{diaz-santos17}.
The ISM conditions leading to such results are still under investigation and several scenarios have been proposed. These include a change in the ionization parameter, various optical depth effects, or galaxy compactness \cite[for details see e.g.,][]{gracia-carpio11,herrera-camus18, rybak19}.

We report several fine structure line-to-FIR ratios of \pisco\ in Fig.~\ref{fig:deficit}, along with a recent study of local luminous infrared galaxies (LIRGs; $L_{\mathrm{TIR}}=10^{10-11}$\,L$_\sun$) performed by \cite{diaz-santos17}. Measurements obtained on our $z=7.5$ quasar host galaxy are consistent within the scatter with trends observed in the local LIRGs.
%A possible indication of an enhanced \nii\ emission can be inferred. %, however this is still at a significance less than $3\sigma$.

\subsubsection{Photo-dissociation regions}
\label{sec:pdr}

\begin{figure}
	\includegraphics[width=\linewidth]{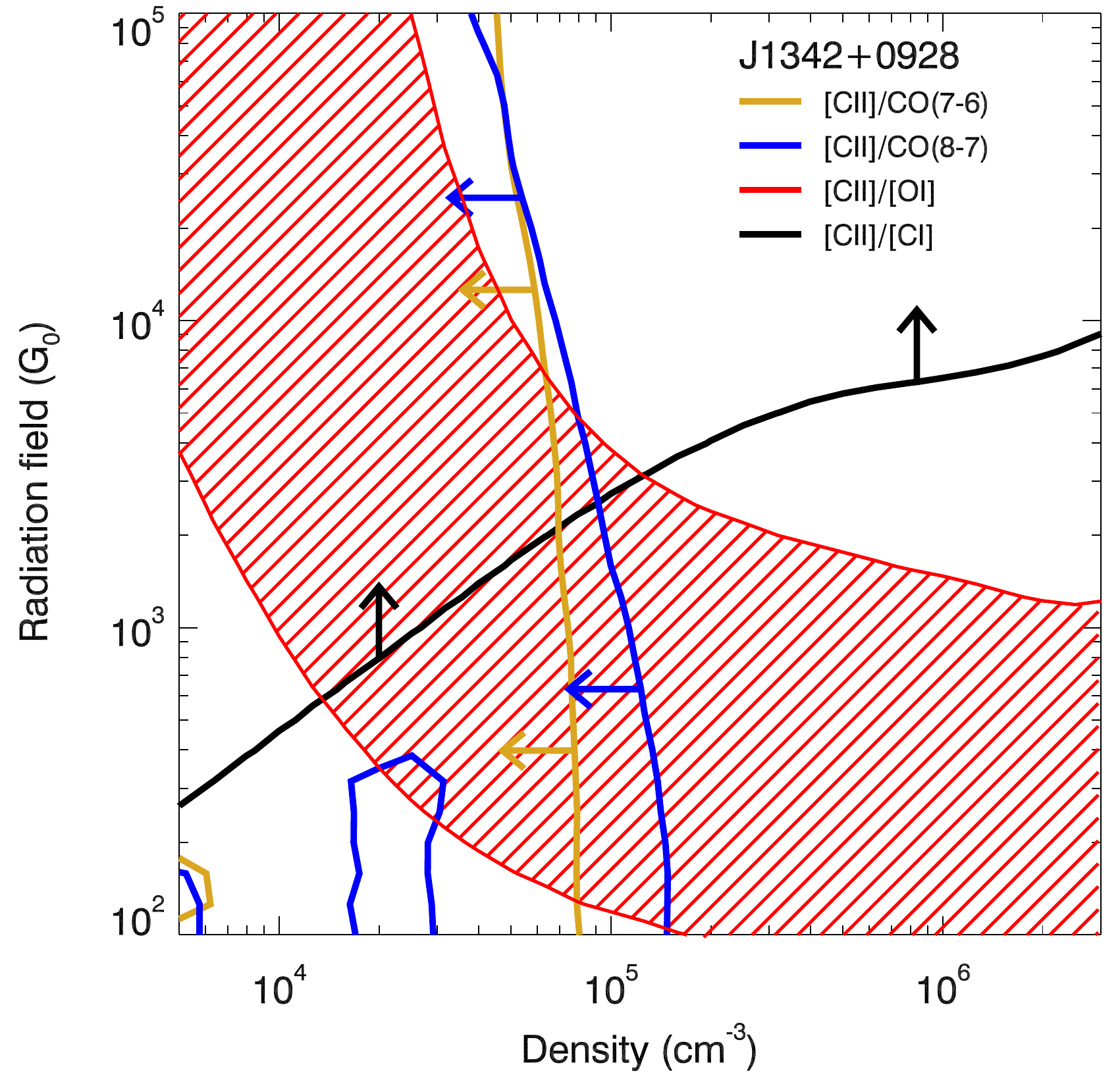}
	\caption{Constraints on the radiation field intensity (in units of the Habing flux, $G_0=1.6\times10^{-3}$\,erg\,cm$^{-2}$\,s$^{-1}$) and the density of the ISM in \pisco\ based on the PDR model grid from \cite{meijerink07}. Colored lines correspond to contours in the parameter space defined by specific line luminosity ratios, namely  $\cii/ {\rm CO} (7-6)>33$ (yellow), $\cii/ {\rm CO} (8-7)>27$ (blue), 	$3<\cii/ \oi<19$ (red hatched area) and $\cii/\ci>34$ (black).
	}
	\label{fig:lineratio}
\end{figure}

The type of radiation field the ISM is subjected to divides its physical and chemical processes into two distinct regimes: the photo-dissociation regions (PDRs; also photon dominated regions) and the X-ray dissociation regions (XDRs; also X-ray dominated regions). The former implies the main source of energy are far ultraviolet photons with energies between 6 and 13.6~eV originating in O and B stars, and is therefore tied with star-formation processes \cite[see e.g.,][]{tielens85}. In the latter regime, the gas is illuminated by hard X-ray radiation (energies above 1~keV), which penetrates much deeper into the molecular clouds before being absorbed \cite[see e.g.,][]{maloney96}. Such hard radiation can originate in super-massive black hole accretion episodes and is therefore indicative of significant AGN activity.

In order to tie our observations to the underlying physical properties, we make use of the PDR and XDR grid models developed by \cite{meijerink05,meijerink07}.  These models include various heating and cooling processes and chemical reactions that reproduce emission line luminosities as a function of  physical parameters, such as the radiation field and density of the ISM.
For the PDR calculations, we may consider only emission arising in neutral medium. The \cii\ can originate in the ionized medium as well, however we will show in Sect.~\ref{sec:ion} that this fraction can be constrained to less than 25\%. Since we can only obtain an upper limit, we do not correct for the emission fraction arising in ionized medium, and assume that the entire \cii\ emission is dominated by the PDR.

The line luminosity ratio $\cii/\ci$ is useful for discerning between the two main regimes (PDR and XDR) as the ratio is expected to be below 6 in XDRs regardless of ISM density or the radiation field \citep[see Fig.~6 in][]{venemans17a}. Although we do not detect the \ci\ line, its upper limit sets a lower limit on the line luminosity ratio $\cii/\ci>34$, which rules out XDR as the dominant regime in \pisco. We can therefore assume pure PDR models for the subsequent analysis. 
In PDRs, the ratio of \cii\ and CO emission lines is strongly dependent on the density, and we can use our limits listed in Table~\ref{tab:lines} to define an area in the density versus radiation field parameter space, which \pisco\ occupies. As shown in Fig.~\ref{fig:lineratio}, PDR model parameters, which are derived from our limits, suggest densities of $\lesssim 5\times10^4$\,cm$^{-3}$ and radiation field intensity of $\gtrsim 10^3\,G_0$, where $G_0$ is the Habing flux ($G_0=1.6\times10^{-3}$\,erg\,cm$^{-2}$\,s$^{-1}$). 
For comparison, similar ISM studies performed in five different $z>6$ quasars \citep[see][and Yang et al., subm.]{wang16,venemans17b} find densities higher than $10^5$\,cm$^{-3}$, and radiation field intensities of the order of $\sim 10^3\,G_0$.  On the other hand, several studies based on CO excitation ladders suggest  densities in the range of $10^{4-5}$\,cm$^{-3}$ \citep[see][]{riechers09, weiss07}.

\subsubsection{Carbon mass}
\label{sec:carbon}

The measured upper limit of the \ci\ emission line can provide constraints on the atomic carbon content of the galaxy. Following \cite{weiss03, weiss05} we can estimate the mass of the neutral carbon as
\begin{equation}
\frac{M_{\text{\rm [CI]}}}{\msol}=4.566\times 10^{-4} Q(T_{\text{ex}})\frac{1}{5}e^{T_2/T_{\text{ex}}}\frac{L^\prime_{\ci}}{\text{K\,km\,s}^{-1}\,\text{pc}^{2}},
\end{equation}
where $Q(T_{\text{ex}})=1+3 e^{-T_1/T_{\text{ex}}} + 5 e^{-T_2/T_{\text{ex}}}$ is the partition function, excitation energies of two different carbon transitions are $T_1=23.6$\,K, $T_2=62.5$\,K,
and the excitation temperature is assumed to be $T_{\text{ex}}=30$\,K \citep[see][]{walter11}. Resulting atomic carbon mass limit is $M_{\text{\rm [CI]}} < 5.3\times10^{6}\,\msol$.
\cite{walter11} derived atomic carbon to molecular hydrogen abundance of $\text{X}[\text{CI}]/\text{X}[\text{H}_2] =M_{\text{[CI]}}/ (6 M_{\text{H}_2})= (8.4 \pm 3.5)\times10^{-5}$ using a sample of $z>2$ submillimeter and quasar host galaxies. If we apply the same scaling relation to \pisco\ we obtain $M_{\text{H}_2}< 10^{10}\,\msol$ implying a gas-to-dust ratio upper limit of $<250$.

For completeness, we mention here the ionized carbon mass of $M_{\text{\rm [CII]}} = 4.9\times10^{6}\,\msol$ as reported by  \cite{venemans17c}, emitted from the outer layers of the PDR assuming $T_{\text{ex}}=100\,$K (temperature range from 75 to 200\,K would give a  20\% difference on the estimated mass).

\subsubsection{Gas mass}
\label{sec:gasmass}

\begin{figure}
	\includegraphics[width=\linewidth]{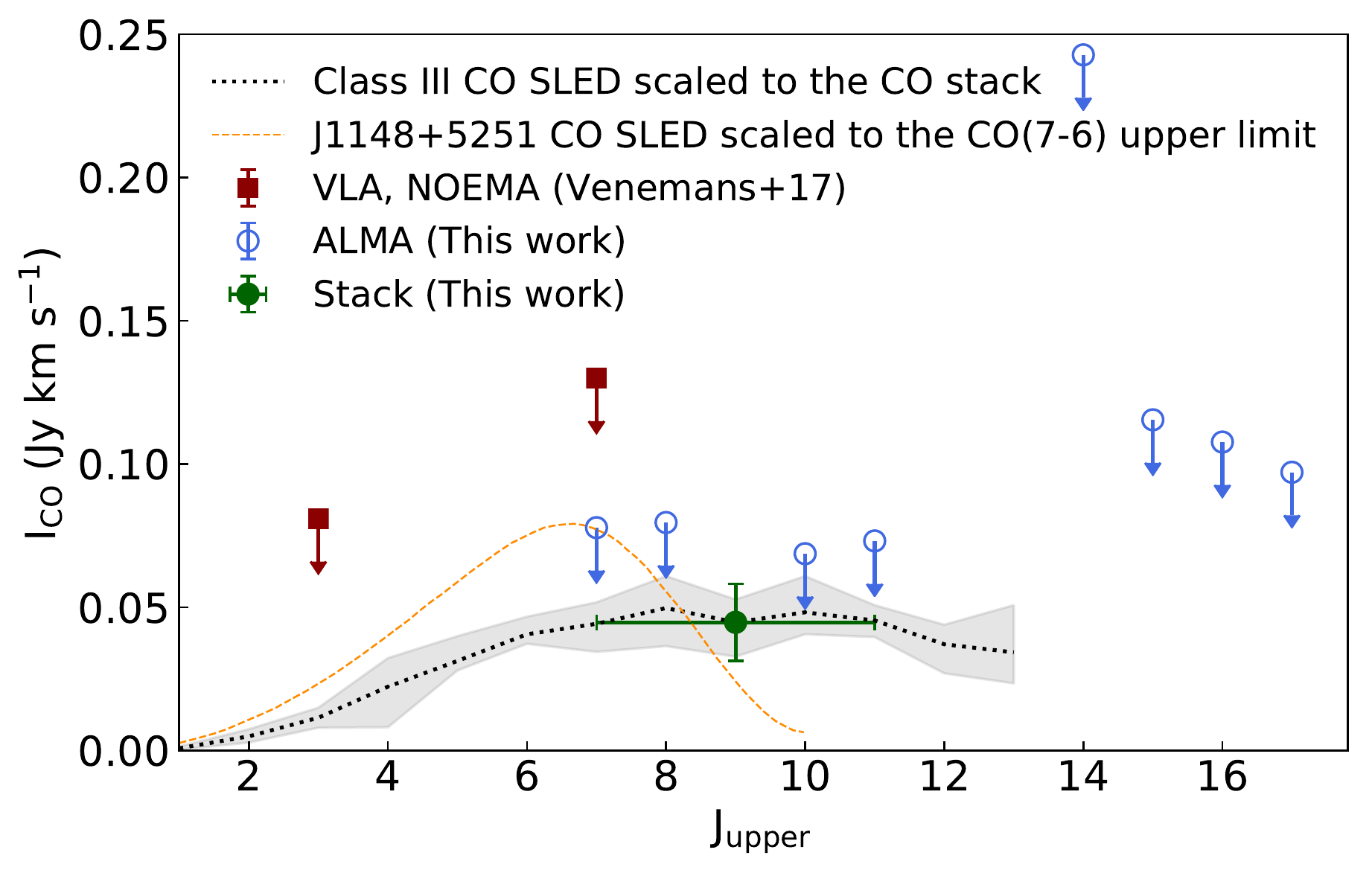}
	\caption{Constraints on the CO spectral line energy distribution of \pisco\ as a function of the rotational quantum number. The dotted black line shows the mean trends observed in local ULIRGs \citep[see][]{rosenberg15}, scaled to our stacked measurement (gray shaded area shows the scatter). The dashed orange line shows the CO SLED model of a $z=6.4$ quasar \citep{stefan15} scaled to our  CO(7-6) upper limit.}
	\label{fig:co_sled}
\end{figure}

\begin{table}
	\caption{The total gas mass of \pisco\ estimated using different proxies (see Sect.~\ref{sec:gasmass} for details). Conversion factor of $\alpha_{\text{CO}}= 0.8\,\text{M}_\sun (\kkmspc)^{-1}$ is used for all CO measurements.}
	\label{tab:gasmass}
	\centering
	%	\scriptsize 
	\begin{tabular}{ccc}
		\hline
		Proxy and extrapolation& Gas mass & Gas-to-dust\\
		  & ($10^8$\,\msol) & ratio\\
		\hline
		CO to FIR continuum correlation & 80 & 230 \\
		CO (7-6) with J1148+5251 SLED & $<35$ & $<100$ \\
		CO stack with Class III SLED& 11 & 30 \\
		\cii\ calibration& 420 & 1200\\
		\hline
	\end{tabular}
\end{table}

In this section we  constrain the total molecular gas mass  of \pisco\ employing a variety of methods. They are listed in Table~\ref{tab:gasmass} and explained below. The most common approach involves using the CO molecule as an H$_2$ tracer.
In order to obtain the gas mass from the CO line emission, we must assume a CO-to-H$_2$ (light-to-mass) conversion factor $\alpha_{\text{CO}}$  \citep[see][]{bolatto13}. The scaling is linear with line luminosity and is defined as $M_{\text{H}_2} [\text{M}_\sun]=\alpha_{\text{CO}} L^\prime_{\text{CO(1-0)}} [\kkmspc]$. 
We employ a value of $\alpha_{\text{CO}}= 0.8\,\text{M}_\sun (\kkmspc)^{-1}$ \citep[see][]{downes98}, which is often used. However, we note that a broader range of $\alpha_{\text{CO}}= 0.3 - 2.5\,\text{M}_\sun (\kkmspc)^{-1}$ is recommended for ULIRGs \citep[see][]{bolatto13}. The $\alpha_{\text{CO}}$  factor is highly uncertain largely due to unknown average surface density of the giant molecular clouds and the kinetic temperature of the gas. 
To use this scaling relation, we  first need to estimate the (unobserved) luminosity of the CO transmission of the rotational ground state.

One approach in estimating the CO(1-0) emission would be to use its empirical correlation with the FIR continuum, $\log(L_{\text{FIR}}[\lsol]/L^\prime_{\text{CO(1-0)}}[\kkmspc])\sim2$ with a scatter of 0.5\,dex, as shown in Fig.~7 of \cite{carilliwalter13}. This correlation would imply
$L^\prime_{\text{CO(1-0)}} \sim 100 \times10^8\,\kkmspc$, and thus a high gas-to-dust mass ratio of $\sim230$.

We have a significant number of CO upper limits as well as a tentative stack detection, as shown in Fig.~\ref{fig:co_sled}.
If the CO(9-8) is thermalized it would imply $L^\prime_{\text{CO(1-0)}}=L^\prime_{\text{CO(9-8)}}$. A correction to this simplified model involves usage of a CO spectral line energy distribution (SLED) of a high-redshift quasar. 
\cite{stefan15} employed large velocity gradients to model the CO SLED of J114816.64+525150.3, a $z=6.4$ quasar, using four observed CO transitions up to J$_{\text{up}}=7$. This model is also shown in Fig.~\ref{fig:co_sled}, scaled to our CO(7-6) upper limit. The model extrapolation beyond J$_{\text{up}}\geq 8$ is consistent with our stacked measurement within $2\sigma$ and predicts $L^\prime_{\text{CO(1-0)}}<44 \times10^8\,\kkmspc$, $M_{\text{H}_2} <35\times10^8\,\msol$,  and a gas-to-dust ratio of $< 100$ , based on our CO(7-6) upper limit.

We can also consider the CO (SLEDs) of the local ULIRGs. In Fig.~\ref{fig:co_sled} we show the Class III CO SLED, which has a turnover at higher J levels, as reported by \cite{rosenberg15}, scaled to our stack measurement. If we employ the CO SLED of ARP~299, a merger induced starburst belonging to the  above mentioned Class III group, the line flux of the CO(1-0) transition is $\sim50$ times weaker than the CO(9-8) transition, which from Eq.~\ref{eq:lprime} implies $L^\prime_{\text{CO(1-0)}}\sim 1.5 L^\prime_{\text{CO(9-8)}}=13.5 \times10^8\,\kkmspc$, $M_{\text{H}_2} = (11\pm3)\times10^8$\,M$_\sun$ and a somewhat small gas-to-dust mass ratio of $\sim30$.

\cite{zanella18} suggests to use the \cii\ emission as a tracer of molecular gas mass via the conversion factor of $M_{\text{H}_2}/L_{\cii}=\alpha_{[\text{CII}]}\sim30\,\msol/\lsol$, with a standard deviation of 0.2\,dex, based on a sample of main sequence galaxies at $z\sim2$. These authors report that the calibration is largely independent of the phase gas metallicity and the starburst behavior of the galaxy. When applied to \pisco\ we obtain molecular mass of   $M_{\text{H}_2}\sim4\times10^{10}\,\msol$. This would imply very large gas-to-dust ratios of $\sim1200$ and is not consistent with our non detection of neutral carbon (see Sect.~\ref{sec:carbon}).
Although large uncertainties are involved in the estimates, our upper limits and the tentative stack detection, paired with different models of the CO SLEDs, do not favor such a large  gas-to-dust ratio. For comparison, standard gas-to-dust ratios observed in nearby galaxies are around  $\sim100$ \citep[e.g.,][]{sandstrom13}.

\subsubsection{Ionized gas}
\label{sec:ion}

Due to different ionization energies of the fine structure lines (for C$^+$, O$^+$, N$^+$, O$^{++}$ they are 11.3, 13.6, 14.5 and 35\,eV respectively), and their critical densities\footnote{De-excitation emission rates are equal to collision rates at the critical density.}, line ratios can further  constrain the properties of the ISM, specifically the H\textsc{ii} regions. 
The \nii\ lines originate exclusively in the ionized medium due to its ionization energy being higher than that of  hydrogen (13.6\,eV). The model provided by \cite{oberst06}, see their Fig.~2, which assumes an electron impact excitation and electron temperature of 8000\,K, allows us to use our measurement of $\nii/\niilow>4.4$ to set a lower limit on the electron density of $n_e>180\,\text{cm}^{-3}$, hence the N$^+$ emission is dominated by the H\textsc{ii} regions and not the diffuse ISM.

Carbon has a lower ionization energy than hydrogen allowing the \cii\ emission line to arise in  neutral and ionized medium. Both transitions \cii\ and \niilow\ have similar critical densities, 40\,cm$^{-3}$ and 44\,cm$^{-3}$, respectively, implying that their line luminosity ratio will depend solely on their abundance ratio.  This fact allows us to use their line ratio in order to disentangle the fraction of emission arising in different phases of the ISM.
At the limit obtained above for the electron density ($n_e>180\,\text{cm}^{-3}$), the expected value of the carbon to nitrogen line ratio inside the ionized medium, according to  \cite{oberst06}, is $3\lesssim\cii/\niilow<4.3$. From this ratio, and our \niilow\ luminosity limit of $<0.8\times10^8$\,\lsol, we can derive 
an upper limit on the \cii\ emission arising from the ionized medium of $ \cii^{\mathrm{ion}}<3.4\times10^8\,\lsol$. This accounts for  less than 25\% of the entire $14\times10^8\,\lsol$ observed \cii\  emission.

Following \cite{ferkinhoff11}, assuming high density and high temperature limit so that all nitrogen is singly ionized (valid around B2 to O8 stars), we can calculate the minimum H$^+$ mass required to produce the observed N$^+$ luminosity as
\begin{equation}
M(\text{H}^+)=\frac{L_{\nii}}{\frac{g_2}{g_t}A_{21}h\nu_{21}}\frac{m_{\text{H}}}{\chi(\text{N}^+)},
\label{eq:hmin}
\end{equation}
where parameters are: statistical weight of the emission line $g_2=5$, partition function $g_t=9$, Einstein coefficient $A_{21}=7.5\times10^{-6}\,\text{s}^{-1}$, relative abundance of nitrogen compared to hydrogen $\chi(\text{N}^+)=9.3\times10^{-5}$ \citep[taken from][]{savage96}, hydrogen atom mass $m_{\text{H}}$, Planck constant $h$ and the frequency of the 122\,$\mu$m line transition $\nu_{21}$. Inserting the luminosity value of 
$L_{\nii}=3.55\times10^8\,\lsol$ (from Table~\ref{tab:lines}) yields $M_{\text{min}}(\text{H}^+)=1.8\times 10^8\,\msol$.
When compared to the molecular gas mass derived from the CO(7-6) upper limit we obtain
a lower limit on the gas ionization percentage $M(\text{H}^+)/M(\text{H}_2)>4\%$, while the estimate from the CO stack measurement would yield  an even greater value of $\sim16\%$.
 For comparison, the ionized gas fraction is observed to be less than $<1\%$ in nearby galaxies (see Fig.~3 in \cite{ferkinhoff11} based on \cite{brauher08} data), while larger ionization percentages point to the increased SFR surface density ($\sim 100 - 1000\, \msolyr\,\text{pc}^{-2}$).

A similar approach can be utilized for the O$^{++}$ line, as described by \cite{ferkinhoff10}. We can insert into Eq.~\ref{eq:hmin} values valid for the \oiii\ line, namely $L_{\oiii}=26.5\times10^8\,\lsol$, $g_2=3$, $g_t=9$, $A_{21}=2.7\times10^{-5}\,\text{s}^{-1}$, $\chi(\text{O}^{++})=5.9\times10^{-4}$ \citep[from][]{savage96}, thus obtaining $M_{\text{min}}(\text{H}^+)=0.7\times 10^8\,\msol$. We do not expect all oxygen to be in the doubly ionized state \cite[see][]{ferkinhoff10}, so the minimum ionized gas we report here
is strictly a lower limit and it can easily be an order of magnitude higher.

Further investigation of ISM physical properties via emission line ratios is available with the Cloudy spectral synthesis code \citep{ferland17}. The model computes line strengths on a grid of changing ISM conditions and we defer the details of the simulation and its setup to an upcoming paper by Decarli et al. (in prep.).
Intensity of the radiation field  can be estimated from the observed line ratio $\oiii/\nii=7.5\pm4.0$. This is possible due to the fact that both ions have similar critical densities, but different ionization energies. Assuming that stars with effective temperature of $\sim40\,000\,K$ are responsible for the ionization radiation, the model predicts intensity of the radiation field of $\log U\sim-2.5$, where $U$ is the ionization parameter defined as a ratio between the ionizing photon density and the total hydrogen density \citep[see e.g.,][]{draine11}: $U=n(h\nu>13.6\,\mathrm{eV})/n_{\mathrm{H}}$. An equivalent way to write the ionization parameter  is $U=\Phi/(n_{\mathrm{H}} c)$, where $\Phi$ is the flux of the hydrogen ionizing photons, and $c$ is the speed of light.

Assuming that the star formation is the main driver of observed atomic line emissions, we can use findings reported in \cite{delooze14} to calculate the SFR of \pisco\ using FIR fine structure lines. For high-redshift galaxies ($z>0.5$) these authors report the following calibration:
\begin{equation}
\log (\text{SFR}/\msol)=\beta+\alpha\log(L_{\text{line}}/\lsol),
\end{equation}
where the parameters are $\alpha=1$ and $\beta=-6.89$  for the \oiii\ emission line, with a dispersion of 0.46\,dex. This calibration yields $\text{SFR}=340\,\msolyr$. Similarly, for \cii\ ($\alpha=1.18$ and $\beta=-8.52$, dispersion 0.4\,dex) we obtain $\text{SFR}=180\,\msolyr$. Both of these values agree within the uncertainties with the SFR inferred from the dust continuum.
 %log SFR [M⊙ yr−1] = β + α × log Lline [L⊙] 
 
As evident from Fig.~\ref{fig:line_det}, the peak of the \oiii\ emission is offset by $0.6\arcsec$ from the peak of the underlying continuum and peaks of other detected lines. Although this might be attributed to possible outflows \citep[see e.g.,][]{bischetti17}, the low signal-to-noise of our detection, and the fact that the separation is less than one resolution element, does not allow for a robust interpretation. 
Theoretical studies performed by \cite{katz17, katz19} predict that \oiii\ would be at an offset from \cii, because it arises from a medium with higher temperatures and ionization parameter. They also predict that \nii\ emission would be slightly more extended compared to the \cii\ emission. 
Deeper data,  preferably at higher resolution comparable to the \cii\ data, is required to investigate this potential offset further.
Finally, we report oxygen-to-carbon line luminosity ratio of $\oiii/\cii=1.9\pm0.7$, similar to other recent  high-redshift observations \citep[see][]{walter18, hashimoto18b}.

\subsubsection{Water}

Water line emission originates in the warm dust ($40-70$\,K) regions where densities are of the order of $10^{5-6}$\,cm$^{-3}$, and higher rotational CO lines (J$_{\text{up}}>8$) are also common \citep[see][]{liu17}. 
We do not detect any significant emission from H$_2$O, OH and OH$^+$ species within ten possible line candidates that lie inside our observed frequency ranges.
Water lines $3_{12}-2_{21}$ at 260\,$\mu$m and  $3_{03}-2_{12}$ at 175\,$\mu$m show tentative detections, however the first one could be dominated by the neighboring  CO(10-9) line, while the second one shows an excess in the spectrum, but a prominent negative peak at the required central frequency (see Appendix~\ref{sec:non_det}).
Our PDR measurements (see Sect.~\ref{sec:pdr}) point toward lower densities, thus non-detection of water lines is consistent with our findings determined from CO non-detections.

\subsubsection{Metallicity}

Due to the high dust content and large star-formation rates, we expect that the ISM of \pisco\ has already been enriched by the yields of supernovae.
The metallicity of a galaxy is usually determined from optical line ratios, which are severely affected by dust extinction, and are not accessible to ground instruments for high-redshift objects. The FIR emission lines can provide an alternative method to measure metallicity, once it is properly calibrated \citep[see e.g.,][]{pereira-santaella17}. 
The N$^{++}$ and O$^{++}$ ions are especially interesting for this study due to their similar ionization structure.
We use the calibration reported by \cite{rigopoulou18}, see their Fig.~5, and assume that the ionization parameter is  $\log U=-2.5$. From the $\oiii/\nii=7.5\pm4$ line ratio we derive gas-phase metallicities in the H\textsc{ii} region of $Z_{\mathrm{gas}}=1.3^{+0.3}_{-0.1}\,Z_\sun$, where $Z_\sun$ is the value for the Solar neighborhood. A range of $\log U$ between $-2$ and $-3$ would yield metallicities of $0.7-2.0\,Z_\sun$, with line flux errors folded inside.

Multiple studies of high-redshift quasars performed using the optical spectra derive metallicities consistent with values observed in the Solar neighborhood \citep[see e.g.,][]{kurk07,derosa11,mazzucchelli17}. These measurements were obtained 
from the Mg{\sc ii} / Fe{\sc ii} line ratio measured in the broad line region, at spatial scales of $\lesssim1\,$kpc. On the other hand, our measurements, also pointing to Solar like metallicities, are averaged across the entire galaxy host.

\cite{remy-ruyer14} have shown that the metallicity is an important contributor to the observed gas-to-dust ratio. These authors report that the observed gas-to-dust ratio in their sample is around $\sim100$ for metallicities around Solar, with a clear negative correlation between the two. 
Our findings of the rich metal content in the host galaxy of \pisco\ are therefore in agreement with the  low estimates on its gas-to-dust ratio derived from the CO limits (or the CO stack).

\section{Conclusions} \label{sec:conc}

We have presented ALMA observations of the dust continuum and the interstellar medium of the host galaxy \pisco, the most distant quasar known to date. The quasar is observed at $z=7.54$ when the universe was only 680 millions years old. 
With eight spectral setups positioned between 93.5 and 412\,GHz, we constrain the Rayleigh-Jeans tail of the dust continuum and show that a modified black body with the canonical dust temperature of 47\,K fits the data well.  At the same time we narrow the dust spectral emissivity coefficient to $\beta=1.85\pm0.3$, derive the  dust mass of $M_{\mathrm{dust}}=0.35\times10^8$\,M$_\sun$ and a high star-formation rate of $\sim150\,\msolyr$.

With detections of atomic fine structure lines \cii, \nii, \oiii; limits on \ci, \niilow, \oi; and multiple CO lines (with a tentative stack detection) we derived the following main results.
Observed line deficits in \pisco\ are comparable to local ULIRGs.
We do not see evidence of X-rays dominated regions, and photo-dissociation region-only model  implies low gas densities ($\lesssim 5\times10^4$\,cm$^{-3}$) and strong radiation fields ($\gtrsim 10^3\,G_0$).
The CO data hints at lower gas-to-dust ratios, less than a hundred. The \cii\ emission originates predominantly inside the neutral medium, and there is evidence of possible spatial offset between \cii\ and \oiii\ emission. However, deeper data is required to confirm and interpret this offset.
A limit on the ratio of the two N$^{++}$ emission lines imply high electron densities of $n_e>180\,\mathrm{cm}^{-3}$.
We also find high percentages ($>4\%$, possibly $\sim16\%$) of ionized to molecular hydrogen, indicating large SFR surface densities. We detect no water lines consistent with upper limits on CO lines and the derived density of the medium. Finally, we estimate that the metallicities are consistent with Solar values, supporting the view of \pisco\ as a galaxy extremely rich in  dust and metals, despite the early epoch it is located in.

In this study of the interstellar medium of a galaxy in the epoch of reionization, we demonstrate the wide range of investigations that are possible with ALMA observations of FIR emission lines. We have shown that such an analysis benefits greatly from observations of multiple lines arising in different states of the ISM. The coming years will provide more high-redshift candidates, and ALMA follow-ups targeting several FIR emission lines in these objects will lead to a deeper understanding of the ISM  at the epoch of reionization. 
%The upcoming James Webb Space Telescope will greatly expand the available wavelength range to study  high redshift objects, and given its synergy with ALMA, we can expect  many fruitful investigations in the future.

\acknowledgments

We thank Dominik Riechers for helpful discussions and suggestions.
MN, FW, BV, MaN, acknowledge support from the ERC Advanced Grant 740246 (Cosmic Gas).
This paper makes use of the following ALMA data: ADS/JAO.ALMA\#2017.1.00396.S. ALMA is a partnership of ESO (representing its member states), NSF (USA) and NINS (Japan), together with NRC (Canada), NSC and ASIAA (Taiwan), and KASI (Republic of Korea), in cooperation with the Republic of Chile. The Joint ALMA Observatory is operated by ESO, AUI/NRAO and NAOJ.
This research made use of Astropy,\footnote{http://www.astropy.org} a community-developed core Python package for Astronomy \citep{astropy:2013, astropy:2018}, and Matplotlib \citep{Hunter:2007}.

\bibliographystyle{aasjournal}
\bibliography{refs}

\begin{thebibliography}{}
\expandafter\ifx\csname natexlab\endcsname\relax\def\natexlab#1{#1}\fi
\providecommand{\url}[1]{\href{#1}{#1}}
\providecommand{\dodoi}[1]{doi:~\href{http://doi.org/#1}{\nolinkurl{#1}}}
\providecommand{\doeprint}[1]{\href{http://ascl.net/#1}{\nolinkurl{http://ascl.net/#1}}}
\providecommand{\doarXiv}[1]{\href{https://arxiv.org/abs/#1}{\nolinkurl{https://arxiv.org/abs/#1}}}

\bibitem[{{Astropy Collaboration} {et~al.}(2013){Astropy Collaboration},
  {Robitaille}, {Tollerud}, {Greenfield}, {Droettboom}, {Bray}, {Aldcroft},
  {Davis}, {Ginsburg}, {Price-Whelan}, {Kerzendorf}, {Conley}, {Crighton},
  {Barbary}, {Muna}, {Ferguson}, {Grollier}, {Parikh}, {Nair}, {Unther},
  {Deil}, {Woillez}, {Conseil}, {Kramer}, {Turner}, {Singer}, {Fox}, {Weaver},
  {Zabalza}, {Edwards}, {Azalee Bostroem}, {Burke}, {Casey}, {Crawford},
  {Dencheva}, {Ely}, {Jenness}, {Labrie}, {Lim}, {Pierfederici}, {Pontzen},
  {Ptak}, {Refsdal}, {Servillat}, \& {Streicher}}]{astropy:2013}
{Astropy Collaboration}, {Robitaille}, T.~P., {Tollerud}, E.~J., {et~al.} 2013,
  \aap, 558, A33, \dodoi{10.1051/0004-6361/201322068}

\bibitem[{{Ba{\~n}ados} {et~al.}(2016){Ba{\~n}ados}, {Venemans}, {Decarli},
  {Farina}, {Mazzucchelli}, {Walter}, {Fan}, {Stern}, {Schlafly}, {Chambers},
  {Rix}, {Jiang}, {McGreer}, {Simcoe}, {Wang}, {Yang}, {Morganson}, {De Rosa},
  {Greiner}, {Balokovi{\'c}}, {Burgett}, {Cooper}, {Draper}, {Flewelling},
  {Hodapp}, {Jun}, {Kaiser}, {Kudritzki}, {Magnier}, {Metcalfe}, {Miller},
  {Schindler}, {Tonry}, {Wainscoat}, {Waters}, \& {Yang}}]{banados16}
{Ba{\~n}ados}, E., {Venemans}, B.~P., {Decarli}, R., {et~al.} 2016, \apjs, 227,
  11, \dodoi{10.3847/0067-0049/227/1/11}

\bibitem[{{Ba{\~n}ados} {et~al.}(2018){Ba{\~n}ados}, {Venemans},
  {Mazzucchelli}, {Farina}, {Walter}, {Wang}, {Decarli}, {Stern}, {Fan},
  {Davies}, {Hennawi}, {Simcoe}, {Turner}, {Rix}, {Yang}, {Kelson}, {Rudie}, \&
  {Winters}}]{banados18}
{Ba{\~n}ados}, E., {Venemans}, B.~P., {Mazzucchelli}, C., {et~al.} 2018, \nat,
  553, 473, \dodoi{10.1038/nature25180}

\bibitem[{{Becker} {et~al.}(2015){Becker}, {Bolton}, \& {Lidz}}]{becker15}
{Becker}, G.~D., {Bolton}, J.~S., \& {Lidz}, A. 2015, \pasa, 32, e045,
  \dodoi{10.1017/pasa.2015.45}

\bibitem[{{Beelen} {et~al.}(2006){Beelen}, {Cox}, {Benford}, {Dowell},
  {Kov{\'a}cs}, {Bertoldi}, {Omont}, \& {Carilli}}]{beelen06}
{Beelen}, A., {Cox}, P., {Benford}, D.~J., {et~al.} 2006, \apj, 642, 694,
  \dodoi{10.1086/500636}

\bibitem[{{Bischetti} {et~al.}(2017){Bischetti}, {Piconcelli}, {Vietri},
  {Bongiorno}, {Fiore}, {Sani}, {Marconi}, {Duras}, {Zappacosta}, {Brusa},
  {Comastri}, {Cresci}, {Feruglio}, {Giallongo}, {La Franca}, {Mainieri},
  {Mannucci}, {Martocchia}, {Ricci}, {Schneider}, {Testa}, \&
  {Vignali}}]{bischetti17}
{Bischetti}, M., {Piconcelli}, E., {Vietri}, G., {et~al.} 2017, \aap, 598,
  A122, \dodoi{10.1051/0004-6361/201629301}

\bibitem[{{Bolatto} {et~al.}(2013){Bolatto}, {Wolfire}, \& {Leroy}}]{bolatto13}
{Bolatto}, A.~D., {Wolfire}, M., \& {Leroy}, A.~K. 2013, \araa, 51, 207,
  \dodoi{10.1146/annurev-astro-082812-140944}

\bibitem[{{Brauher} {et~al.}(2008){Brauher}, {Dale}, \& {Helou}}]{brauher08}
{Brauher}, J.~R., {Dale}, D.~A., \& {Helou}, G. 2008, The Astrophysical Journal
  Supplement Series, 178, 280, \dodoi{10.1086/590249}

\bibitem[{{Carilli} \& {Walter}(2013)}]{carilliwalter13}
{Carilli}, C.~L., \& {Walter}, F. 2013, Annual Review of Astronomy and
  Astrophysics, 51, 105, \dodoi{10.1146/annurev-astro-082812-140953}

\bibitem[{{Carniani} {et~al.}(2017){Carniani}, {Maiolino}, {Pallottini},
  {Vallini}, {Pentericci}, {Ferrara}, {Castellano}, {Vanzella}, {Grazian},
  {Gallerani}, {Santini}, {Wagg}, \& {Fontana}}]{carniani17}
{Carniani}, S., {Maiolino}, R., {Pallottini}, A., {et~al.} 2017, \aap, 605,
  A42, \dodoi{10.1051/0004-6361/201630366}

\bibitem[{{Carniani} {et~al.}(2019){Carniani}, {Gallerani}, {Vallini},
  {Pallottini}, {Tazzari}, {Ferrara}, {Maiolino}, {Cicone}, {Feruglio}, {Neri},
  {D'Odorico}, {Wang}, \& {Li}}]{carniani19}
{Carniani}, S., {Gallerani}, S., {Vallini}, L., {et~al.} 2019, arXiv e-prints.
\newblock \doarXiv{1902.01413}

\bibitem[{{Chabrier}(2003)}]{chabrier03}
{Chabrier}, G. 2003, \pasp, 115, 763, \dodoi{10.1086/376392}

\bibitem[{{da Cunha} {et~al.}(2013){da Cunha}, {Groves}, {Walter}, {Decarli},
  {Weiss}, {Bertoldi}, {Carilli}, {Daddi}, {Elbaz}, {Ivison}, {Maiolino},
  {Riechers}, {Rix}, {Sargent}, \& {Smail}}]{dacunha13}
{da Cunha}, E., {Groves}, B., {Walter}, F., {et~al.} 2013, \apj, 766, 13,
  \dodoi{10.1088/0004-637X/766/1/13}

\bibitem[{{Daddi} {et~al.}(2015){Daddi}, {Dannerbauer}, {Liu}, {Aravena},
  {Bournaud}, {Walter}, {Riechers}, {Magdis}, {Sargent}, {B{\'e}thermin},
  {Carilli}, {Cibinel}, {Dickinson}, {Elbaz}, {Gao}, {Gobat}, {Hodge}, \&
  {Krips}}]{daddi15}
{Daddi}, E., {Dannerbauer}, H., {Liu}, D., {et~al.} 2015, \aap, 577, A46,
  \dodoi{10.1051/0004-6361/201425043}

\bibitem[{{De Looze} {et~al.}(2014){De Looze}, {Cormier}, {Lebouteiller},
  {Madden}, {Baes}, {Bendo}, {Boquien}, {Boselli}, {Clements}, {Cortese},
  {Cooray}, {Galametz}, {Galliano}, {Graci{\'a}-Carpio}, {Isaak}, {Karczewski},
  {Parkin}, {Pellegrini}, {R{\'e}my-Ruyer}, {Spinoglio}, {Smith}, \&
  {Sturm}}]{delooze14}
{De Looze}, I., {Cormier}, D., {Lebouteiller}, V., {et~al.} 2014, \aap, 568,
  A62, \dodoi{10.1051/0004-6361/201322489}

\bibitem[{{De Rosa} {et~al.}(2011){De Rosa}, {Decarli}, {Walter}, {Fan},
  {Jiang}, {Kurk}, {Pasquali}, \& {Rix}}]{derosa11}
{De Rosa}, G., {Decarli}, R., {Walter}, F., {et~al.} 2011, \apj, 739, 56,
  \dodoi{10.1088/0004-637X/739/2/56}

\bibitem[{{De Rosa} {et~al.}(2014){De Rosa}, {Venemans}, {Decarli}, {Gennaro},
  {Simcoe}, {Dietrich}, {Peterson}, {Walter}, {Frank}, {McMahon}, {Hewett},
  {Mortlock}, \& {Simpson}}]{derosa14}
{De Rosa}, G., {Venemans}, B.~P., {Decarli}, R., {et~al.} 2014, \apj, 790, 145,
  \dodoi{10.1088/0004-637X/790/2/145}

\bibitem[{{Decarli} {et~al.}(2018){Decarli}, {Walter}, {Venemans},
  {Ba{\~n}ados}, {Bertoldi}, {Carilli}, {Fan}, {Farina}, {Mazzucchelli},
  {Riechers}, {Rix}, {Strauss}, {Wang}, \& {Yang}}]{decarli18}
{Decarli}, R., {Walter}, F., {Venemans}, B.~P., {et~al.} 2018, \apj, 854, 97,
  \dodoi{10.3847/1538-4357/aaa5aa}

\bibitem[{{D{\'{\i}}az-Santos} {et~al.}(2013){D{\'{\i}}az-Santos}, {Armus},
  {Charmandaris}, {Stierwalt}, {Murphy}, {Haan}, {Inami}, {Malhotra},
  {Meijerink}, {Stacey}, {Petric}, {Evans}, {Veilleux}, {van der Werf}, {Lord},
  {Lu}, {Howell}, {Appleton}, {Mazzarella}, {Surace}, {Xu}, {Schulz},
  {Sanders}, {Bridge}, {Chan}, {Frayer}, {Iwasawa}, {Melbourne}, \&
  {Sturm}}]{diaz-santos13}
{D{\'{\i}}az-Santos}, T., {Armus}, L., {Charmandaris}, V., {et~al.} 2013, \apj,
  774, 68, \dodoi{10.1088/0004-637X/774/1/68}

\bibitem[{{D{\'\i}az-Santos} {et~al.}(2017){D{\'\i}az-Santos}, {Armus},
  {Charmandaris}, {Lu}, {Stierwalt}, {Stacey}, {Malhotra}, {van der Werf},
  {Howell}, {Privon}, {Mazzarella}, {Goldsmith}, {Murphy}, {Barcos-Mu{\~n}oz},
  {Linden}, {Inami}, {Larson}, {Evans}, {Appleton}, {Iwasawa}, {Lord},
  {Sanders}, \& {Surace}}]{diaz-santos17}
{D{\'\i}az-Santos}, T., {Armus}, L., {Charmandaris}, V., {et~al.} 2017, \apj,
  846, 32, \dodoi{10.3847/1538-4357/aa81d7}

\bibitem[{{Downes} \& {Solomon}(1998)}]{downes98}
{Downes}, D., \& {Solomon}, P.~M. 1998, \apj, 507, 615, \dodoi{10.1086/306339}

\bibitem[{{Draine}(2011)}]{draine11}
{Draine}, B.~T. 2011, \apj, 732, 100, \dodoi{10.1088/0004-637X/732/2/100}

\bibitem[{{Dunne} {et~al.}(2003){Dunne}, {Eales}, \& {Edmunds}}]{dunne03}
{Dunne}, L., {Eales}, S.~A., \& {Edmunds}, M.~G. 2003, \mnras, 341, 589,
  \dodoi{10.1046/j.1365-8711.2003.06440.x}

\bibitem[{{Fan} {et~al.}(2006){Fan}, {Strauss}, {Becker}, {White}, {Gunn},
  {Knapp}, {Richards}, {Schneider}, {Brinkmann}, \& {Fukugita}}]{fan06}
{Fan}, X., {Strauss}, M.~A., {Becker}, R.~H., {et~al.} 2006, \aj, 132, 117,
  \dodoi{10.1086/504836}

\bibitem[{{Ferkinhoff} {et~al.}(2010){Ferkinhoff}, {Hailey-Dunsheath},
  {Nikola}, {Parshley}, {Stacey}, {Benford}, \& {Staguhn}}]{ferkinhoff10}
{Ferkinhoff}, C., {Hailey-Dunsheath}, S., {Nikola}, T., {et~al.} 2010, \apj,
  714, L147, \dodoi{10.1088/2041-8205/714/1/L147}

\bibitem[{{Ferkinhoff} {et~al.}(2011){Ferkinhoff}, {Brisbin}, {Nikola},
  {Parshley}, {Stacey}, {Phillips}, {Falgarone}, {Benford}, {Staguhn}, \&
  {Tucker}}]{ferkinhoff11}
{Ferkinhoff}, C., {Brisbin}, D., {Nikola}, T., {et~al.} 2011, \apj, 740, L29,
  \dodoi{10.1088/2041-8205/740/1/L29}

\bibitem[{{Ferland} {et~al.}(2017){Ferland}, {Chatzikos}, {Guzm{\'a}n},
  {Lykins}, {van Hoof}, {Williams}, {Abel}, {Badnell}, {Keenan}, {Porter}, \&
  {Stancil}}]{ferland17}
{Ferland}, G.~J., {Chatzikos}, M., {Guzm{\'a}n}, F., {et~al.} 2017, \rmxaa, 53,
  385.
\newblock \doarXiv{1705.10877}

\bibitem[{{Fischer} {et~al.}(2010){Fischer}, {Sturm}, {Gonz{\'a}lez-Alfonso},
  {Graci{\'a}-Carpio}, {Hailey-Dunsheath}, {Poglitsch}, {Contursi}, {Lutz},
  {Genzel}, {Sternberg}, {Verma}, \& {Tacconi}}]{fischer10}
{Fischer}, J., {Sturm}, E., {Gonz{\'a}lez-Alfonso}, E., {et~al.} 2010, \aap,
  518, L41, \dodoi{10.1051/0004-6361/201014676}

\bibitem[{{Gonz{\'a}lez-Alfonso} {et~al.}(2013){Gonz{\'a}lez-Alfonso},
  {Fischer}, {Bruderer}, {M{\"u}ller}, {Graci{\'a}-Carpio}, {Sturm}, {Lutz},
  {Poglitsch}, {Feuchtgruber}, {Veilleux}, {Contursi}, {Sternberg},
  {Hailey-Dunsheath}, {Verma}, {Christopher}, {Davies}, {Genzel}, \&
  {Tacconi}}]{gonzales13}
{Gonz{\'a}lez-Alfonso}, E., {Fischer}, J., {Bruderer}, S., {et~al.} 2013, \aap,
  550, A25, \dodoi{10.1051/0004-6361/201220466}

\bibitem[{{Graci{\'a}-Carpio} {et~al.}(2011){Graci{\'a}-Carpio}, {Sturm},
  {Hailey-Dunsheath}, {Fischer}, {Contursi}, {Poglitsch}, {Genzel},
  {Gonz{\'a}lez-Alfonso}, {Sternberg}, {Verma}, {Christopher}, {Davies},
  {Feuchtgruber}, {de Jong}, {Lutz}, \& {Tacconi}}]{gracia-carpio11}
{Graci{\'a}-Carpio}, J., {Sturm}, E., {Hailey-Dunsheath}, S., {et~al.} 2011,
  \apjl, 728, L7, \dodoi{10.1088/2041-8205/728/1/L7}

\bibitem[{{Hashimoto} {et~al.}(2018{\natexlab{a}}){Hashimoto}, {Inoue},
  {Tamura}, {Matsuo}, {Mawatari}, \& {Yamaguchi}}]{hashimoto18b}
{Hashimoto}, T., {Inoue}, A.~K., {Tamura}, Y., {et~al.} 2018{\natexlab{a}},
  ArXiv e-prints, arXiv:1811.00030.
\newblock \doarXiv{1811.00030}

\bibitem[{{Hashimoto} {et~al.}(2018{\natexlab{b}}){Hashimoto}, {Inoue},
  {Mawatari}, {Tamura}, {Matsuo}, {Furusawa}, {Harikane}, {Shibuya}, {Knudsen},
  {Kohno}, {Ono}, {Zackrisson}, {Okamoto}, {Kashikawa}, {Oesch}, {Ouchi},
  {Ota}, {Shimizu}, {Taniguchi}, {Umehata}, \& {Watson}}]{hashimoto18a}
{Hashimoto}, T., {Inoue}, A.~K., {Mawatari}, K., {et~al.} 2018{\natexlab{b}},
  ArXiv e-prints, arXiv:1806.00486.
\newblock \doarXiv{1806.00486}

\bibitem[{{Herrera-Camus} {et~al.}(2016){Herrera-Camus}, {Bolatto}, {Smith},
  {Draine}, {Pellegrini}, {Wolfire}, {Croxall}, {de Looze}, {Calzetti},
  {Kennicutt}, {Crocker}, {Armus}, {van der Werf}, {Sandstrom}, {Galametz},
  {Brandl}, {Groves}, {Rigopoulou}, {Walter}, {Leroy}, {Boquien}, {Tabatabaei},
  \& {Beirao}}]{herrera-camus16}
{Herrera-Camus}, R., {Bolatto}, A., {Smith}, J.~D., {et~al.} 2016, \apj, 826,
  175, \dodoi{10.3847/0004-637X/826/2/175}

\bibitem[{{Herrera-Camus} {et~al.}(2018){Herrera-Camus}, {Sturm},
  {Graci{\'a}-Carpio}, {Lutz}, {Contursi}, {Veilleux}, {Fischer},
  {Gonz{\'a}lez-Alfonso}, {Poglitsch}, {Tacconi}, {Genzel}, {Maiolino},
  {Sternberg}, {Davies}, \& {Verma}}]{herrera-camus18}
{Herrera-Camus}, R., {Sturm}, E., {Graci{\'a}-Carpio}, J., {et~al.} 2018, \apj,
  861, 95, \dodoi{10.3847/1538-4357/aac0f9}

\bibitem[{Hunter(2007)}]{Hunter:2007}
Hunter, J.~D. 2007, Computing In Science \& Engineering, 9, 90,
  \dodoi{10.1109/MCSE.2007.55}

\bibitem[{{Jiang} {et~al.}(2016){Jiang}, {McGreer}, {Fan}, {Strauss},
  {Ba{\~n}ados}, {Becker}, {Bian}, {Farnsworth}, {Shen}, {Wang}, {Wang},
  {Wang}, {White}, {Wu}, {Wu}, {Yang}, \& {Yang}}]{jiang16}
{Jiang}, L., {McGreer}, I.~D., {Fan}, X., {et~al.} 2016, \apj, 833, 222,
  \dodoi{10.3847/1538-4357/833/2/222}

\bibitem[{{Jorsater} \& {van Moorsel}(1995)}]{jorsater95}
{Jorsater}, S., \& {van Moorsel}, G.~A. 1995, \aj, 110, 2037,
  \dodoi{10.1086/117668}

\bibitem[{{Katz} {et~al.}(2017){Katz}, {Kimm}, {Sijacki}, \&
  {Haehnelt}}]{katz17}
{Katz}, H., {Kimm}, T., {Sijacki}, D., \& {Haehnelt}, M.~G. 2017, \mnras, 468,
  4831, \dodoi{10.1093/mnras/stx608}

\bibitem[{{Katz} {et~al.}(2019){Katz}, {Galligan}, {Kimm}, {Rosdahl},
  {Haehnelt}, {Blaizot}, {Devriendt}, {Slyz}, {Laporte}, \& {Ellis}}]{katz19}
{Katz}, H., {Galligan}, T.~P., {Kimm}, T., {et~al.} 2019, arXiv e-prints,
  arXiv:1901.01272.
\newblock \doarXiv{1901.01272}

\bibitem[{{Kennicutt}(1998)}]{kennicutt98}
{Kennicutt}, Robert~C., J. 1998, Annual Review of Astronomy and Astrophysics,
  36, 189, \dodoi{10.1146/annurev.astro.36.1.189}

\bibitem[{{Kennicutt} \& {Evans}(2012)}]{kennicutt12}
{Kennicutt}, R.~C., \& {Evans}, N.~J. 2012, \araa, 50, 531,
  \dodoi{10.1146/annurev-astro-081811-125610}

\bibitem[{{Kurk} {et~al.}(2007){Kurk}, {Walter}, {Fan}, {Jiang}, {Riechers},
  {Rix}, {Pentericci}, {Strauss}, {Carilli}, \& {Wagner}}]{kurk07}
{Kurk}, J.~D., {Walter}, F., {Fan}, X., {et~al.} 2007, \apj, 669, 32,
  \dodoi{10.1086/521596}

\bibitem[{{Liu} {et~al.}(2017){Liu}, {Wei{\ss}}, {Perez-Beaupuits},
  {G{\"u}sten}, {Liu}, {Gao}, {Menten}, {van der Werf}, {Israel}, {Harris},
  {Martin-Pintado}, {Requena-Torres}, \& {Stutzki}}]{liu17}
{Liu}, L., {Wei{\ss}}, A., {Perez-Beaupuits}, J.~P., {et~al.} 2017, \apj, 846,
  5, \dodoi{10.3847/1538-4357/aa81b4}

\bibitem[{{Malhotra} {et~al.}(1997){Malhotra}, {Helou}, {Stacey}, {Hollenbach},
  {Lord}, {Beichman}, {Dinerstein}, {Hunter}, {Lo}, {Lu}, {Rubin},
  {Silbermann}, {Thronson}, \& {Werner}}]{malhotra97}
{Malhotra}, S., {Helou}, G., {Stacey}, G., {et~al.} 1997, \apjl, 491, L27,
  \dodoi{10.1086/311044}

\bibitem[{{Maloney} {et~al.}(1996){Maloney}, {Hollenbach}, \&
  {Tielens}}]{maloney96}
{Maloney}, P.~R., {Hollenbach}, D.~J., \& {Tielens}, A.~G.~G.~M. 1996, \apj,
  466, 561, \dodoi{10.1086/177532}

\bibitem[{{Marrone} {et~al.}(2018){Marrone}, {Spilker}, {Hayward}, {Vieira},
  {Aravena}, {Ashby}, {Bayliss}, {B{\'e}thermin}, {Brodwin}, {Bothwell},
  {Carlstrom}, {Chapman}, {Chen}, {Crawford}, {Cunningham}, {De Breuck},
  {Fassnacht}, {Gonzalez}, {Greve}, {Hezaveh}, {Lacaille}, {Litke}, {Lower},
  {Ma}, {Malkan}, {Miller}, {Morningstar}, {Murphy}, {Narayanan}, {Phadke},
  {Rotermund}, {Sreevani}, {Stalder}, {Stark}, {Strandet}, {Tang}, \&
  {Wei{\ss}}}]{marrone18}
{Marrone}, D.~P., {Spilker}, J.~S., {Hayward}, C.~C., {et~al.} 2018, \nat, 553,
  51, \dodoi{10.1038/nature24629}

\bibitem[{{Matsuoka} {et~al.}(2018){Matsuoka}, {Strauss}, {Kashikawa}, {Onoue},
  {Iwasawa}, {Tang}, {Lee}, {Imanishi}, {Nagao}, {Akiyama}, {Asami}, {Bosch},
  {Furusawa}, {Goto}, {Gunn}, {Harikane}, {Ikeda}, {Izumi}, {Kawaguchi},
  {Kato}, {Kikuta}, {Kohno}, {Komiyama}, {Lupton}, {Minezaki}, {Miyazaki},
  {Murayama}, {Niida}, {Nishizawa}, {Noboriguchi}, {Oguri}, {Ono}, {Ouchi},
  {Price}, {Sameshima}, {Schulze}, {Shirakata}, {Silverman}, {Sugiyama},
  {Tait}, {Takada}, {Takata}, {Tanaka}, {Toba}, {Utsumi}, {Wang}, \&
  {Yamashita}}]{matsuoka18}
{Matsuoka}, Y., {Strauss}, M.~A., {Kashikawa}, N., {et~al.} 2018, \apj, 869,
  150, \dodoi{10.3847/1538-4357/aaee7a}

\bibitem[{{Mazzucchelli} {et~al.}(2017){Mazzucchelli}, {Ba{\~n}ados},
  {Venemans}, {Decarli}, {Farina}, {Walter}, {Eilers}, {Rix}, {Simcoe},
  {Stern}, {Fan}, {Schlafly}, {De Rosa}, {Hennawi}, {Chambers}, {Greiner},
  {Burgett}, {Draper}, {Kaiser}, {Kudritzki}, {Magnier}, {Metcalfe}, {Waters},
  \& {Wainscoat}}]{mazzucchelli17}
{Mazzucchelli}, C., {Ba{\~n}ados}, E., {Venemans}, B.~P., {et~al.} 2017, \apj,
  849, 91, \dodoi{10.3847/1538-4357/aa9185}

\bibitem[{{McMullin} {et~al.}(2007){McMullin}, {Waters}, {Schiebel}, {Young},
  \& {Golap}}]{mcmullin07}
{McMullin}, J.~P., {Waters}, B., {Schiebel}, D., {Young}, W., \& {Golap}, K.
  2007, in Astronomical Data Analysis Software and Systems XVI, ed. R.~A.
  {Shaw}, F.~{Hill}, \& D.~J. {Bell}, Vol. 376, 127

\bibitem[{{Meijerink} \& {Spaans}(2005)}]{meijerink05}
{Meijerink}, R., \& {Spaans}, M. 2005, \aap, 436, 397,
  \dodoi{10.1051/0004-6361:20042398}

\bibitem[{{Meijerink} {et~al.}(2007){Meijerink}, {Spaans}, \&
  {Israel}}]{meijerink07}
{Meijerink}, R., {Spaans}, M., \& {Israel}, F.~P. 2007, \aap, 461, 793,
  \dodoi{10.1051/0004-6361:20066130}

\bibitem[{{Oberst} {et~al.}(2006){Oberst}, {Parshley}, {Stacey}, {Nikola},
  {L{\"o}hr}, {Harnett}, {Tothill}, {Lane}, {Stark}, \& {Tucker}}]{oberst06}
{Oberst}, T.~E., {Parshley}, S.~C., {Stacey}, G.~J., {et~al.} 2006, \apjl, 652,
  L125, \dodoi{10.1086/510289}

\bibitem[{{Pereira-Santaella} {et~al.}(2017){Pereira-Santaella}, {Rigopoulou},
  {Farrah}, {Lebouteiller}, \& {Li}}]{pereira-santaella17}
{Pereira-Santaella}, M., {Rigopoulou}, D., {Farrah}, D., {Lebouteiller}, V., \&
  {Li}, J. 2017, \mnras, 470, 1218, \dodoi{10.1093/mnras/stx1284}

\bibitem[{{Price-Whelan} {et~al.}(2018){Price-Whelan}, {Sip{\H{o}}cz},
  {G{\"u}nther}, {Lim}, {Crawford}, {Conseil}, {Shupe}, {Craig}, {Dencheva},
  {Ginsburg}, {VanderPlas}, {Bradley}, {P{\'e}rez-Su{\'a}rez}, {de Val-Borro},
  {Paper Contributors}, {Aldcroft}, {Cruz}, {Robitaille}, {Tollerud},
  {Coordination Committee}, {Ardelean}, {Babej}, {Bach}, {Bachetti}, {Bakanov},
  {Bamford}, {Barentsen}, {Barmby}, {Baumbach}, {Berry}, {Biscani}, {Boquien},
  {Bostroem}, {Bouma}, {Brammer}, {Bray}, {Breytenbach}, {Buddelmeijer},
  {Burke}, {Calderone}, {Cano Rodr{\'\i}guez}, {Cara}, {Cardoso}, {Cheedella},
  {Copin}, {Corrales}, {Crichton}, {D{\textquoteright}Avella}, {Deil},
  {Depagne}, {Dietrich}, {Donath}, {Droettboom}, {Earl}, {Erben}, {Fabbro},
  {Ferreira}, {Finethy}, {Fox}, {Garrison}, {Gibbons}, {Goldstein}, {Gommers},
  {Greco}, {Greenfield}, {Groener}, {Grollier}, {Hagen}, {Hirst}, {Homeier},
  {Horton}, {Hosseinzadeh}, {Hu}, {Hunkeler}, {Ivezi{\'c}}, {Jain}, {Jenness},
  {Kanarek}, {Kendrew}, {Kern}, {Kerzendorf}, {Khvalko}, {King}, {Kirkby},
  {Kulkarni}, {Kumar}, {Lee}, {Lenz}, {Littlefair}, {Ma}, {Macleod},
  {Mastropietro}, {McCully}, {Montagnac}, {Morris}, {Mueller}, {Mumford},
  {Muna}, {Murphy}, {Nelson}, {Nguyen}, {Ninan}, {N{\"o}the}, {Ogaz}, {Oh},
  {Parejko}, {Parley}, {Pascual}, {Patil}, {Patil}, {Plunkett}, {Prochaska},
  {Rastogi}, {Reddy Janga}, {Sabater}, {Sakurikar}, {Seifert}, {Sherbert},
  {Sherwood-Taylor}, {Shih}, {Sick}, {Silbiger}, {Singanamalla}, {Singer},
  {Sladen}, {Sooley}, {Sornarajah}, {Streicher}, {Teuben}, {Thomas},
  {Tremblay}, {Turner}, {Terr{\'o}n}, {van Kerkwijk}, {de la Vega}, {Watkins},
  {Weaver}, {Whitmore}, {Woillez}, {Zabalza}, \& {Contributors}}]{astropy:2018}
{Price-Whelan}, A.~M., {Sip{\H{o}}cz}, B.~M., {G{\"u}nther}, H.~M., {et~al.}
  2018, \aj, 156, 123, \dodoi{10.3847/1538-3881/aabc4f}

\bibitem[{{R{\'e}my-Ruyer} {et~al.}(2014){R{\'e}my-Ruyer}, {Madden},
  {Galliano}, {Galametz}, {Takeuchi}, {Asano}, {Zhukovska}, {Lebouteiller},
  {Cormier}, {Jones}, {Bocchio}, {Baes}, {Bendo}, {Boquien}, {Boselli},
  {DeLooze}, {Doublier-Pritchard}, {Hughes}, {Karczewski}, \&
  {Spinoglio}}]{remy-ruyer14}
{R{\'e}my-Ruyer}, A., {Madden}, S.~C., {Galliano}, F., {et~al.} 2014, \aap,
  563, A31, \dodoi{10.1051/0004-6361/201322803}

\bibitem[{{Riechers} {et~al.}(2009){Riechers}, {Walter}, {Bertoldi}, {Carilli},
  {Aravena}, {Neri}, {Cox}, {Wei{\ss}}, \& {Menten}}]{riechers09}
{Riechers}, D.~A., {Walter}, F., {Bertoldi}, F., {et~al.} 2009, \apj, 703,
  1338, \dodoi{10.1088/0004-637X/703/2/1338}

\bibitem[{{Riechers} {et~al.}(2013){Riechers}, {Bradford}, {Clements},
  {Dowell}, {P{\'e}rez-Fournon}, {Ivison}, {Bridge}, {Conley}, {Fu}, {Vieira},
  {Wardlow}, {Calanog}, {Cooray}, {Hurley}, {Neri}, {Kamenetzky}, {Aguirre},
  {Altieri}, {Arumugam}, {Benford}, {B{\'e}thermin}, {Bock}, {Burgarella},
  {Cabrera-Lavers}, {Chapman}, {Cox}, {Dunlop}, {Earle}, {Farrah}, {Ferrero},
  {Franceschini}, {Gavazzi}, {Glenn}, {Solares}, {Gurwell}, {Halpern},
  {Hatziminaoglou}, {Hyde}, {Ibar}, {Kov{\'a}cs}, {Krips}, {Lupu}, {Maloney},
  {Martinez-Navajas}, {Matsuhara}, {Murphy}, {Naylor}, {Nguyen}, {Oliver},
  {Omont}, {Page}, {Petitpas}, {Rangwala}, {Roseboom}, {Scott}, {Smith},
  {Staguhn}, {Streblyanska}, {Thomson}, {Valtchanov}, {Viero}, {Wang},
  {Zemcov}, \& {Zmuidzinas}}]{riechers13}
{Riechers}, D.~A., {Bradford}, C.~M., {Clements}, D.~L., {et~al.} 2013, \nat,
  496, 329, \dodoi{10.1038/nature12050}

\bibitem[{{Rigopoulou} {et~al.}(2018){Rigopoulou}, {Pereira-Santaella},
  {Magdis}, {Cooray}, {Farrah}, {Marques-Chaves}, {Perez-Fournon}, \&
  {Riechers}}]{rigopoulou18}
{Rigopoulou}, D., {Pereira-Santaella}, M., {Magdis}, G.~E., {et~al.} 2018,
  \mnras, 473, 20, \dodoi{10.1093/mnras/stx2311}

\bibitem[{{Rosenberg} {et~al.}(2015){Rosenberg}, {van der Werf}, {Aalto},
  {Armus}, {Charmandaris}, {D{\'{\i}}az-Santos}, {Evans}, {Fischer}, {Gao},
  {Gonz{\'a}lez-Alfonso}, {Greve}, {Harris}, {Henkel}, {Israel}, {Isaak},
  {Kramer}, {Meijerink}, {Naylor}, {Sanders}, {Smith}, {Spaans}, {Spinoglio},
  {Stacey}, {Veenendaal}, {Veilleux}, {Walter}, {Wei{\ss}}, {Wiedner}, {van der
  Wiel}, \& {Xilouris}}]{rosenberg15}
{Rosenberg}, M.~J.~F., {van der Werf}, P.~P., {Aalto}, S., {et~al.} 2015, \apj,
  801, 72, \dodoi{10.1088/0004-637X/801/2/72}

\bibitem[{{Rybak} {et~al.}(2019){Rybak}, {Calistro Rivera}, {Hodge}, {Smail},
  {Walter}, {van der Werf}, {da Cunha}, {Chen}, {Dannerbauer}, {Ivison},
  {Karim}, {Simpson}, {Swinbank}, \& {Wardlow}}]{rybak19}
{Rybak}, M., {Calistro Rivera}, G., {Hodge}, J.~A., {et~al.} 2019, arXiv
  e-prints.
\newblock \doarXiv{1901.10027}

\bibitem[{{Sandstrom} {et~al.}(2013){Sandstrom}, {Leroy}, {Walter}, {Bolatto},
  {Croxall}, {Draine}, {Wilson}, {Wolfire}, {Calzetti}, {Kennicutt}, {Aniano},
  {Donovan Meyer}, {Usero}, {Bigiel}, {Brinks}, {de Blok}, {Crocker}, {Dale},
  {Engelbracht}, {Galametz}, {Groves}, {Hunt}, {Koda}, {Kreckel}, {Linz},
  {Meidt}, {Pellegrini}, {Rix}, {Roussel}, {Schinnerer}, {Schruba}, {Schuster},
  {Skibba}, {van der Laan}, {Appleton}, {Armus}, {Brandl}, {Gordon}, {Hinz},
  {Krause}, {Montiel}, {Sauvage}, {Schmiedeke}, {Smith}, \&
  {Vigroux}}]{sandstrom13}
{Sandstrom}, K.~M., {Leroy}, A.~K., {Walter}, F., {et~al.} 2013, \apj, 777, 5,
  \dodoi{10.1088/0004-637X/777/1/5}

\bibitem[{{Savage} \& {Sembach}(1996)}]{savage96}
{Savage}, B.~D., \& {Sembach}, K.~R. 1996, \araa, 34, 279,
  \dodoi{10.1146/annurev.astro.34.1.279}

\bibitem[{{Sijacki} {et~al.}(2015){Sijacki}, {Vogelsberger}, {Genel},
  {Springel}, {Torrey}, {Snyder}, {Nelson}, \& {Hernquist}}]{sijacki15}
{Sijacki}, D., {Vogelsberger}, M., {Genel}, S., {et~al.} 2015, \mnras, 452,
  575, \dodoi{10.1093/mnras/stv1340}

\bibitem[{{Solomon} {et~al.}(1997){Solomon}, {Downes}, {Radford}, \&
  {Barrett}}]{solomon97}
{Solomon}, P.~M., {Downes}, D., {Radford}, S.~J.~E., \& {Barrett}, J.~W. 1997,
  \apj, 478, 144, \dodoi{10.1086/303765}

\bibitem[{{Stefan} {et~al.}(2015){Stefan}, {Carilli}, {Wagg}, {Walter},
  {Riechers}, {Bertoldi}, {Green}, {Fan}, {Menten}, \& {Wang}}]{stefan15}
{Stefan}, I.~I., {Carilli}, C.~L., {Wagg}, J., {et~al.} 2015, \mnras, 451,
  1713, \dodoi{10.1093/mnras/stv1108}

\bibitem[{{Tadaki} {et~al.}(2019){Tadaki}, {Iono}, {Hatsukade}, {Kohno}, {Lee},
  {Matsuda}, {Michiyama}, {Nakanishi}, {Nagao}, {Saito}, {Tamura}, {Ueda}, \&
  {Umehata}}]{tadaki19}
{Tadaki}, K.-i., {Iono}, D., {Hatsukade}, B., {et~al.} 2019, arXiv e-prints.
\newblock \doarXiv{1903.11234}

\bibitem[{{Tielens} \& {Hollenbach}(1985)}]{tielens85}
{Tielens}, A.~G.~G.~M., \& {Hollenbach}, D. 1985, \apj, 291, 722,
  \dodoi{10.1086/163111}

\bibitem[{{van der Vlugt} \& {Costa}(2019)}]{vlugt19}
{van der Vlugt}, D., \& {Costa}, T. 2019, arXiv e-prints.
\newblock \doarXiv{1903.04544}

\bibitem[{{van der Werf} {et~al.}(2011){van der Werf}, {Berciano Alba},
  {Spaans}, {Loenen}, {Meijerink}, {Riechers}, {Cox}, {Wei{\ss}}, \&
  {Walter}}]{vanderwerf11}
{van der Werf}, P.~P., {Berciano Alba}, A., {Spaans}, M., {et~al.} 2011, \apjl,
  741, L38, \dodoi{10.1088/2041-8205/741/2/L38}

\bibitem[{{Venemans} {et~al.}(2017{\natexlab{a}}){Venemans}, {Walter},
  {Decarli}, {Ferkinhoff}, {Wei{\ss}}, {Findlay}, {McMahon}, {Sutherland}, \&
  {Meijerink}}]{venemans17b}
{Venemans}, B.~P., {Walter}, F., {Decarli}, R., {et~al.} 2017{\natexlab{a}},
  \apj, 845, 154, \dodoi{10.3847/1538-4357/aa81cb}

\bibitem[{{Venemans} {et~al.}(2017{\natexlab{b}}){Venemans}, {Walter},
  {Decarli}, {Ba{\~n}ados}, {Carilli}, {Winters}, {Schuster}, {da Cunha},
  {Fan}, {Farina}, {Mazzucchelli}, {Rix}, \& {Weiss}}]{venemans17c}
---. 2017{\natexlab{b}}, \apj, 851, L8, \dodoi{10.3847/2041-8213/aa943a}

\bibitem[{{Venemans} {et~al.}(2017{\natexlab{c}}){Venemans}, {Walter},
  {Decarli}, {Ba{\~n}ados}, {Hodge}, {Hewett}, {McMahon}, {Mortlock}, \&
  {Simpson}}]{venemans17a}
---. 2017{\natexlab{c}}, \apj, 837, 146, \dodoi{10.3847/1538-4357/aa62ac}

\bibitem[{{Volonteri}(2012)}]{volonteri12}
{Volonteri}, M. 2012, Science, 337, 544, \dodoi{10.1126/science.1220843}

\bibitem[{{Walter} \& {Brinks}(1999)}]{walter99}
{Walter}, F., \& {Brinks}, E. 1999, \aj, 118, 273, \dodoi{10.1086/300906}

\bibitem[{{Walter} {et~al.}(2008){Walter}, {Brinks}, {de Blok}, {Bigiel},
  {Kennicutt}, {Thornley}, \& {Leroy}}]{walter08}
{Walter}, F., {Brinks}, E., {de Blok}, W.~J.~G., {et~al.} 2008, \aj, 136, 2563,
  \dodoi{10.1088/0004-6256/136/6/2563}

\bibitem[{{Walter} {et~al.}(2011){Walter}, {Wei{\ss}}, {Downes}, {Decarli}, \&
  {Henkel}}]{walter11}
{Walter}, F., {Wei{\ss}}, A., {Downes}, D., {Decarli}, R., \& {Henkel}, C.
  2011, \apj, 730, 18, \dodoi{10.1088/0004-637X/730/1/18}

\bibitem[{{Walter} {et~al.}(2018){Walter}, {Riechers}, {Novak}, {Decarli},
  {Ferkinhoff}, {Venemans}, {Ba{\~n}ados}, {Bertoldi}, {Carilli}, {Fan},
  {Farina}, {Mazzucchelli}, {Neeleman}, {Rix}, {Strauss}, {Uzgil}, \&
  {Wang}}]{walter18}
{Walter}, F., {Riechers}, D., {Novak}, M., {et~al.} 2018, \apjl, 869, L22,
  \dodoi{10.3847/2041-8213/aaf4fa}

\bibitem[{{Wang} {et~al.}(2013){Wang}, {Wagg}, {Carilli}, {Walter}, {Fan},
  {Bertoldi}, {Riechers}, {Omont}, {Menten}, {Cox}, {Strauss}, \&
  {Narayanan}}]{wang13}
{Wang}, R., {Wagg}, J., {Carilli}, C.~L., {et~al.} 2013, in Molecular Gas,
  Dust, and Star Formation in Galaxies, ed. T.~{Wong} \& J.~{Ott}, Vol. 292,
  184--187

\bibitem[{{Wang} {et~al.}(2016){Wang}, {Wu}, {Neri}, {Fan}, {Walter},
  {Carilli}, {Momjian}, {Bertoldi}, {Strauss}, {Li}, {Wang}, {Riechers},
  {Jiang}, {Omont}, {Wagg}, \& {Cox}}]{wang16}
{Wang}, R., {Wu}, X.-B., {Neri}, R., {et~al.} 2016, \apj, 830, 53,
  \dodoi{10.3847/0004-637X/830/1/53}

\bibitem[{{Wei{\ss}} {et~al.}(2005){Wei{\ss}}, {Downes}, {Henkel}, \&
  {Walter}}]{weiss05}
{Wei{\ss}}, A., {Downes}, D., {Henkel}, C., \& {Walter}, F. 2005, \aap, 429,
  L25, \dodoi{10.1051/0004-6361:200400085}

\bibitem[{{Wei{\ss}} {et~al.}(2007){Wei{\ss}}, {Downes}, {Neri}, {Walter},
  {Henkel}, {Wilner}, {Wagg}, \& {Wiklind}}]{weiss07}
{Wei{\ss}}, A., {Downes}, D., {Neri}, R., {et~al.} 2007, \aap, 467, 955,
  \dodoi{10.1051/0004-6361:20066117}

\bibitem[{{Wei{\ss}} {et~al.}(2003){Wei{\ss}}, {Henkel}, {Downes}, \&
  {Walter}}]{weiss03}
{Wei{\ss}}, A., {Henkel}, C., {Downes}, D., \& {Walter}, F. 2003, \aap, 409,
  L41, \dodoi{10.1051/0004-6361:20031337}

\bibitem[{{Willott} {et~al.}(2015){Willott}, {Bergeron}, \&
  {Omont}}]{willott15}
{Willott}, C.~J., {Bergeron}, J., \& {Omont}, A. 2015, \apj, 801, 123,
  \dodoi{10.1088/0004-637X/801/2/123}

\bibitem[{{Zanella} {et~al.}(2018){Zanella}, {Daddi}, {Magdis}, {Diaz Santos},
  {Cormier}, {Liu}, {Cibinel}, {Gobat}, {Dickinson}, {Sargent}, {Popping},
  {Madden}, {Bethermin}, {Hughes}, {Valentino}, {Rujopakarn}, {Pannella},
  {Bournaud}, {Walter}, {Wang}, {Elbaz}, \& {Coogan}}]{zanella18}
{Zanella}, A., {Daddi}, E., {Magdis}, G., {et~al.} 2018, \mnras, 481, 1976,
  \dodoi{10.1093/mnras/sty2394}

\end{thebibliography}

\appendix

\section{Measuring resolved emission in interferometric maps} \label{sec:residual_scaling}

The difficulty of properly interpreting the area of the synthesized beam in interferometric maps can have a detrimental effect on the accuracy of the measured flux density, especially for significantly resolved emission. A Fourier transform of measured visibilities is called a dirty map, where each pixel has a unit of Jy per dirty beam. The integral under the entire dirty beam is zero, because an integral of a single sine wave is zero, and the dirty beam is just a sum of various sine waves.
During the cleaning process, the dirty beam is deconvolved from the map and replaced with a Gaussian beam with a properly defined integral. The cleaning process progresses down to some chosen threshold, and the final cleaned interferometric map is then a combination of the residual given in original units of Jy per dirty beam, and a cleaned component in units of Jy per clean beam (chosen to be a Gaussian). The end result is that units in the map are ill defined, and if the observed source is spread over multiple beams, the cumulative effect can be significant.

Several choices are available to tackle this problem. The first one is to model the source using visibilities directly, without invoking the Fourier transform and going into the image plane. This works well for objects exhibiting some sort of symmetry, but is less useful for morphologically complex systems such as mergers, where a lot of details are hidden in the data coming from high-resolution/longer baselines. 
The second choice is to clean very deep, which comes at a cost of treating every noise peak (both positive and negative) as emission. The third choice, which we employ in this paper, involves estimating the proper area of the dirty beam in the region of interest, as the dirty beam area is zero when integrated across all space, but is different from zero in a smaller region. This method is called the residual scaling and is outlined in appendix A.2 of \cite{jorsater95}, see also \cite{walter99, walter08}. We summarize it here briefly.

We can consider the true flux $G$ of a source as a sum of cleaned flux $C$ in proper Jy per beam units, and the residual $R$, which has to be scaled to proper units in order to make the sum viable: $G=C+\epsilon R$. The parameter $\epsilon$ corresponds to the clean beam to dirty beam area ratio. If the same map is cleaned down to two different thresholds, we have $G=C_1+\epsilon R_1=C_2+\epsilon R_2$, because the true flux must be recovered independent of the cleaning threshold. We can also choose not to clean, in which case $C_2=0$ and $R_2$ is the dirty map. Solving equations in this scenario yields $\epsilon=C_1/(R_2-R_1)$ and $G=\epsilon R_2$. These calculations are performed within some aperture, and use the dirty map $R_2$, the clean flux map $C_1$ and the residual map $R_1$ to measure the flux density $G$.
A drawback of the residual scaling method arises in situations when $R_1\approx R_2$, or when there is not enough clean flux $C_1$, and the solution becomes numerically unstable.

\begin{figure}
	\centering
	\includegraphics[width=0.45\linewidth]{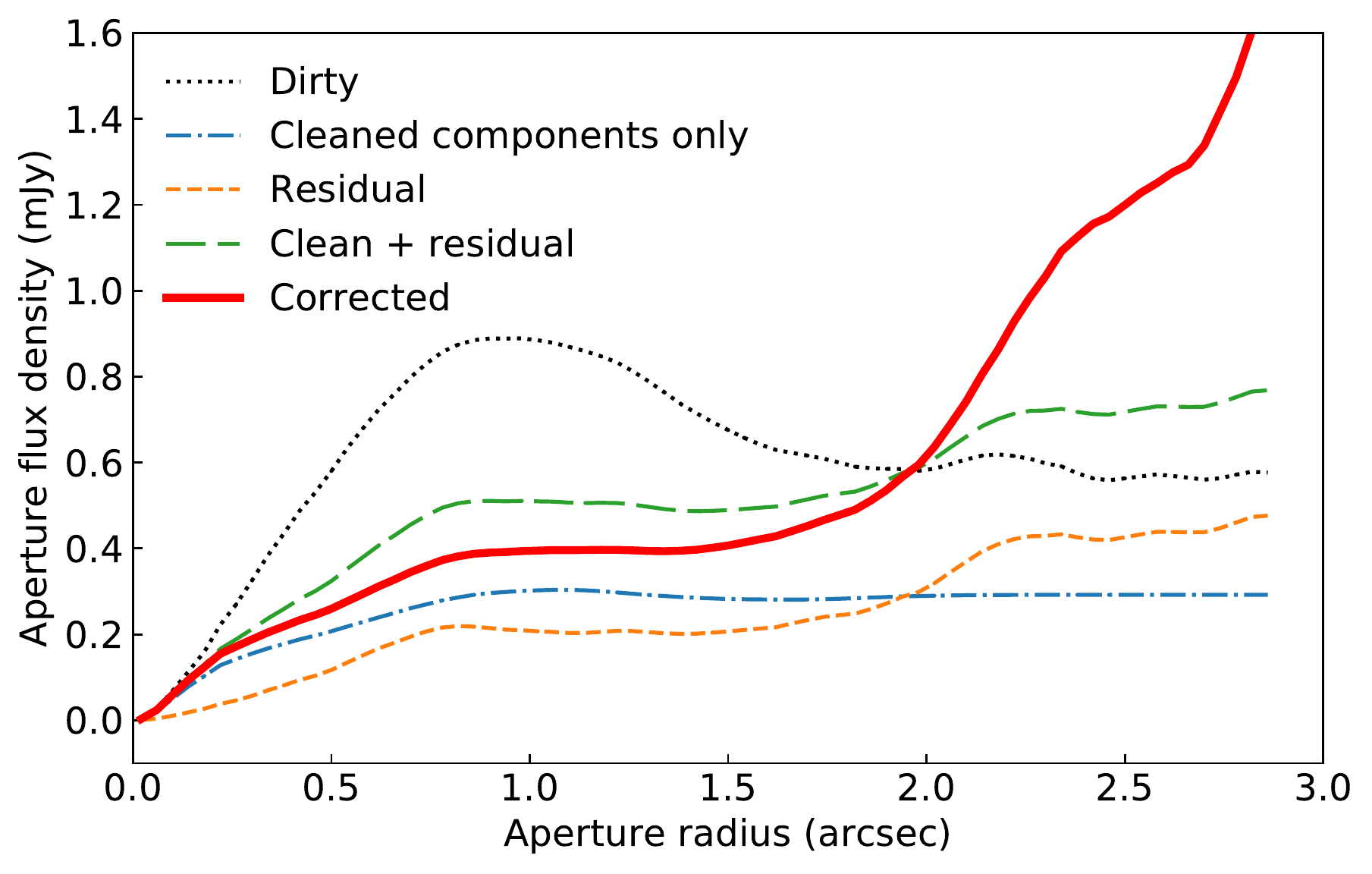}
	\includegraphics[width=0.45\linewidth]{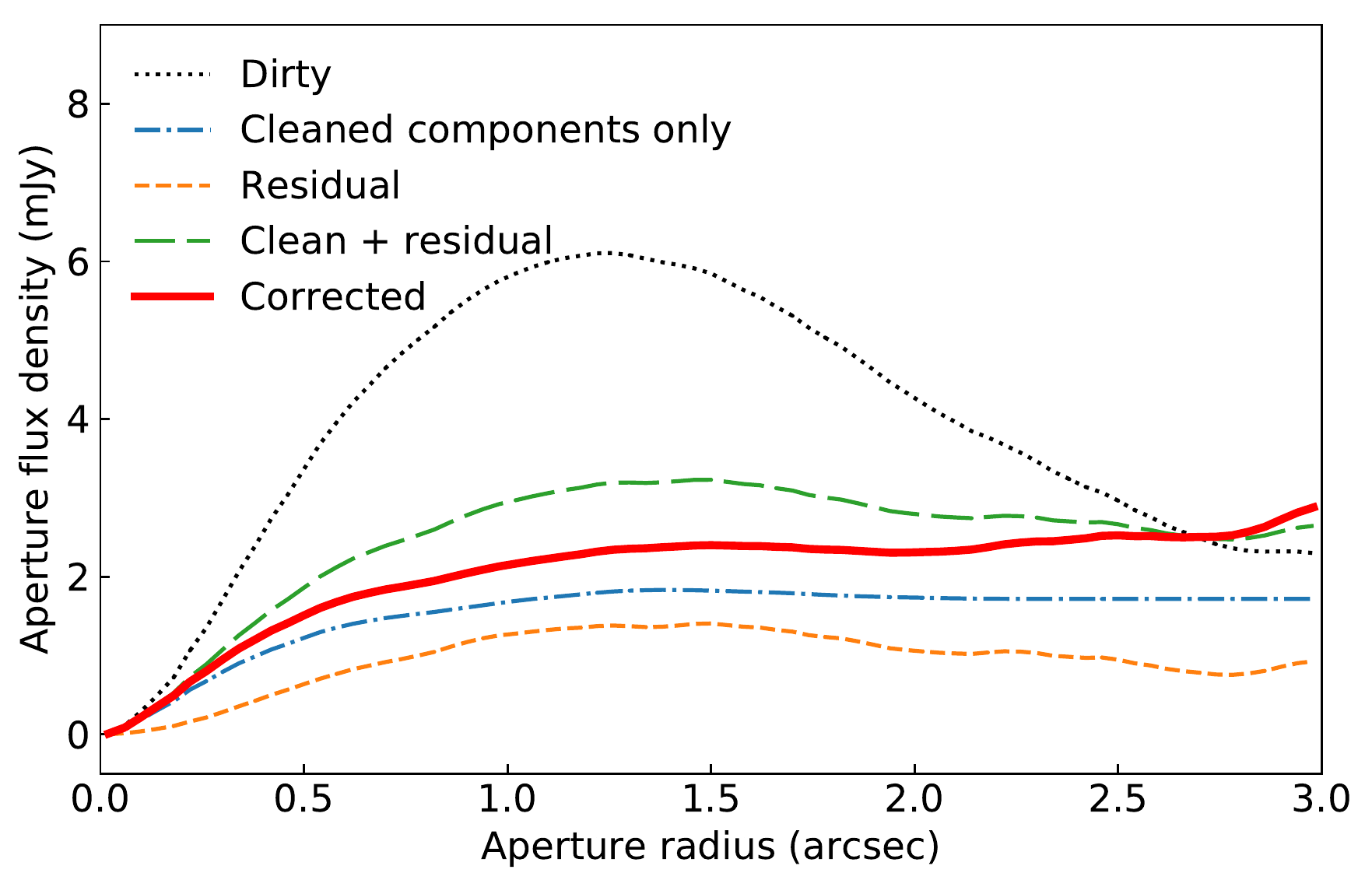}
	\caption{Aperture flux density of \pisco\ observed with high resolution measured in various interferometric maps. Cleaning was performed down to $2\sigma$ in a $2\arcsec$ radius circle. The clean beam size is used to convert from Jy per beam into Jy for all measurements. 
		The green line represent flux density measured in the usual cleaned interferometric map. The red line demonstrates the effect of residual scaling correction. \textit{Left: } Continuum measurement at 231.5\,GHz. \textit{Right: } Moment zero measurement of \cii\ integrated over 455\,\kms.}
	\label{fig:aper_hres}
\end{figure}

\begin{figure}
	\centering
	\includegraphics[width=0.45\linewidth]{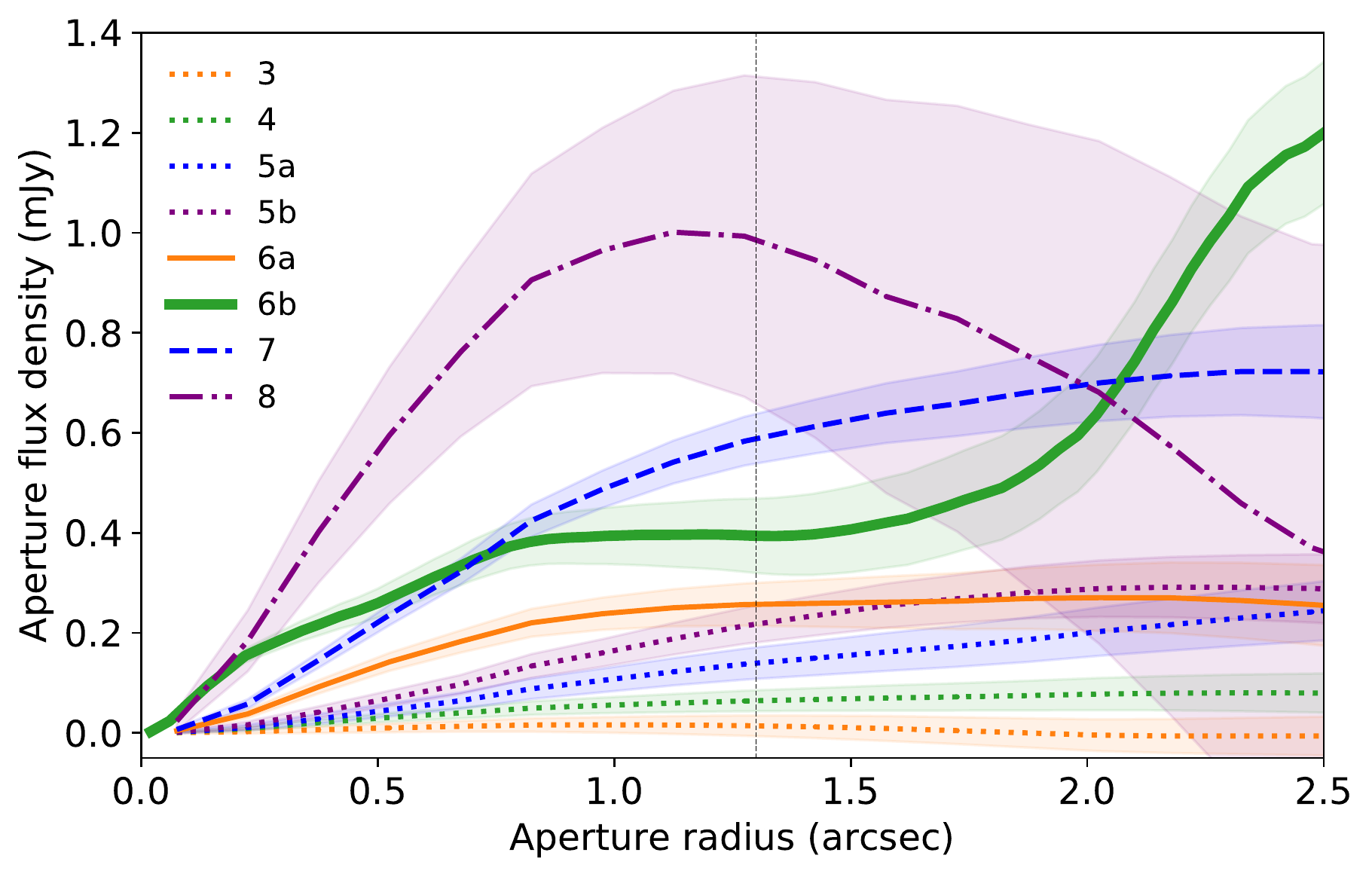}
	\includegraphics[width=0.45\linewidth]{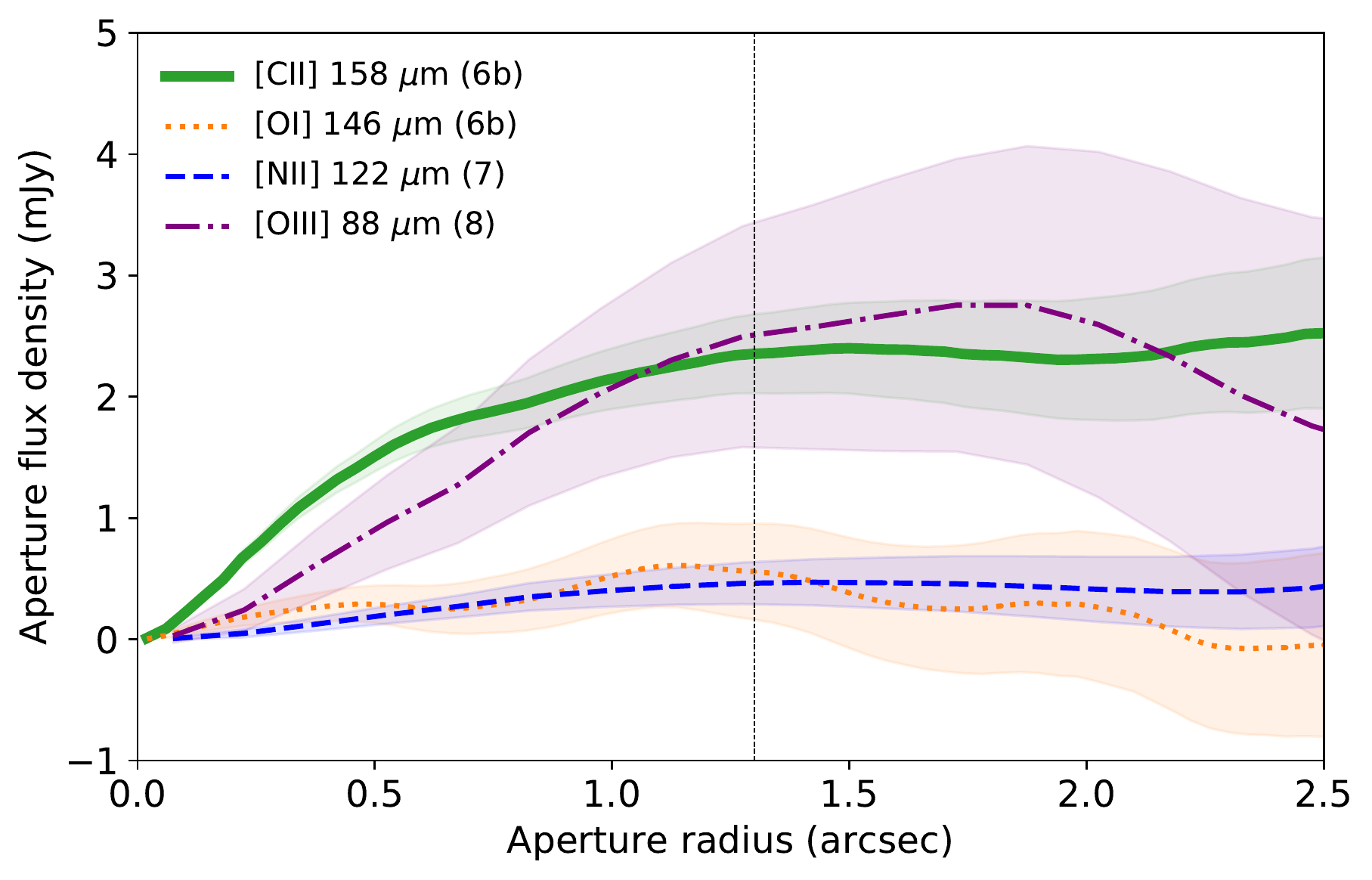}
	\caption{Flux density integrated inside an aperture as a function of its radius corrected with residual scaling. \textit{Left:} Continuum bands (see Table~\ref{tab:obs}). \textit{Right:} Atomic fine structure lines.
		The vertical line shows the aperture used throughout the paper.}
	\label{fig:aper}
\end{figure}

In Fig.~\ref{fig:aper_hres} we demonstrate this method on our high resolution ($\sim0.2\arcsec$) observations of \pisco\ at 231.5\,GHz and the \cii\ line. The map was cleaned down to $2\sigma$ in a $2\arcsec$ radius circle centered at the source. The flux density measured inside an aperture of $r=1.3\arcsec$ from the usual map (clean components added on top of the residual) is 25\% and 35\% larger than the corrected one for the continuum and the line, respectively. The effect is more pronounced in sources with more extended emission and lower surface brightness, where a lot of flux remains in the residual. The numerical instability is evident in the continuum measurements at radii above $2\arcsec$, as both the dirty map and the residual map measurements approach the same value.

In Fig.~\ref{fig:aper} we show the residual scaling corrected aperture fluxes for different continuum bands and atomic fine structure lines. Here we motivate the aperture radius with a diameter of $2.6\arcsec$ (radius of $1.3\arcsec$). The \cii emission is detected at the highest S/N and is the most extended, the flux density reaches a plateau at $r=1.3\arcsec$.
The chosen aperture size recovers the flux density measured at the plateau within the error bars in all of the maps.
We therefore opt for the same aperture in all measurements for the sake of consistent interpretation of probed spatial scales of line ratios, despite observations being at different resolutions.

\section{Spectra and non-detections of \pisco} \label{sec:non_det}
\begin{figure}
	\centering
	\includegraphics[width=0.49\linewidth]{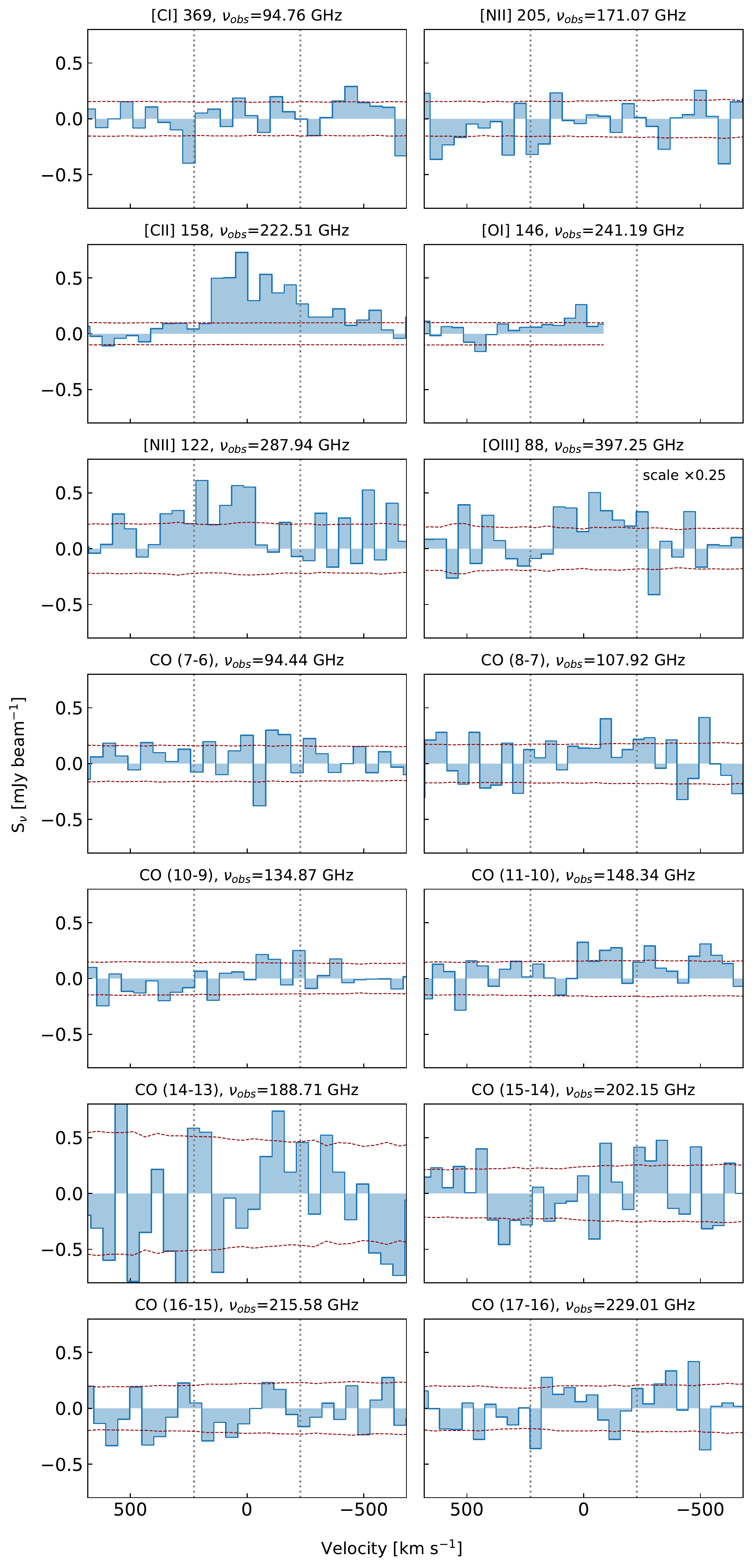} % height=\textheight
	\includegraphics[width=0.49\linewidth]{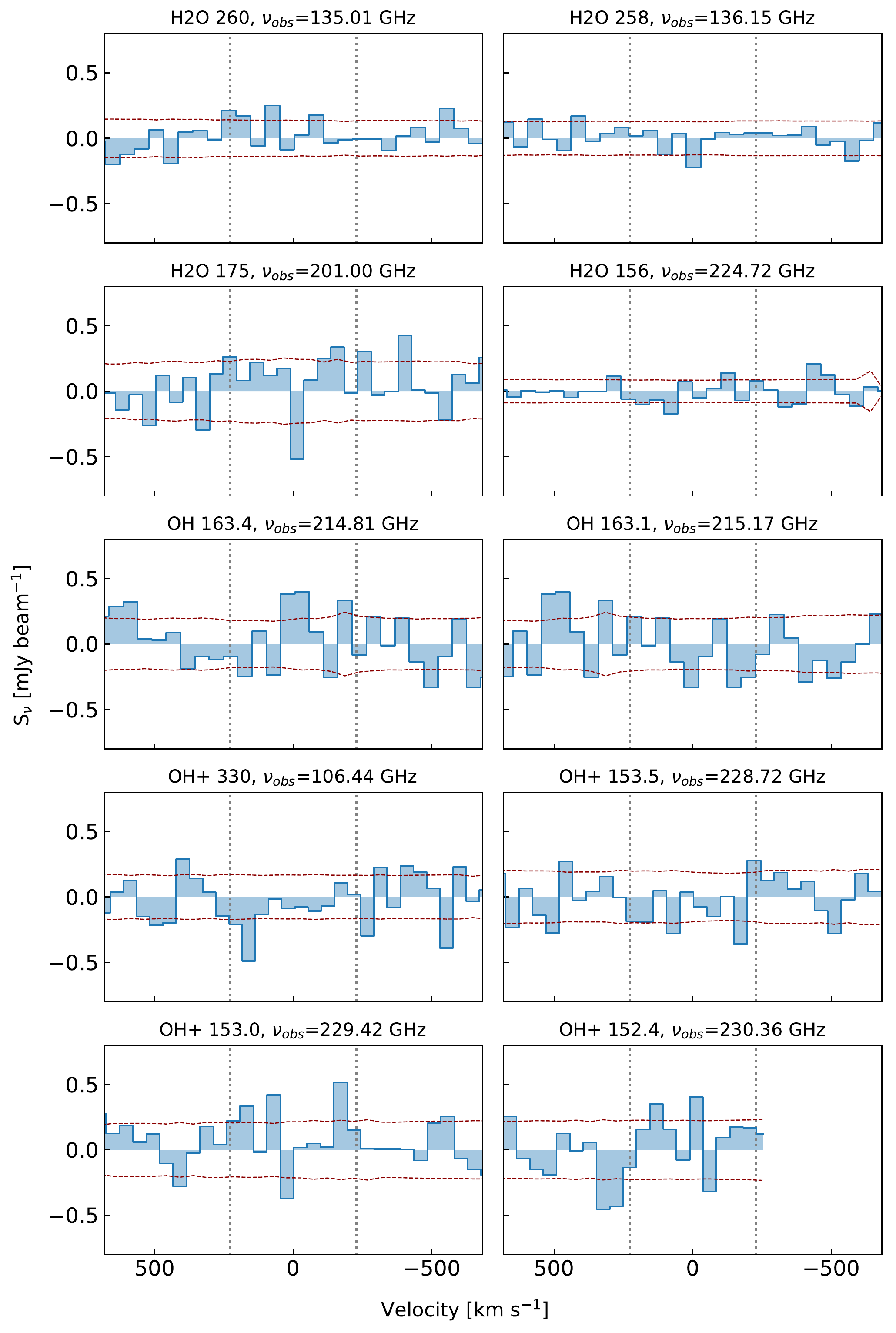} 
	
	\caption{Single pixel spectra of atomic fine structure lines and CO lines (left two columns) and  H$_2$O, OH and OH$^+$ (right two columns) of \pisco, extracted from the continuum subtracted cubes at the coordinate of the high resolution  231.5\,GHz  continuum peak. The red lines show the rms measured inside each, approximately $50$\,km\,s$^{-1}$ wide, channel. Vertical dashed lines encompass the 455\,km\,s$^{-1}$ width used to measure the line flux (see text for details). Zero velocity corresponds to the redshifted line emission frequency ($z=7.5413$).}
	\label{fig:spectra}
\end{figure}

In order to properly constrain the width of a spectral line, at least $6\sigma$ is required in the central channel, so that the FWHM can be computed at a $3\sigma$ significance. All of our lines, except \cii, have less than $5\sigma$ in the moment zero map integrated over 455\,\kms, and therefore have insufficient S/N to fit a line profile in a meaningful way. However, since this work is a multi-line spectral survey of an object, we show spectra of all considered lines in Fig.~\ref{fig:spectra} for completeness. The main goal of these spectra is to demonstrate the excess emission in individual channels. To maximize the S/N they are extracted from a single pixel.
%This measurement does not collect all the flux in the high resolution map, however the resolved \cii\ emission is a topic of an accompanying paper (\banados\ et al. in prep.).

\begin{figure}

\includegraphics[width=0.5\linewidth]{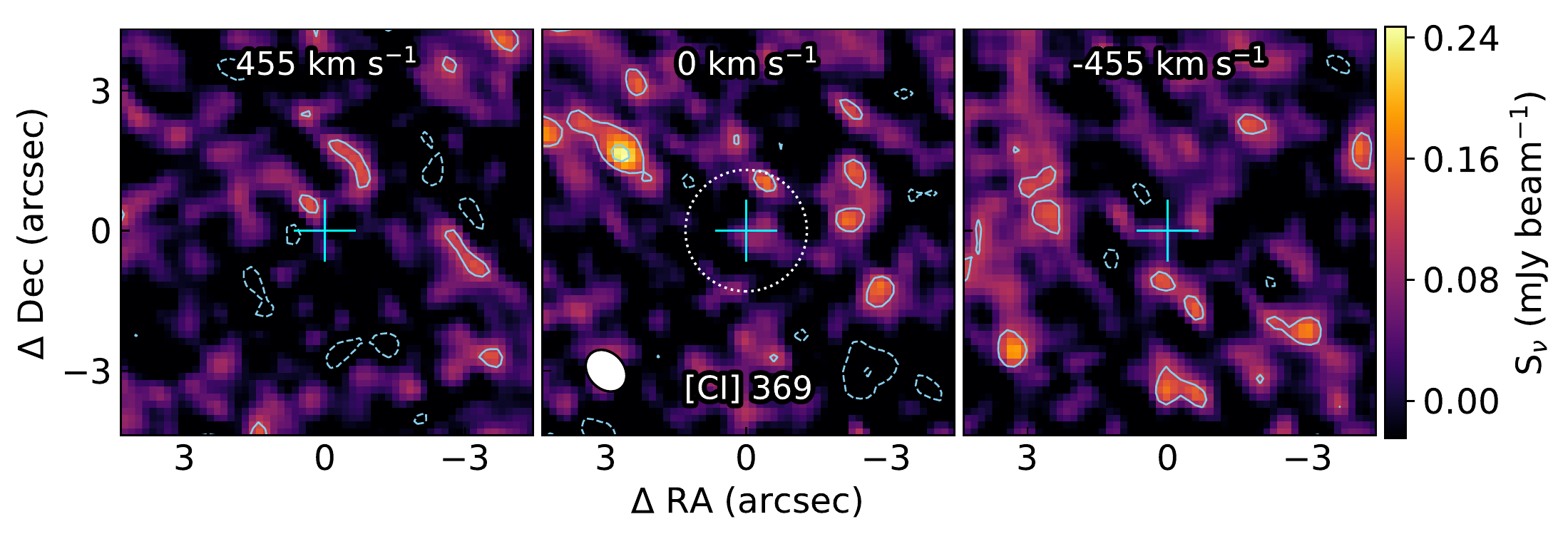}
\includegraphics[width=0.5\linewidth]{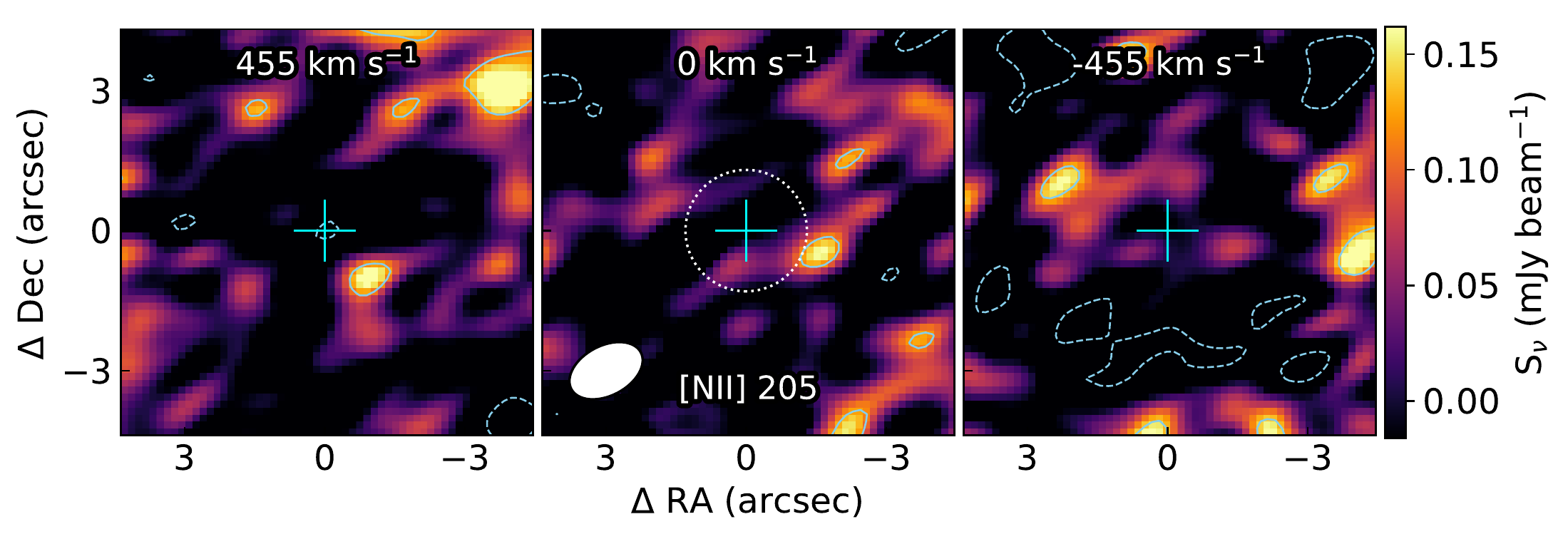}
\includegraphics[width=0.5\linewidth]{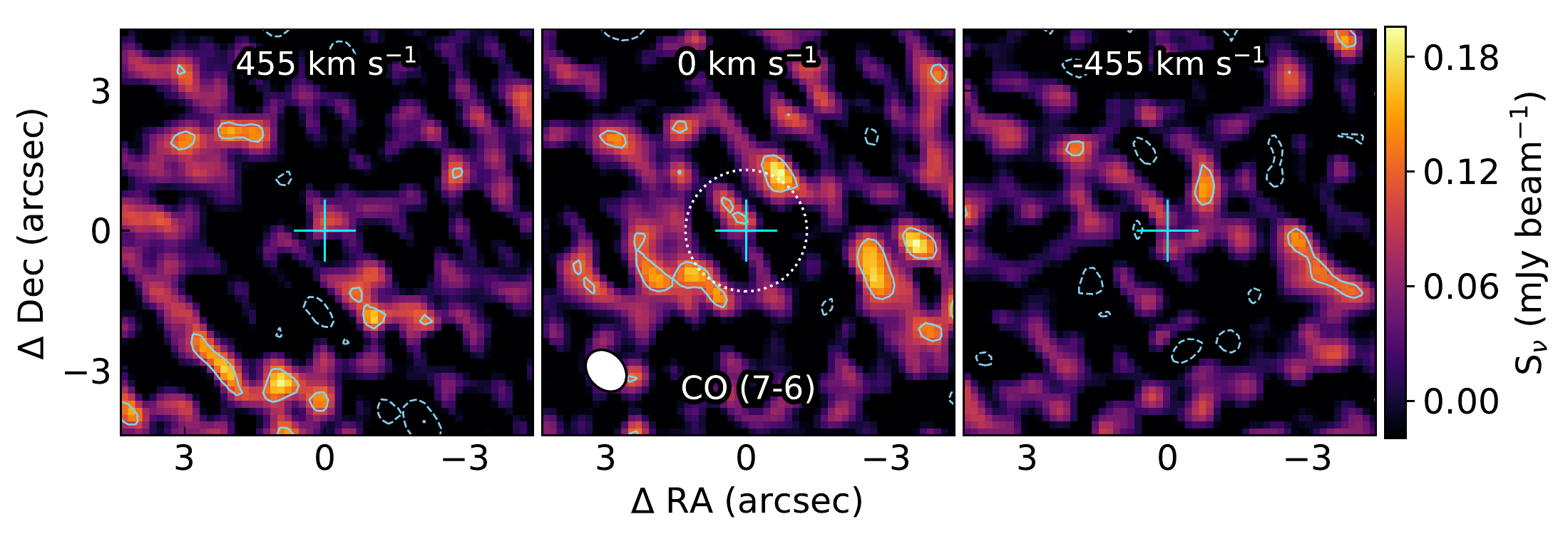}
\includegraphics[width=0.5\linewidth]{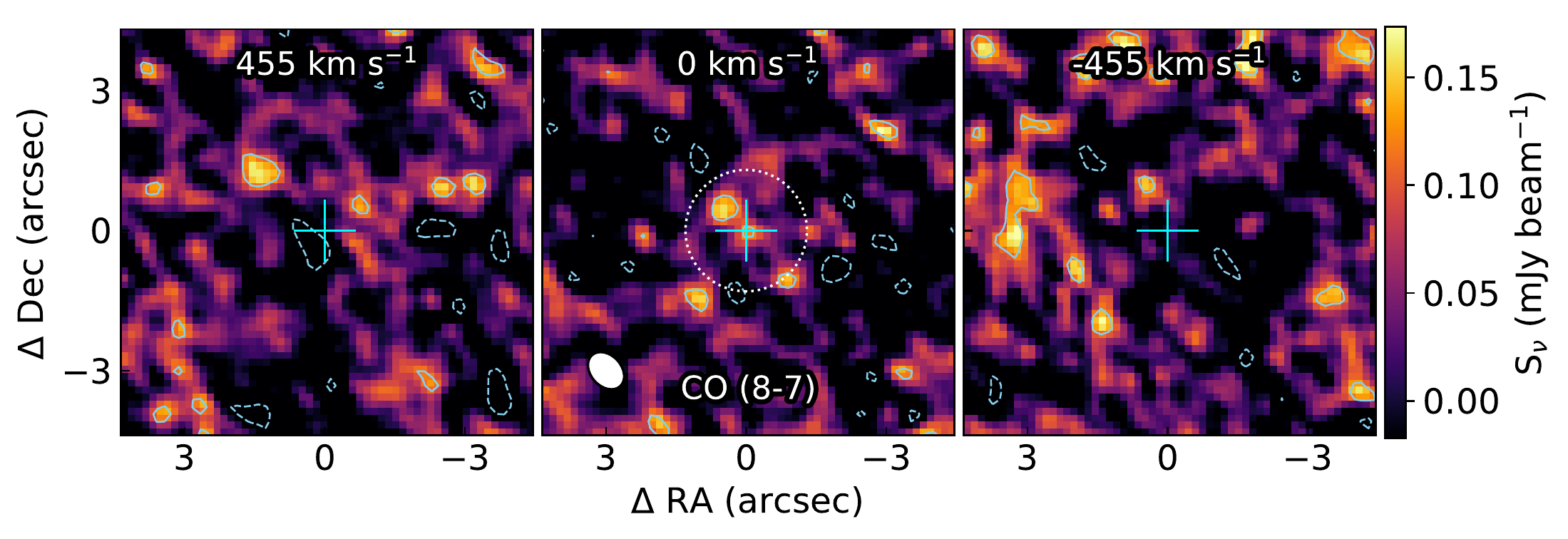}
\includegraphics[width=0.5\linewidth]{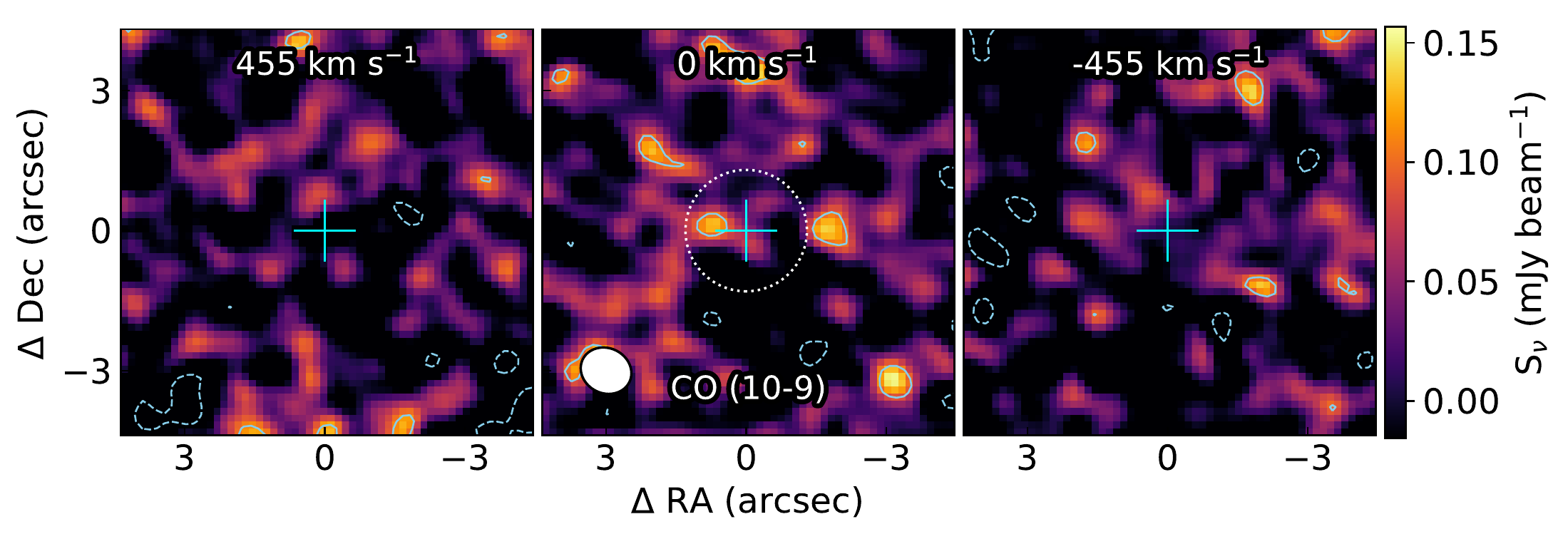}
\includegraphics[width=0.5\linewidth]{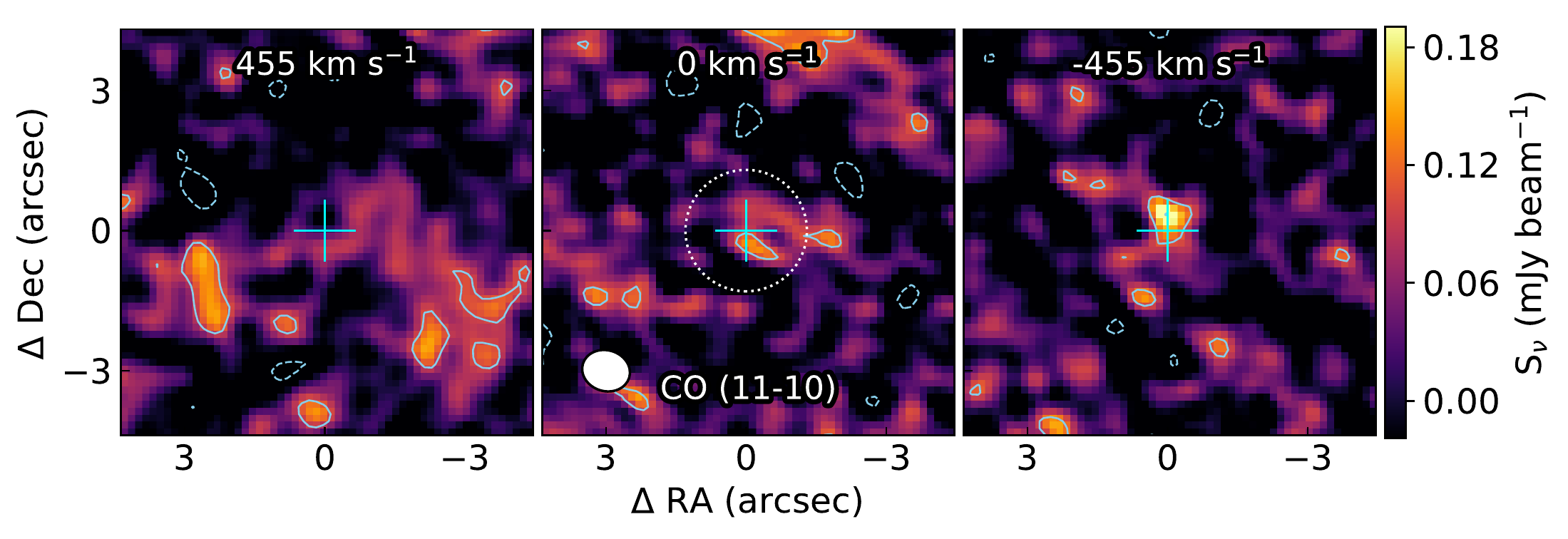}
\includegraphics[width=0.5\linewidth]{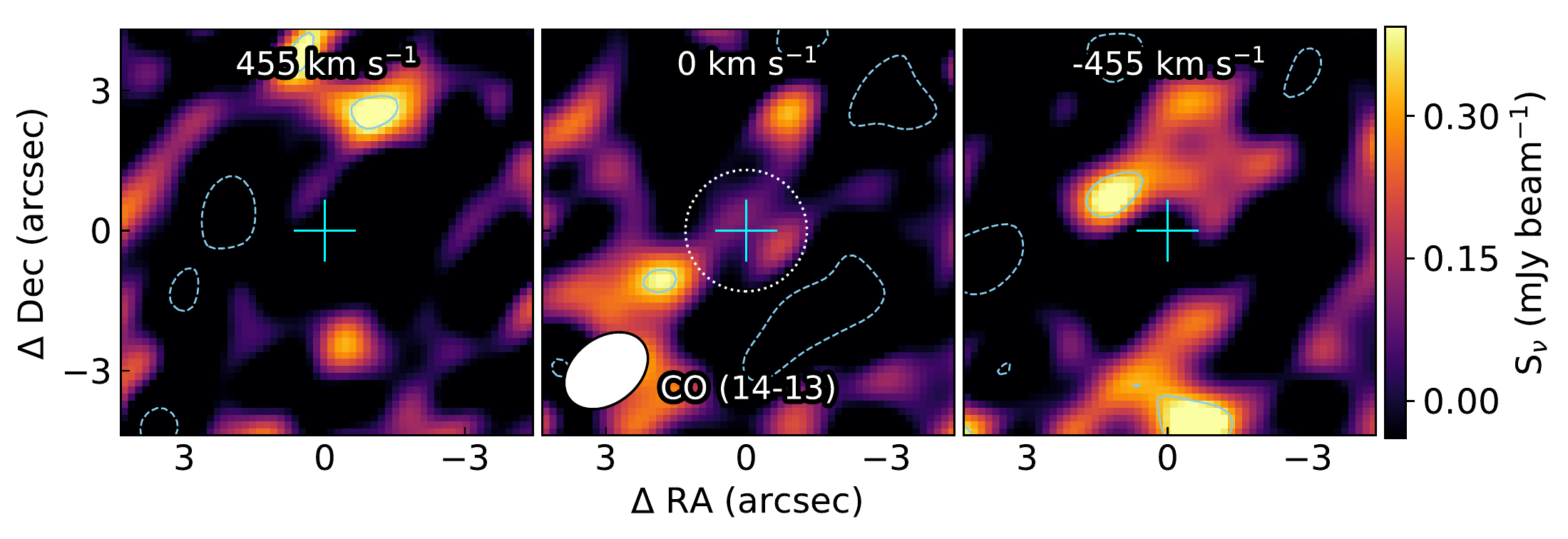}
\includegraphics[width=0.5\linewidth]{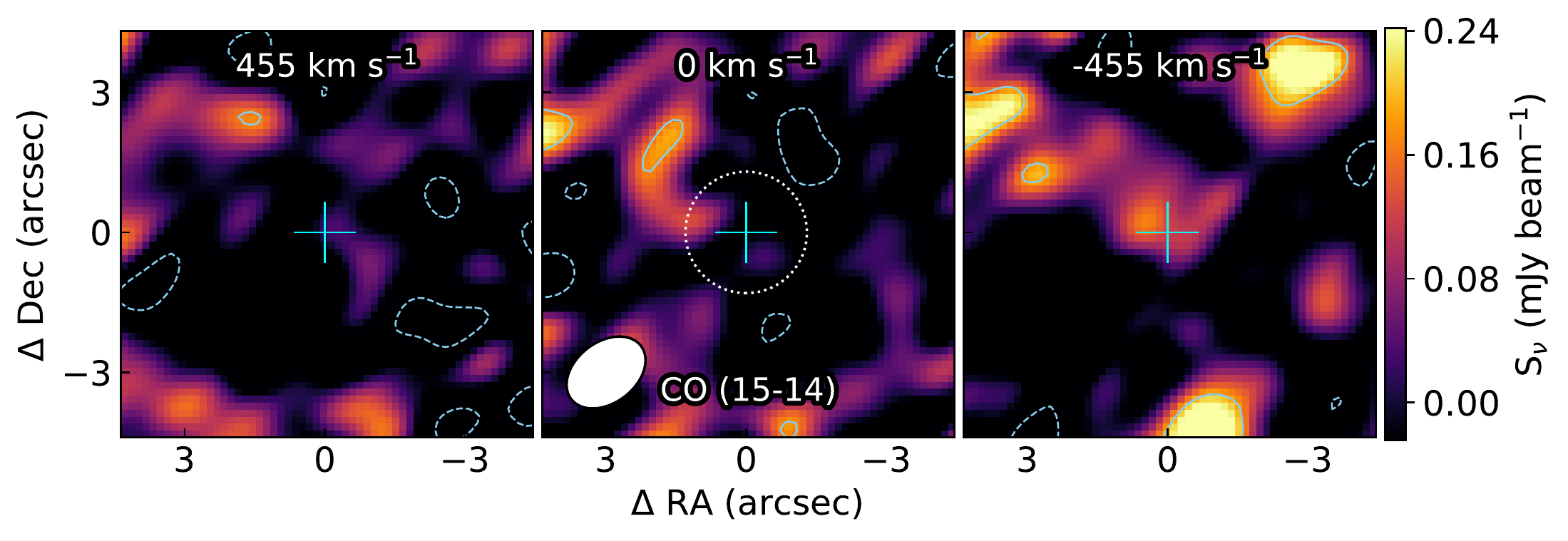}
\includegraphics[width=0.5\linewidth]{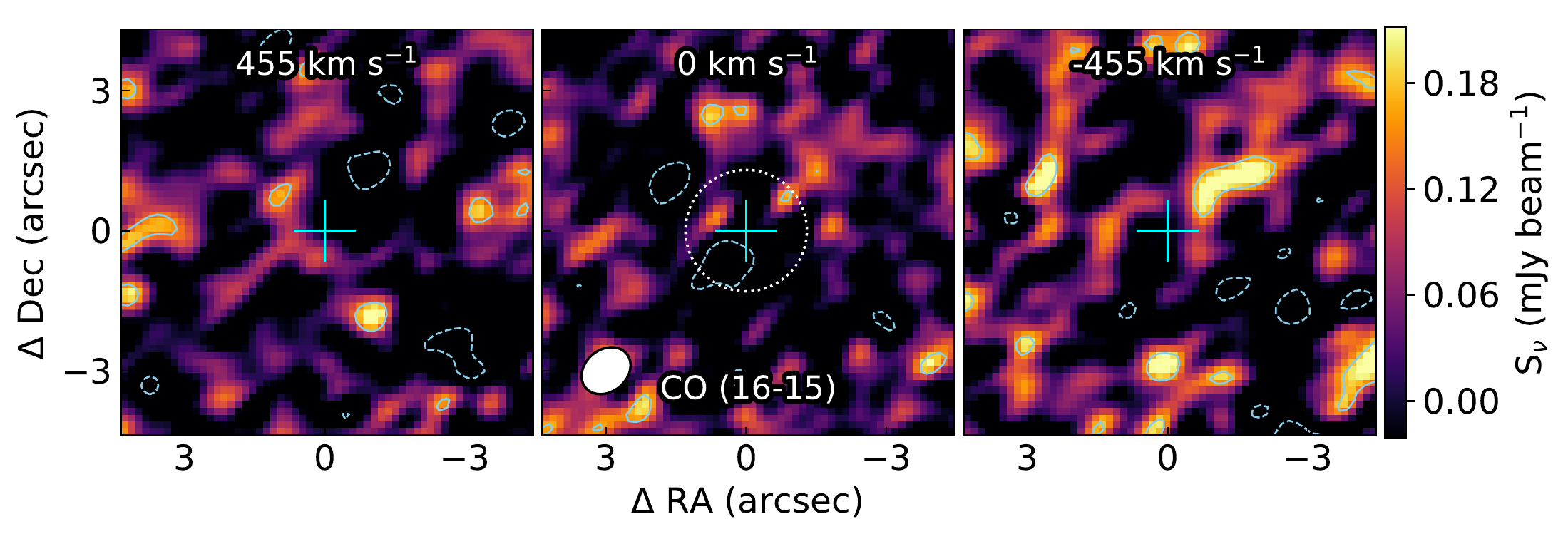}
\includegraphics[width=0.5\linewidth]{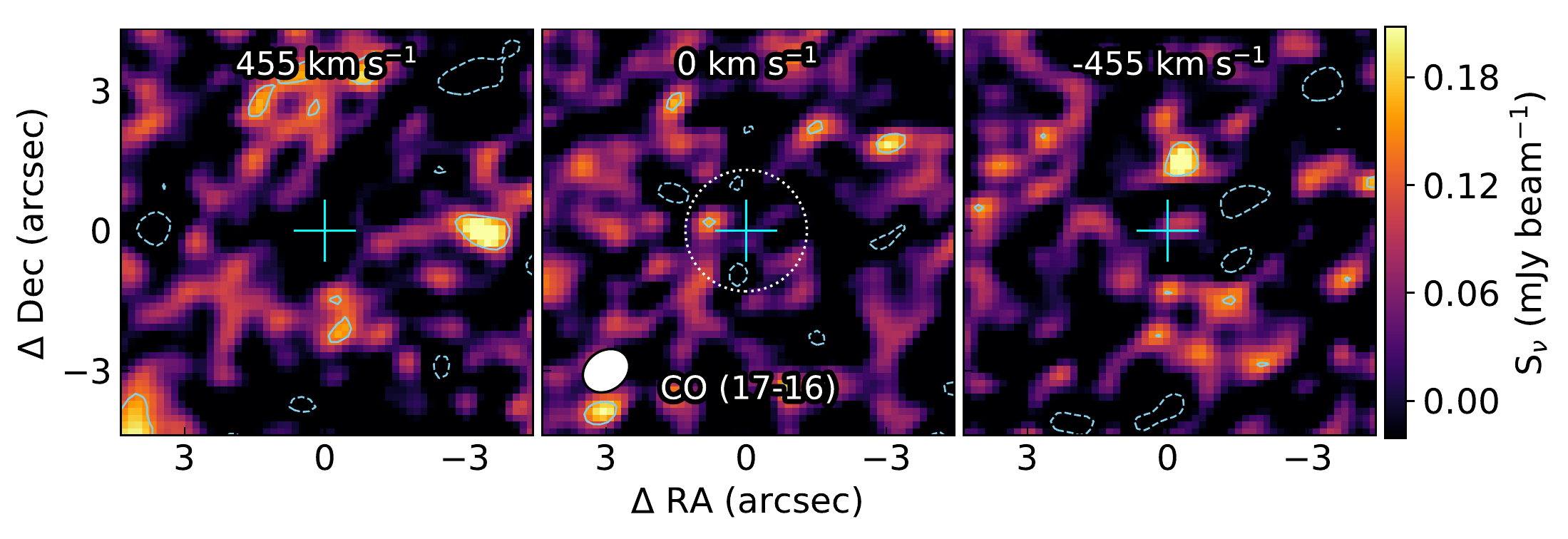}

	\caption{Non detected atomic fine structure lines and CO lines in \pisco. 
			Full (dashed) contours represent the +(-) $2\sigma$, $4\sigma$ emission significance.  The cross marks the dust continuum emission center in the high resolution data (same as in Fig. \ref{fig:line_det}).}
	\label{fig:line_nondet}
\end{figure}

\begin{figure}
	\includegraphics[width=0.5\linewidth]{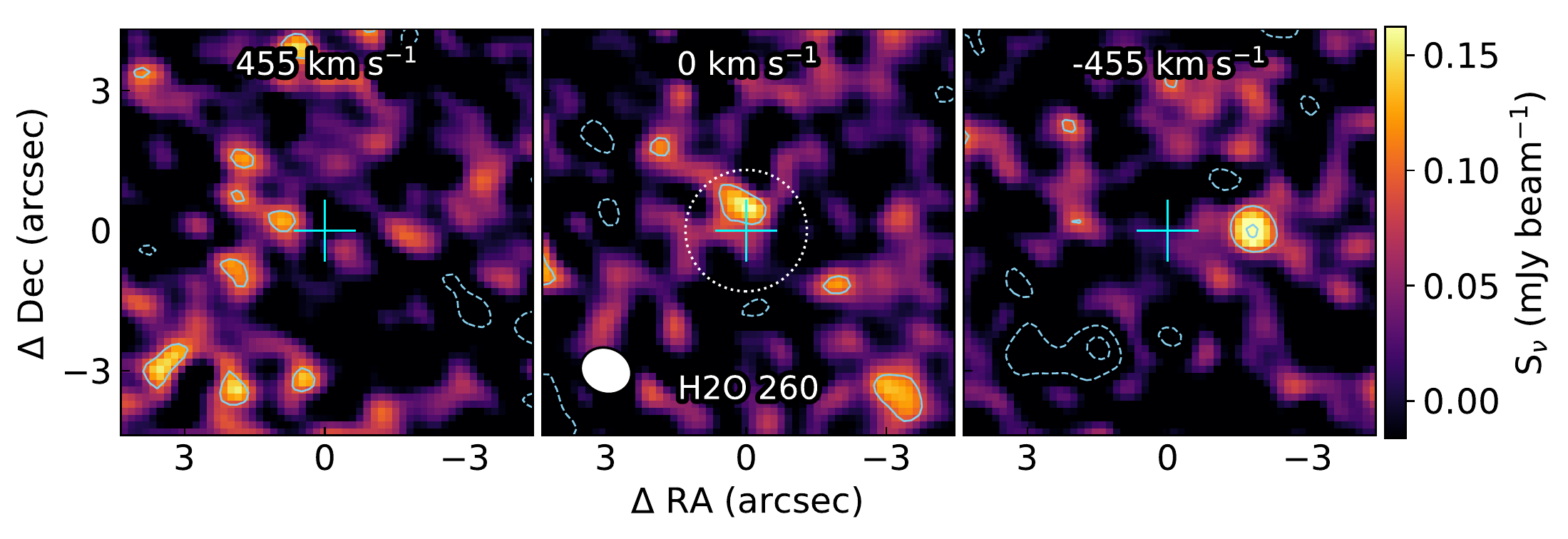}
	\includegraphics[width=0.5\linewidth]{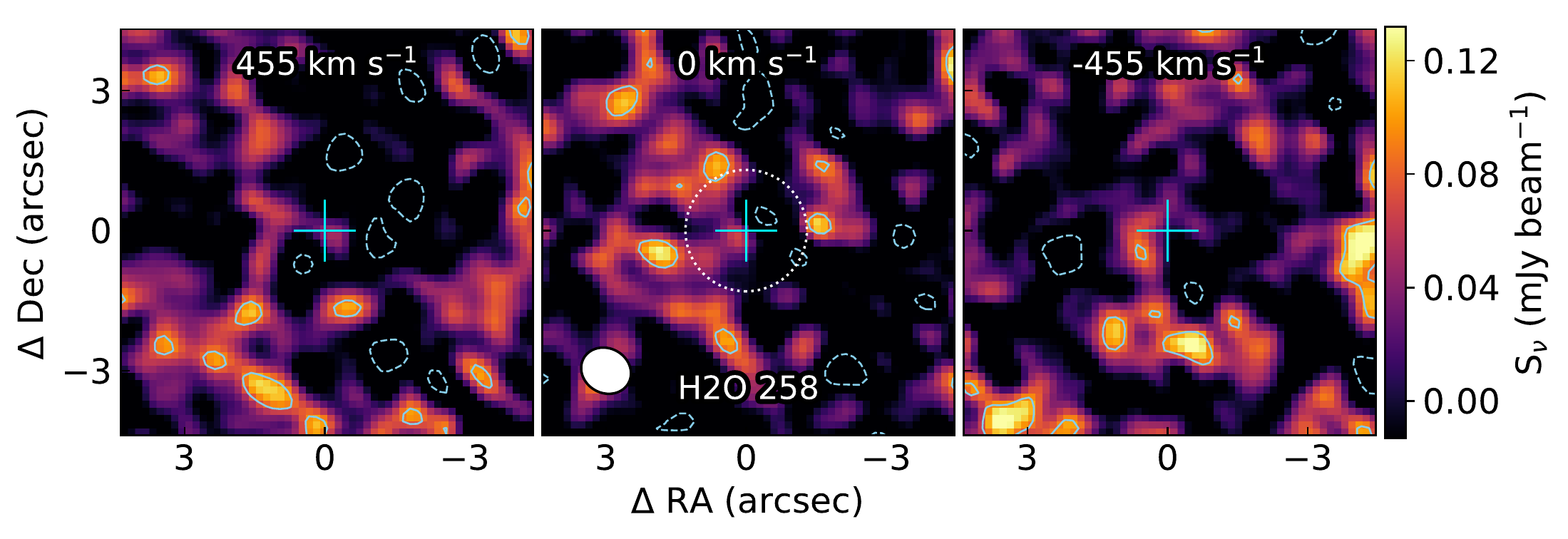}
	\includegraphics[width=0.5\linewidth]{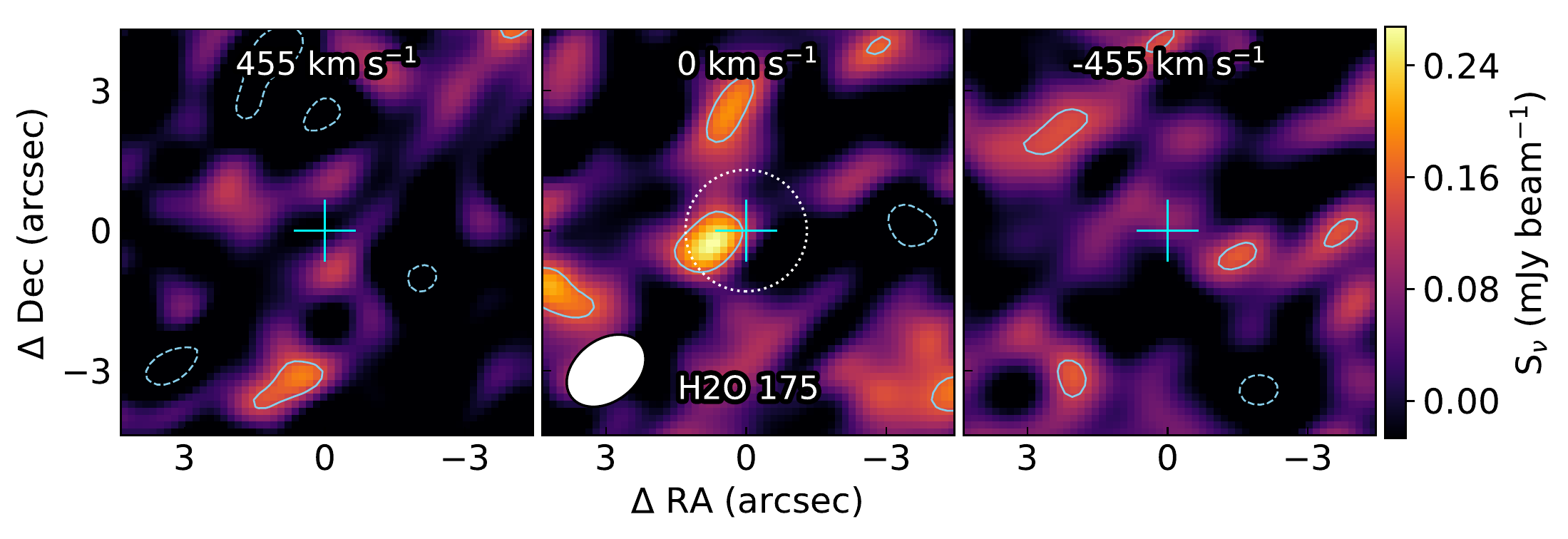}
	\includegraphics[width=0.5\linewidth]{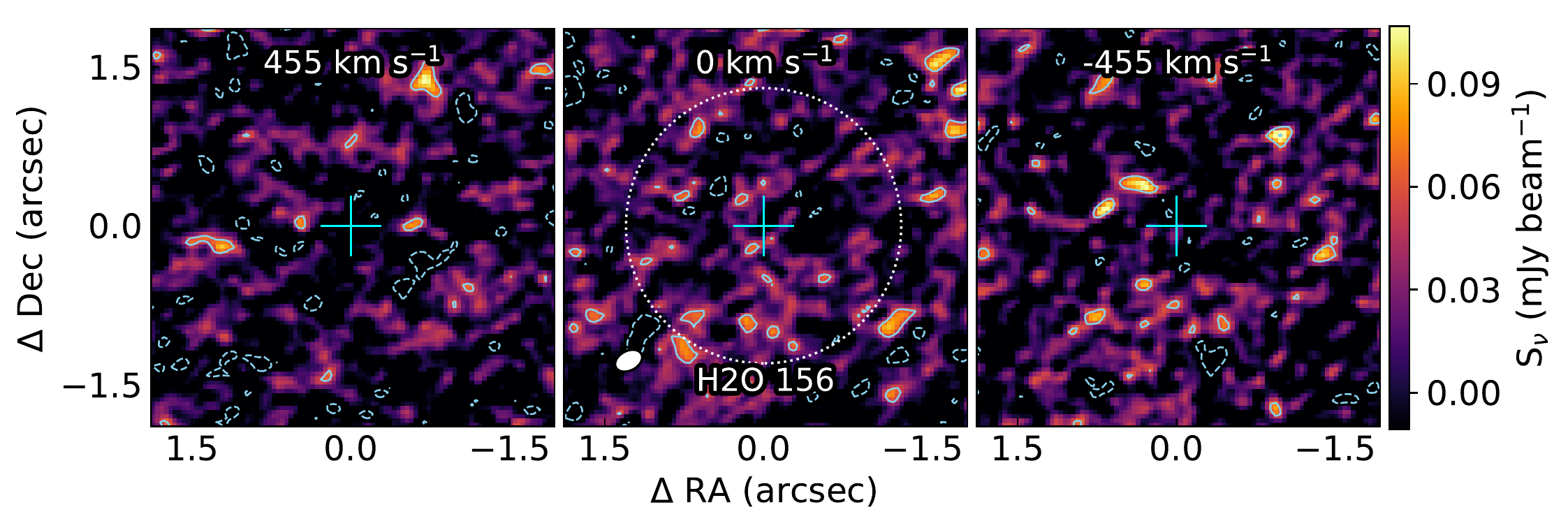}
	\includegraphics[width=0.5\linewidth]{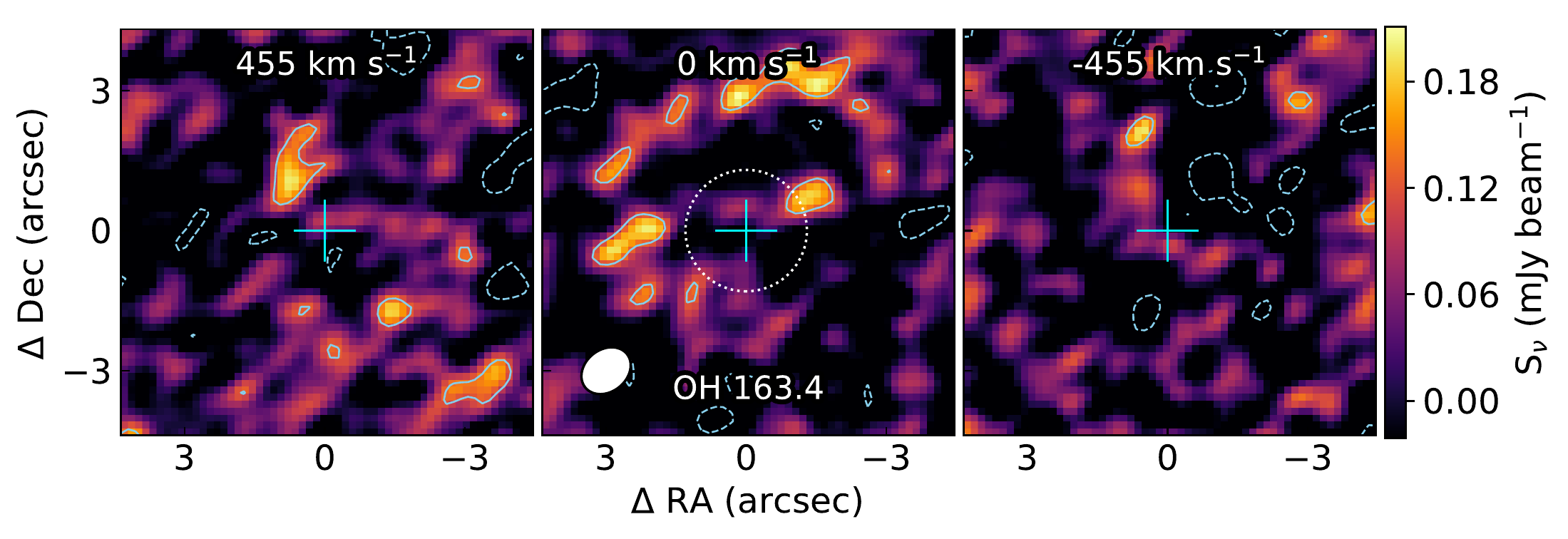}
	\includegraphics[width=0.5\linewidth]{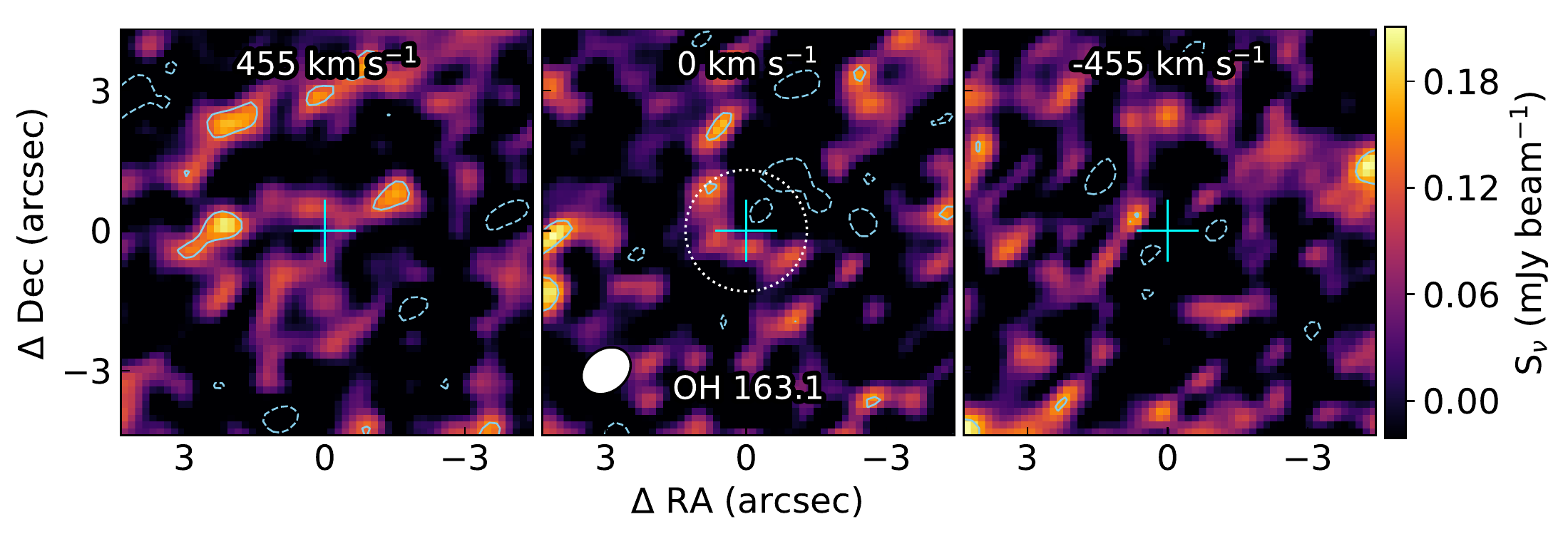}
	\includegraphics[width=0.5\linewidth]{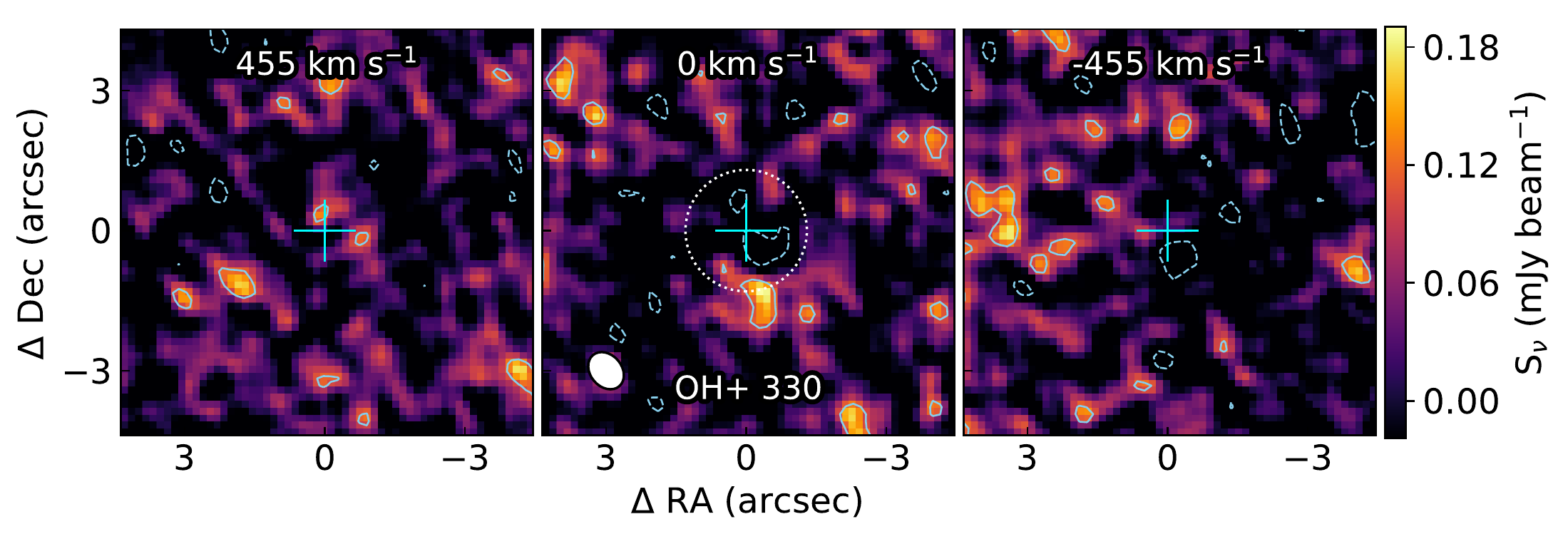}
	\includegraphics[width=0.5\linewidth]{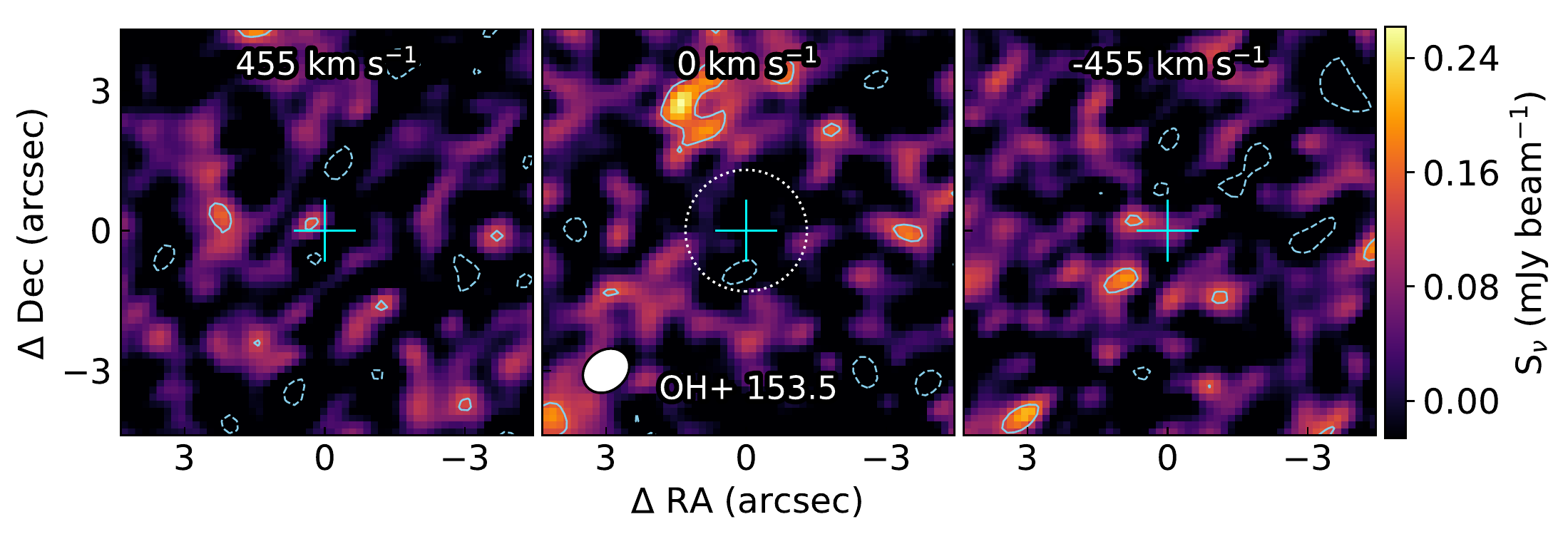}
	\includegraphics[width=0.5\linewidth]{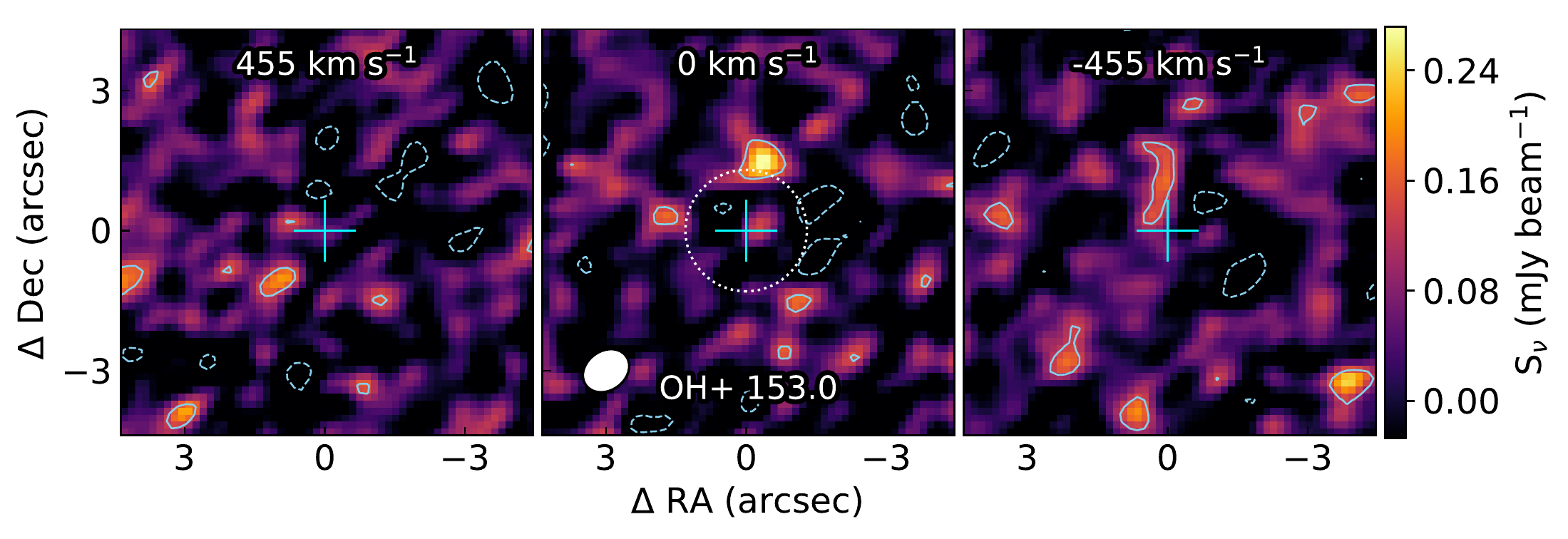}
	\includegraphics[width=0.5\linewidth]{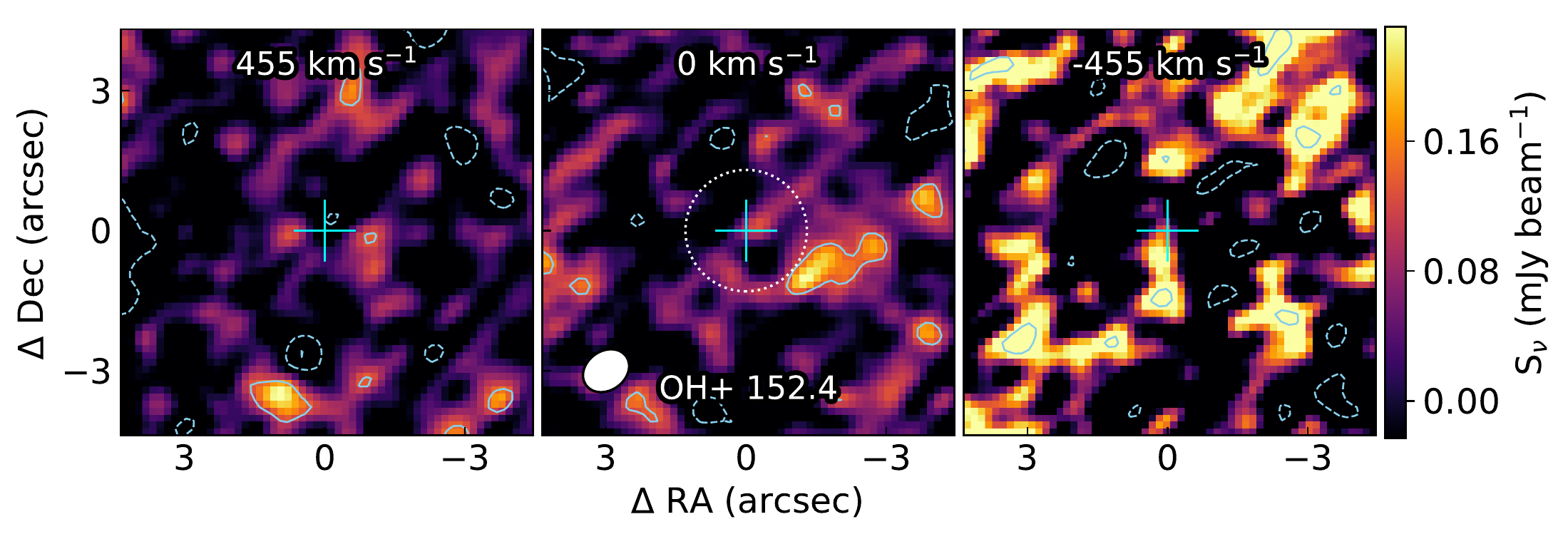}
	
	\caption{Non detected H$_2$O, OH and OH$^+$ emission lines in \pisco. 	Full (dashed) contours represent the +(-) $2\sigma$, $4\sigma$ emission significance.  The cross marks the dust continuum emission center in the high resolution data (same as in Fig. \ref{fig:line_det}). }
	\label{fig:line_nondet2}
\end{figure}

Also for completeness, we show integrated moment zero maps of our  emission line non-detections in Figs.~\ref{fig:line_nondet} and \ref{fig:line_nondet2}.

\end{document}